\documentclass[aps,prb,reprint,superscriptaddress,nofootinbib,floatfix,longbibliography]{revtex4-2}

\usepackage{amsmath}
\usepackage{amsfonts}
\usepackage{newtxtext}
\usepackage{newtxmath}
\usepackage{mathtools}
\usepackage[mathcal]{eucal}
\usepackage{dsfont}
\usepackage{bm}
\usepackage{braket}

\usepackage[percent]{overpic}

\usepackage{tabularray}
\UseTblrLibrary{booktabs}

\usepackage[version=4]{mhchem}

\usepackage{xcolor}
\usepackage[bookmarksnumbered]{hyperref}
\definecolor{cblue}{RGB}{55,126,184}
\hypersetup{colorlinks=true,linkcolor=cblue,citecolor=cblue,urlcolor=cblue}

\usepackage[capitalize]{cleveref}

\newcommand{\bq}{\bm{q}}
\DeclareMathOperator{\Tr}{Tr}

\newcommand\Tstrut{\rule{0pt}{2.6ex}}
\newcommand\Bstrut{\rule[-0.9ex]{0pt}{0pt}}

\begin{document}


\title{
    Geometrically Frustrated Quadrupoles on the Pyrochlore Lattice and Generalized Spin Liquids
}

\author{Kristian Tyn Kai Chung}
\affiliation{Max Planck Institute for the Physics of Complex Systems, N\"othnitzer Strasse 38, 01187 Dresden, Germany}

\author{Sylvain Petit}
\affiliation{Laboratoire L\'{e}on Brillouin, CEA, CNRS, Universit\'{e} Paris-Saclay, CEA-Saclay, 91191 Gif-sur-Yvette, France}

\author{Julien Robert}
\affiliation{Universit\'{e} Grenoble Alpes, CNRS, Institut N\'{e}el, 38000, Grenoble, France}

\author{Paul McClarty}
\affiliation{Laboratoire L\'{e}on Brillouin, CEA, CNRS, Universit\'{e} Paris-Saclay, CEA-Saclay, 91191 Gif-sur-Yvette, France}

\begin{abstract}
As an instance of geometrical frustration with non-magnetic degrees of freedom, we explore the physics of local spin $S\geq 1$ moments on the pyrochlore lattice that interact via pure quadrupolar couplings. In the presence of spin-orbit coupling, there are nine allowed couplings between nearest neighbor quadrupoles. We determine the semi-classical phases and survey the phase diagram of the model. One may view the Hamiltonian as being composed of two copies of the well-studied dipolar model with couplings between the copies, and we find that each easy-plane dipolar phase has two quadrupolar counterparts. As geometrical frustration is important over broad swathes of the parameter space, there are many classical quadrupolar liquids and regions with order-by-disorder selection of discrete states. Order-by-disorder with quadrupoles admits cubic terms in the Landau theory whose effects appear in Monte Carlo simulations and flavor wave calculations for fixed spin~$S$. We showcase a number of examples of quadrupolar spin liquids, including one realizing a rank-3 symmetric tensor gauge theory exhibiting 6-fold pinch point singularities. We discuss remarkable differences between the quadrupolar physics of different spin quantum number. We also discuss connections to the non-Kramers rare earth pyrochlore materials. 
\end{abstract}

\date{\today}

\maketitle

\tableofcontents

\section{Introduction}

Geometrical frustration is the generation of unsatisfiable classical energetic constraints through a combination of suitable local couplings and lattice connectivity. Classic examples of geometrically frustrated models include Heisenberg antiferromagnets on kagome and pyrochlore lattices. In practice, geometrical frustration provides a route to generate a large low-lying density of strongly coupled states from which exotic phases may arise, especially in quantum many-body systems. Antiferromagnetism on suitable lattices used as a criterion for materials searches has been an extremely fruitful path to new and interesting physical phenomena \cite{lacroixIntroductionFrustratedMagnetism2011}. 

The pyrochlore lattice is the most prominent of the three-dimensional crystal structures where geometrical magnetic frustration can be studied in the absence of disorder. The significance of this structure was recognized in the early modern era of condensed matter physics \cite{andersonOrderingAntiferromagnetismFerrites1956} but a systematic exploration of the magnetism of pyrochlores in conjunction with experiment did not start until the 1990s. From then until now the field has generated a steady stream of insights \cite{gardnerMagneticPyrochloreOxides2010,rauFrustratedQuantumRareEarth2019}. An early notable event that has played a considerable role in shaping the field was the discovery of the spin ice materials \cite{harrisGeometricalFrustrationFerromagnetic1997,bramwellSpinIceState2001,castelnovoSpinIceFractionalization2012,udagawaSpinIce2021} which realize Ising classical spin liquids at finite temperature. However, geometrical frustration combined with spin-orbit coupling and quantum spins are responsible for an even richer spectrum of phenomena that has been explored extensively, especially among the family of rare earth pyrochlore materials. We refer the reader to various excellent reviews of the subject for further details \cite{gardnerMagneticPyrochloreOxides2010,rauFrustratedQuantumRareEarth2019}. 

A Maxwellian argument first applied to classical spin models in Ref.~\cite{moessnerLowtemperaturePropertiesClassical1998} seeks to maximize non-trivial ground state degeneracy by putting $n$-component degrees of freedom (with Heisenberg spins having $n=3$) on a corner-sharing lattice of $q$ site clusters. The origin of the constraint is the condition from the Heisenberg Hamiltonian that the lowest energy states have total moment on each cluster equal to zero. 
The number of degrees of freedom minus the number of constraints may have an extensive contribution going like $q(n-1)/b -n$ where $b$ is the number of clusters connected to each site with $b=2$ on the pyrochlore lattice. 
From this point of view it might seem attractive to consider local degrees of freedom with more components. 
Physics supplies them in the form of multipolar degrees of freedom originating from angular momenta $J\geq 1$. 

The study of interacting multipoles has a long history in condensed matter. 
One classic model in this context is the bilinear-biquadratic model on some lattice \cite{blume1969,matveev1974,andreev1984,papanicolaou1988,chubukov1990,chubukov1991,ivanov2003,lauchli2006,tsunetsugu2006,bhattacharjee2006,li2007,PhysRevLett.98.190404,fridman2011,penc2011,zhao2012,ivanov2013,shindou2013,smerald2013,muniz2014,smerald2015,smerald2016,wang2016,niesen2017,lai2017,niesen2018,hu2020,lima2021,mao2021,remund2022,szasz2022,pohle2023,pohle2024,momoi2024,mashiko2024,chojnacki2024,pohleAbundanceSpinLiquids2025}:
\begin{equation}
H = \sum_{\langle i,j \rangle} \left( \cos \theta \,\mathbf{S}_i \cdot \mathbf{S}_j + \sin \theta \left( \mathbf{S}_i \cdot \mathbf{S}_j \right)^2 \right).
\label{eq:BBQ}
\end{equation}
Implicitly the second term couples local quadrupoles and for some range of parameters this model leads to a nematic ground state where the quadrupoles acquire a non-zero expectation value while the spins have vanishing expectation value. This is a lattice model realization of the nematic phase of liquid crystals. Evidence has accumulated over the years that  
the isotropic bilinear-biquadratic model is the correct leading order description of certain materials \cite{bhattacharjee2006,stoudenmire2009} including spin-one pyrochlore \ce{NaCaNi2F7}~\cite{zhangDynamicalStructureFactor2019,pohleAbundanceSpinLiquids2025}. Nematic phases have long been sought in quantum materials and the situation now is that such phases are now well established in certain instances albeit often at high magnetic fields where the microscopic description is more naturally phrased in terms of a condensate of bound state excitations \cite{ishikawa2015,janson2016,yoshida2017,orlova2017,povarov2019,bhartiya2019,skoulatos2019,kohama2019,kim2024,sheng2025a,shengUnveilingQuadrupoleWaves2025}.

In parallel with these advances, investigations into heavy electron systems such as the actinides have revealed materials where multipolar order is known or at least strongly suspected in several materials \cite{santini2009}. In these systems, spin-orbit coupling is strong and interactions between multipoles have low symmetry. A schematic model that generalizes the bilinear-biquadratic model to such cases is
\begin{equation}
H = H_{\rm CEF} + \sum_{\langle ij \rangle} \sum_{L,L',M,M'}  [O_L^M]_i [J_{LL'}^{MM'}]_{ij} [O_{L'}^{M'}]_{j}
\end{equation}
where $O_L^M$ are Steven's operator equivalents and $H_{\rm CEF}$ is the single-ion crystal field Hamiltonian. Superexchange in actinides can generate $L$ up to rank $7$ \cite{santini2009}.

Multipolar interactions in rare earth pyrochlores at the bare ion level are similarly anisotropic and quantum chemistry calculation suggest that couplings between multipoles of all possible ranks are of comparable magnitude. This forbidding situation is simplified greatly by the crystal field spectrum that often leads to a single ion ground state doublet that is separated from the first excited state on a scale much larger than the scale of interactions. In such cases, the low energy physics is well described by an effective spin-1/2 model, albeit one perhaps of mixed character. Many years of intensive investigation of pyrochlore magnets have clarified our understanding of several materials and sharpened the problems. An all-too-brief summary of the current situation is that, of all pyrochlore magnets, the most puzzling have mixed character crystal field ground states. 

Of particular relevance to this work is the set of non-Kramers rare earths. Among the pyrochlores with rare earth ions $R^{3+}$ with, for example, R$=$Pr, Tb. Where the crystal field ground state is a doublet this is an effective spin one-half of mixed character that can be thought of as an Ising spin with component along $z$ and with transverse components that are quadrupolar in nature. 
Among the non-Kramers pyrochlores, Tb$_2$Ti$_2$O$_7$ stands out. It exhibits no signs of magnetic order at low temperatures well below the Curie scale \cite{enjalranTheoryParamagneticScattering2004,enjalranSpinLiquidState2004,fritschAntiferromagneticSpinIce2013,fritschTemperatureMagneticField2014,kadowakiCompositeSpinQuadrupole2015,takatsuQuadrupoleOrderFrustrated2016,zhangLowenergyMagnetoopticsTb2Ti2O72021}. This material is complicated by the presence of low-lying crystal field levels and significant magneto-elastic coupling. Similarly puzzling are praesodymium pyrochlores with M$=$Hf,Sn,Zr \cite{zhouDynamicSpinIce2008,kimuraQuantumFluctuationsSpinicelike2013,petitAntiferroquadrupolarCorrelationsQuantum2016,sibilleCandidateQuantumSpin2016,anandPhysicalPropertiesCandidate2016}. Accounting for the observed properties of these materials is among the central problems in the field of frustrated magnetism and a significant literature has built up in this direction. 

Given the mixed character of the local degrees of freedom, it is natural to consider the activity of quadrupoles and possible quadrupolar long-range order in these systems. But, while thermodynamic anomalies are expected at the transition temperature, the usual probes available to experimentalists present challenges in definitively demonstrating quadrupolar long-range order. In short, there is no way to measure the multipolar analogue of magnetic Bragg peaks that both signal and characterize the type of dipolar order. Multipolar order has not been seen definitively in the rare earth pyrochlores, though quadrupolar order has been suggested as a possible explanation for features of the weakly site-mixed Tb$_{2-x}$Ti$_{2+x}$O$_7$ \cite{takatsuQuadrupoleOrderFrustrated2016,kadowakiCompositeSpinQuadrupole2015}.

In this paper, we have two main motivations. One is to extend our understanding of magnetic frustration to a new domain of non-magnetic degrees of freedom and, in particular, to quadrupoles that naturally have a larger local configuration space than spins one-half or vector spins. Our second motivation is to bring a new perspective to the set of non-Kramers pyrochlores which are extremely poorly understood. These materials are quadrupolar active and our hope is by mapping out the landscape of quadrupolar states we might be able to stimulate progress on these systems. Concretely, we take microscopic quadrupoles living on sites of the pyrochlore lattice with couplings between nearest neighbors (\cref{sec:quadrupole_model}). The couplings are strongly constrained by the symmetry of the crystal and we explore all of them together. Although there are $9$ couplings, there are considerable simplifications and we may benefit from known results for the dipoles. Among the semi-classical phases (\cref{sec:irrep_multipoles}) are quadrupolar analogues of the ice states, the all-in/all-out states, splayed ferromagnet, $\Gamma_5$ and Palmer-Chalker phases familiar from the magnetic case. In fact, we can show that the easy plane states are in a sense doubled in passing from magnetism to quadrupoles (\cref{sec:phase-diagram}). This means that the order-by-disorder regime is doubled. This, along with the peculiar time reversal properties of quadrupoles, leads to rich and unusual order-by-disorder physics (\cref{sec:obd}). Much of the paper dwells on the range of semi-classical quadrupolar liquids that arise which includes all of the types seen in the magnetic case and more besides (\cref{sec:spin_liquids}).

So far, we have said little about the underlying angular momentum degrees of freedom that are the precursors to the quadrupoles. It turns out that the physics very much depends on effective spin. Spin one, which is the usual choice for studies of spin nematic phases, is highly constrained such that the full space of quadrupolar states is not available. Increasing the total spin relaxes this constraint but introduces new features that we discuss at length in \cref{sec:quantum_quadrupoles}.

\section{The Anisotropic Bilinear Quadrupole Model}
\label{sec:quadrupole_model}

For spin $S\geq 1$, multipoles higher than the dipole moment may be active. 
In this paper, we consider the quadrupole moments, encoded in the tensor operator
\begin{equation}
    {Q}^{\alpha\beta} = \frac{1}{2}\left( {S}^\alpha {S}^\beta + {S}^\beta {S}^\alpha \right) - \frac{1}{3}\vert\bm{S}\vert^2 \delta^{\alpha\beta} \quad \alpha,\beta \in \{x,y,z\}
    \label{eq:quadrupole_operator}
\end{equation}
as the dominant terms in the Hamiltonian. 
These spin operators could arise either as physical angular momenta of electronic degrees of freedom in an ion, or as effective pseudo-spin operators acting in a restricted Hilbert space of low-energy crystal field states. 
Many previous works have studied the effects of quadrupole interactions  in combination with dipolar couplings on various lattices: the most well-known example being the rotationally symmetric bilinear-biquadratic model, \cref{eq:BBQ}.
This model exhibits quadrupolar order for some range of the free parameter $\theta$, as the biquadratic term is the isotropic quadrupole-quadrupole interaction, 
\begin{equation}
    \Tr[Q_i Q_j] \equiv \sum_{\alpha,\beta} Q_i^{\alpha\beta}Q_j^{\beta\alpha} = (\bm{S}_i\cdot\bm{S}_j)^2 + \frac{1}{2}(\bm{S}_i \cdot \bm{S}_j) - \frac{1}{3}\vert \bm{S}\vert^4.
\end{equation}
For spins arising from light elements in the period table it is often the case that quadrupolar couplings are very much a sub-leading contribution to the magnetic exchange.  In general, when spin-orbit coupling is significant at the bare ion level, which is the case especially for magnetic materials with heavy elements, such a rotationally-symmetric form of the Hamiltonian is fine-tuned. Also, especially in rare earth and actinide systems, multipolar operators may be the primary couplings between the local degrees of freedom. In the case of dipolar and quadrupole active spin-orbit coupled systems, one should expect all symmetry-allowed anisotropic interactions between dipole operators $S_i^\alpha$ and quadrupole operators $\smash{Q_i^{\alpha\beta}}$ to be present. The most general Hamiltonian of dipolar-quadrupolar character allowed by spatial and time-reversal symmetry has the form\footnote{
    We assume throughout this work that the operators $S_i^\alpha$ transform as dipoles. As we discuss in Section~\ref{sec:materials}, in the rare earth pyrochlores, it is frequently useful to use pseudo-spins to address the low energy physics. These have mixed dipolar-quadrupolar character in non-Kramers ions~\cite{rauFrustratedQuantumRareEarth2019}. 
}
\begin{equation}
    H = H_{\rm CEF} + \sum_{ij} \sum_{\substack{\alpha\beta \\ \alpha'\beta'}}\left(S_i^\alpha [\mathcal{J}_{S}]_{ij}^{\alpha\beta} S_j^\beta +  Q_i^{\alpha\beta} [\mathcal{J}_{Q}]_{ij}^{\alpha\beta,\alpha'\beta'} Q_j^{\alpha'\beta'}\right)
    \label{eq:Hamiltonian-S-Q}
\end{equation}
Such anisotropies are well-studied in the context of dipoles on the pyrochlore lattice, giving rise to a rich phase diagram featuring a plethora of phenomena such as order by disorder and exotic spin liquids~\cite{rauFrustratedQuantumRareEarth2019,chungMappingPhaseDiagram2024,lozano-gomezAtlasClassicalPyrochlore2024,yanTheoryMultiplephaseCompetition2017,hickeyOrderbydisorderQuantumZeropoint2025,javanparastOrderbydisorderCriticalityXY2015,wongGroundStatePhase2013,lozano-gomezCompetingGaugeFields2024,franciniHigherrankSpinLiquids2025,noculakClassicalQuantumPhases2023,yanRank2U1Spin2020,bentonSpinliquidPinchlineSingularities2016,taillefumierCompetingSpinLiquids2017,hallasExperimentalInsightsGroundState2018}. 
As the dipolar sector is relatively well understood, here, we study the physics of the model with purely quadrupolar couplings. 

\subsection{Symmetry-Allowed Interactions: Global Frame}

\begin{figure}
    \centering
    \includegraphics[width=\columnwidth]{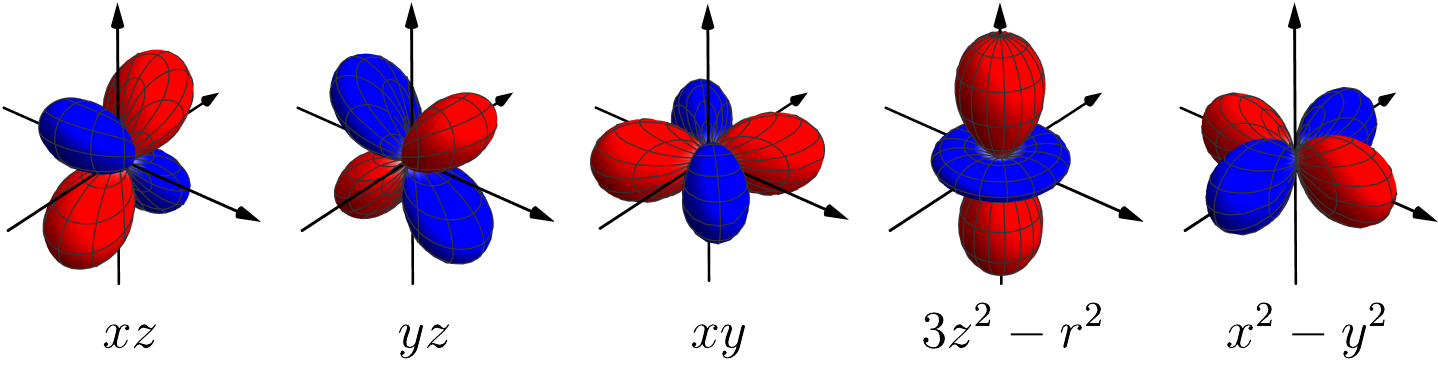}
    \caption{(a) Shapes of the quadrupolar orbitals, corresponding to an orthonormal basis in the space of symmetric trace-free tensors $q^{\alpha\beta}$. Surfaces are plotted as a function of angular positions $\hat{\bm{r}}$ on a unit sphere, with radius $\hat{r}^\alpha q^{\alpha\beta} \hat{r}^\beta$, and with positive radii colored red and negative ones colored blue. 
    }
    \label{fig:orbitals}
\end{figure}

The symmetric trace-free tensor operators $Q^{\alpha\beta}$ transform in a five-dimensional irreducible representation (hereafter {\it irrep}) of $O(3)$.
In an anisotropic crystalline environment the symmetry is lowered and this five-dimensional representation becomes reducible.\footnote{
    Because crystalline symmetries are discrete, the largest irreducible representations of point groups are three-dimensional, so all multipoles higher than dipoles must become reducible.
}. 
For a lattice with cubic point group $O_h$, such as the pyrochlore, it decomposes as $t_{2g} \oplus e_g$, corresponding to the symmetric off-diagonal ($t_{2g}$) and trace-free diagonal ($e_g$) components of $Q^{\alpha\beta}$.
They can be organized into a 5-component vector as
\begin{align}
    \left(
    \renewcommand{\arraystretch}{1.1}
    \begin{array}{c}
        Q_{yz} \\[1ex]
        Q_{zx} \\[1ex]
        Q_{xy} \\[1ex]
        Q_{3z^2 - r^2} \\[1ex]
        Q_{x^2 - y^2} 
    \end{array}
    \right)
    \equiv
    \left(
    \renewcommand{\arraystretch}{1.1}
    \begin{array}{c}
        Q_{t_{2g}}^{x} \\[1ex]
        Q_{t_{2g}}^{y} \\[1ex]
        Q_{t_{2g}}^{z} \\[1ex]
        \hline
        \Tstrut 
        Q_{e_g}^{\psi_2} \\[1ex]
        Q_{e_g}^{\psi_3} 
    \end{array}
    \right)
    \,\,
    &= 
    \,\,
    \left(
    \renewcommand{\arraystretch}{1.1}
    \begin{array}{c}
        \frac{1}{\sqrt{2}}\left(Q^{yz} + Q^{zy}\right)\\[1ex]
        \frac{1}{\sqrt{2}}\left(Q^{zx} + Q^{xz}\right)\\[1ex]
        \frac{1}{\sqrt{2}}\left(Q^{xy} + Q^{yx}\right)\\[1ex]
        \hline
        \Tstrut
        \frac{1}{\sqrt{6}}(2Q^{zz} - Q^{xx}  - Q^{yy}) \\[1ex]
        \frac{1}{\sqrt{2}}\left(Q^{xx} - Q^{yy}\right)
    \end{array}
    \right).
    \label{eq:quadrupole_5_vector}
\end{align}
This corresponds to choosing a basis in the 5-dimensional space of symmetric $3\times 3$ matrices, and each such basis matrix can be visualized by plotting its ``orbital structure'', as shown in \cref{fig:orbitals}.
The transformation of these five degrees of freedom under rotations is given by the Wigner D-matrix in the basis of real spherical harmonics (or equivalently, by projecting the transformation $Q \to RQR^{-1}$ to the five-dimensional basis). 
The quadrupole-quadrupole interaction term in \cref{eq:Hamiltonian-S-Q} can be written as
\begin{equation}
    H = \frac{1}{2}\sum_{ij}\sum_{a,b=1}^5 Q_i^a \mathcal{J}_{ij}^{ab}Q_j^b
    \label{eq:Hamiltonian-Q-5-generic}
\end{equation}
where we suppress the subscript on the quadrupole interaction matrix, $\mathcal{J}_Q \equiv \mathcal{J}$, which must be symmetric, $\smash{\mathcal{J}_{ij}^{ab} = \mathcal{J}_{ji}^{ba}}$.
Single-site anisotropies $\mathcal{J}_{ii}^{ab}$ are allowed by symmetry and given in \cref{apx:single_ion_anisotropy}, but for this work we will set all such terms to be zero.

From here on we focus on studying the Hamiltonian \cref{eq:Hamiltonian-Q-5-generic} with all symmetry-allowed nearest-neighbor interactions on the pyrochlore lattice of corner-sharing tetrahedra.
Restricting to nearest-neighbor interactions, the matrix $\mathcal{J}_{ij}^{ab}$ is the same on every tetrahedron;
for each tetrahedron it is a $20 \times 20$ matrix $\mathcal{J}_{\mu\nu}^{ab}$, where $\mu,\nu\in\{1\ldots 4\}$ label the four corners of a tetrahedron, and the Hamiltonian simplifies to
\begin{equation}
    H = \frac{1}{2} \sum_{t} \sum_{\mu,\nu=1}^4\sum_{a,b=1}^5 Q_{t,\mu}^{a} \mathcal{J}_{\mu\nu}^{ab} Q_{t,\nu}^{b}.
    \label{eq:Hamiltonian-Q-5-pyrochlore}
\end{equation}
where $t$ labels tetrahedra.
The matrix elements are restricted by the $T_d$ subgroup of the $O_h$ point-group symmetries, which map a single tetrahedron back to itself. 
For each $g \in T_d$ the four five-dimensional $Q$ vectors transform collectively in a 20-dimensional representation $\rho_{\bm{20}}$ of $T_d$, while the matrix $\mathcal{J}$ (after restricting it to be symmetric and have zero on-site elements) transforms in the 150-dimensional $\rho_{\bm{20}}\otimes \rho_{\bm{20}}^T$ representation.
Symmetry demands that the matrix elements of $\mathcal{J}$ are restricted such that it transforms trivially. 
Direct decomposition shows that the 150-dimensional representation contains nine copies of the trivial representation, indicating that there are nine symmetry-allowed couplings. 
Directly solving the equations $[\rho_{\bm{20}}(g),\mathcal{J}] = 0$ for all $g \in T_d$ allows us to express the matrix elements coupling sublattices 1 and 2 in the global basis \cref{eq:quadrupole_5_vector} as
\begin{align}
    \mathcal{J}_{12}^{ab} &= 
    \left(
    \begin{array}{r r r | r r}
        J_1 & J_3 & -J_4 & -J_6' & -J_7 \\[1ex]
        J_3 & J_1 & -J_4 & -J_6' &  J_7 \\[1ex]
        J_4 & J_4 &  J_2 &  J_5' & 0 \\[1ex]
        \hline
        \vphantom{\Big(} J_6' &  J_6' & J_5' & J_8  & 0 \\[1ex]
        J_7 & -J_7 & 0 & 0 & J_9
    \end{array}
    \right) 
    \label{eq:interaction_matrix_Jglo}
    \\[-2ex]
    &
    \hspace{6mm}
    \underbrace{\hspace{18mm}}_{t_{2g}}
    \hspace{2mm}
    \underbrace{\hspace{13mm}}_{e_g}
    \nonumber
\end{align}
where we define $J_5' \equiv -2J_5/\sqrt{3}$ and $J_6' \equiv (2J_6 - J_7)/\sqrt{3}$ for convenience. Lines indicate the separation of the $t_{2g}$ and $e_g$ components of the 5-component quadrupoles. 
The $t_{2g}$ sector has the exact same structure as that of the dipolar Hamiltonian~\cite{Curnoe2008,mcclartyEnergeticSelectionOrdered2009,rossQuantumExcitationsQuantum2011,yanTheoryMultiplephaseCompetition2017,rauFrustratedQuantumRareEarth2019}.
The two $e_g$-sector components are coupled to the three $t_{2g}$-sector components through the $J_5$, $J_6$, and $J_7$ couplings.

\subsection{Local Frame and Mixed Basis}
\label{sec:mixed_frame}

The Hamiltonian may be expressed in a more symmetric manner by changing basis to a local symmetry-adapted frame.
We define a local frame on the first sublattice,
\begin{equation}
    \hat{\bm{x}}_1 \propto [\bar{1}\bar{1}2],\quad
    \hat{\bm{y}}_1 \propto [1\bar{1}0],\quad
    \hat{\bm{z}}_1 \propto [\bar{1}\bar{1}\bar{1}] 
    \label{eq:local_xyz_basis}
\end{equation}
and similarly for the other four sublattices, where the $\hat{\bm{z}}_i$ lie along the local three-fold easy-axis, the $\hat{\bm{y}}_i$ lie along a local two-fold axis, and the $\hat{\bm{x}}_i$ lie in a mirror plane. 
We present first a ``mixed'' local-global frame which respects the separation of each quadrupole into a 3D $t_{2g}$ component and a 2D $e_g$ component, \cref{eq:quadrupole_5_vector}.
We exploit the fact that the $t_{2g}$ component transforms as a vector, while the $e_g$ component (as we know from the dipolar case) transforms as an easy-plane XY spin, defining the following two vector multiplets
\begin{align}
    \bm{Q}_i^{(t_{2g})} 
    &= 
    (Q_{t_{2g}}^x)_i
    \, 
    \hat{\bm{x}} 
    + 
    (Q_{t_{2g}}^y)_i
    \,
    \hat{\bm{y}} 
    + 
    (Q_{t_{2g}}^z)_i
    \,
    \hat{\bm{z}},
    \nonumber 
    \\
    \bm{Q}_i^{(e_g)} 
    &= 
    (Q_{e_{g}}^{\psi_2})_i
    \,
    \hat{\bm{x}}_i 
    +
    (Q_{e_{g}}^{\psi_3})_i
    \,
    \hat{\bm{y}}_i,
    \label{eq:t2g_eg_vector_mapping}
\end{align}
where for the $t_{2g}$ components we utilize the global Cartesian frame basis vectors and for the $e_g$ components we utilize the local easy-plane basis vectors.
We then rotate the $t_{2g}$ multiplet into the local frame, i.e. decompose it as 
\begin{equation}
    \bm{Q}_i^{(t_{2g})} 
    = 
    Q_{t_{2g},i}^{x_{\text{loc}}} \hat{\bm{x}}_i + 
    Q_{t_{2g},i}^{y_{\text{loc}}} \hat{\bm{y}}_i + 
    Q_{t_{2g},i}^{z_{\text{loc}}} \hat{\bm{z}}_i
\end{equation}
Lastly, we rotate into the eigenbasis of rotations in the local $xy$ planes, creating complex linear combinations of the local $x$ and $y$ degrees of freedom on each site for both the $t_{2g}$ and $e_g$ components,
\begin{equation}
    Q^\pm_{t_{2g}} = Q^{x_{\text{loc}}}_{t_{2g}}\pm i Q^{y_{\text{loc}}}_{t_{2g}}
    \quad\text{and}\quad
    Q^\pm_{e_{g}} = Q^{\psi_2}_{e_{g}}\pm i Q^{\psi_3}_{e_{g}}.
\end{equation}
The end result is a remarkable simplification of the interaction matrix into a highly symmetric form,
\begin{align}
    \mathcal{J}_{\mu\nu}^{\tilde{a}\tilde{b}} 
    &= 
    \omega_{\mu\nu}^{\tilde{a}\tilde{b}}
    \left(
    \begin{array}{c c | c | c c}
        -J_{\pm}^{t_{2g}}    &  J_{\pm\pm}^{t_{2g}} & J_{z\pm}^{t_{2g}}  & J'_{\pm} & J_{\pm\pm}' \\[1ex]
         J_{\pm\pm}^{t_{2g}} & -J_{\pm}^{t_{2g}}    & J_{z\pm}^{t_{2g}}  & J_{\pm\pm}' & J'_{\pm} \\[1ex]
        \hline
         \vphantom{\Big(}J_{z\pm}^{t_{2g}}   &  J_{z\pm}^{t_{2g}}   & J_{zz}    & J_{z\pm}^{e_g} & J_{z\pm}^{e_g} \\[1ex]
        \hline
         \vphantom{\Big(} J_{\pm}'         &       J_{\pm\pm}'     & J_{z\pm}^{e_g} & -J_{\pm}^{e_g} & J_{\pm\pm}^{e_g} \\[1ex]
         J_{\pm\pm}'          &       J_{\pm}'    & J_{z\pm}^{e_g} & J_{\pm\pm}^{e_g} & -J_{\pm}^{e_g} 
    \end{array}
    \right)^{\tilde{a}\tilde{b}}
    \label{eq:interaction_matrix_Jloc}
    \\[-2ex]
    &
    \hspace{12mm}
    \underbrace{\hspace{18mm}}_{Q_{t_{2g}}^\pm}
    \underbrace{\hspace{5mm}}_{Q_{t_{2g}}^{z_{\text{loc}}}}
    \underbrace{\hspace{17mm}}_{Q_{e_g}^\pm}
    \nonumber
\end{align}
where $\omega_{\mu\nu}^{\tilde{a}\tilde{b}}$ is a matrix of phase factors. 
We have used tildes to differentiate these indices from the global frame ones. 
For sublattices 1 and 2 they are all 1, while for sublattices 1 and 3 they have the banded matrix form 
\begin{equation}
    \omega_{13}^{\tilde{a}\tilde{b}} = 
        \left(
        \begin{array}{c c | c | c c}
        1 & -\omega & \omega^2 & 1 & -\omega \\
        \omega^2  & 1 & -\omega & \omega^2 & 1 \\ 
        \hline
        \Tstrut -\omega &  \omega^2 & 1 & -\omega & \omega^2   \\ 
        \hline
        \Tstrut 1 & -\omega & \omega^2 & 1 & -\omega \\
        \omega^2 & 1 & -\omega & \omega^2 & 1 
    \end{array} 
    \right)^{\tilde{a}\tilde{b}}
    \text{with}
    \quad
    \omega = e^{i\pi /3}.
    \label{eq:phase_factors}
\end{equation}
In this basis the Hamiltonian may be written in a form which directly parallels to the well-known dipolar Hamiltonian in the local frame, as $H = \sum_t\sum_{\langle ij \rangle\in t} H_{ij}$ with
\begin{align}
	H_{ij} 
	=& 
    \sum_{r\in\{e_g,t_{2g}\}} 
	\Big[
    - \,
    J_{\pm}^r \,(Q_{r,i}^+ Q_{r,j}^- + Q_{r,i}^- Q_{r,j}^+)
 	\nonumber 
    \\
    &\qquad\qquad
    +\,
    J_{\pm\pm}^r \,\gamma_{ij} (Q_{r,i}^+ Q_{r,j}^+ + Q_{r,i}^- Q_{r,j}^-)
    \nonumber
    \\
    &\qquad\qquad
    +\,
    J_{z\pm}^r\, \left(\zeta_{ij} (Q_{r,i}^+ Q_{t_{2g},j}^{z_{\text{loc}}} + Q_{t_{2g},i}^{z_{\text{loc}}} Q_{r,j}^+) + \text{h.c.}\right)
	\Big] 
 	\nonumber 
    \\
	&\quad
    +\,
	J_{zz}\, Q_{t_{2g},i}^{z_{\text{loc}}} Q_{t_{2g},j}^{z_{\text{loc}}}
    \nonumber
    \\
    &\quad
    +\, 
    J_{\pm}'\, (Q_{t_{2g},i}^+ Q_{e_g,j}^- + Q_{e_g,i}^+ Q_{t_{2g},j}^- + \text{h.c.})
    \nonumber
    \\
    &\quad
    +\,
    J_{\pm\pm}'\, \Big(\gamma_{ij}\, (Q_{t_{2g},i}^+ Q_{e_g,j}^+ + Q_{e_g,i}^- Q_{t_{2g},j}^-) + \text{h.c.}\Big)
    ,
	\label{eq:H_local_Jpm}
\end{align}
We can compare term-by-term to the dipolar Hamiltonian~\cite{rauFrustratedQuantumRareEarth2019,rossQuantumExcitationsQuantum2011}: the first three terms are identical to terms involving $S^+$ and $S^-$ in the dipolar Hamiltonian, but they each appear twice, once for each irrep.
The local $z$-$z$ term of course also appears in the dipolar Hamiltonian.
The matrices $\gamma_{ij}$ and $\zeta_{ij}$ contain the phase factors arising from \cref{eq:phase_factors}, and they are identical to those in the standard presentation of the dipolar Hamiltonian. 
The last two terms are new---they couple the $e_g$ and $t_{2g}$ sectors explicitly (they are also coupled implicitly through their respective couplings to the $Q_i^z$ via the two $J_{z\pm}^r$ terms).
Notice that if we turn off all $e_g$ components and the last two couplings we recover precisely the form of the dipolar Hamiltonian. 

In this form it is clear that there is an effective doubling of the easy-plane degrees of freedom compared to the dipolar case.
The technical reason for this is that in the pyrochlore lattice each quadrupole transforms locally under the site symmetry group $D_{3d}\subseteq O_h\subseteq O(3)$, which has only 1D and 2D irreps.
The five-dimensional quadrupolar irrep of $O(3)$ reduces to $e_g \oplus t_{2g}$ when restricting to the $O_h$ subgroup.
Further restricting to $D_{3d}$, the three-dimensional $t_{2g}$ representation reduces to $a_{1g}\oplus e_g$,
such that the five-dimensional quadrupolar representation of $O(3)$ decomposes as $a_{1g}\oplus 2e_g \in \mathrm{Rep}(D_{3d})$, corresponding to the easy-axis $Q^z$ components and the easy-plane $Q^{\pm}_{t_{2g}}$ and $Q^{\pm}_{e_g}$ components (whose subscripts refer to irreps of $O_h$).

\subsection{Local Frame Spherical Harmonics Basis}
\label{sec:local_basis}

Due to the doubling of the easy-plane degrees of freedom, there is a freedom to make a change of basis between them, albeit at the cost of changing the phase factors appearing in the Hamiltonian. 
There is another natural basis to consider: the basis of spherical harmonics in the local frame. 
We can write the quadrupole operator components in the local frame, 
\begin{align}
    Q^{\bar{\alpha}\bar{\beta}}_{\text{loc}} 
    &= 
    [\hat{e}_{i,\bar{\alpha}}^{\text{loc}}]^{\alpha} 
    \,
    Q_{\text{glo}}^{\alpha\beta} 
    \,
    [\hat{e}_{i,\bar{\beta}}^{\text{loc}}]^{\beta}
    \nonumber
    \\
     &= 
     \frac{1}{2}\left( {S}^{\bar{\alpha}} {S}^{\bar{\beta}} + {S}^{\bar{\beta}} {S}^{\bar{\alpha}} \right) - \frac{1}{3}\vert\bm{S}\vert^2 \delta^{\bar{\alpha}\bar{\beta}}.
    \label{eq:Q_tensor_local}
\end{align}
where $\hat{\bm{e}}_{i,\bar{\alpha}}^{\text{loc}} = \{\hat{\bm{x}}_i,\hat{\bm{y}}_i,\hat{\bm{z}}_i\}_{\tilde{\alpha}}$ are the local frame basis vectors such as given by \cref{eq:local_xyz_basis}, and we use a bar on the local frame indices.
This is equivalent to performing a rotation---$[\hat{e}_{i,\bar{\alpha}}^{\text{loc}}]^{\alpha}$ are the matrix elements of the SO(3) rotation which sends the global Cartesian axes to the local ones at site $i$.
Equivalently, the five quadrupole operators \cref{eq:quadrupole_5_vector} rotate as
\begin{equation}
    Q^{\bar{a}}_\text{loc} = D_{i}^{\bar{a}b} Q^b_\text{glo}
    \label{eq:local_basis_transformation}
\end{equation}
where $D_i$ is the Wigner D-matrix corresponding to the SO(3) rotation. 
The operators $\smash{Q^{\bar{a}}_\text{loc}}$ are the same as those operators in \cref{eq:quadrupole_5_vector} but with $Q^{\alpha\beta}$ replaced by the local frame components $Q^{\bar{\alpha}\bar{\beta}}$ in \cref{eq:Q_tensor_local}. 
This is the basis of cubic harmonics in the local frame, explicitly given in \cref{apx:irrepslocal}.
In this basis, the interaction matrix takes a nice symmetric form,
\begin{align}
    \mathcal{J}_{12}^{\bar{a}\bar{b}} &= 
    \left(
    \begin{array}{cc|c|cc}
     K_1-\sqrt{3} K_2 & 0 & 0  & 0 & K_3+\sqrt{3} K_5 \\[1ex]
     0 & K_1+\frac{K_2}{\sqrt{3}} & -\frac{2 K_4}{\sqrt{3}} &  \frac{K_5}{\sqrt{3}}-K_3 & 0 \\[1ex]
     \hline
     \Tstrut
     0 & -\frac{2 K_4}{\sqrt{3}} & K_9 & \frac{2 K_7}{\sqrt{3}}  & 0\\[1ex]
     \hline
     \Tstrut 
     0 & \frac{K_5}{\sqrt{3}}-K_3 & \frac{2 K_7}{\sqrt{3}} & K_6-\frac{K_8}{\sqrt{3}}  & 0 \\[1ex]
     K_3+\sqrt{3} K_5 & 0 & 0 & 0 & K_6+\sqrt{3} K_8
    \end{array}
    \right)
    \nonumber
    \\[-2.5ex]
    &
    \hspace{7mm}
    \underbrace{\hspace{15mm}}_{\bar{y}\bar{z}}
    \hspace{1mm}
    \underbrace{\hspace{12mm}}_{\bar{z}\bar{x}}
    \hspace{1mm}
    \underbrace{\hspace{8mm}}_{3\bar{z}^2-r^2}
    \hspace{1mm}
    \underbrace{\hspace{12mm}}_{\bar{x}^2-\bar{y}^2}
    \hspace{1mm}
    \underbrace{\hspace{15mm}}_{\bar{x}\bar{y}}
    \label{eq:interaction_matrix_Kloc}
\end{align}
where for convenience of presentation we have changed the order of the quadrupole components.
We can then make a complex rotation to the local frame spherical harmonic basis by defining the linear combinations
\begin{align}
    Q_{\text{loc}}^{0} &= Q_{3\bar{z}^2 - r^2}  
    \nonumber
    \\
    Q_{\text{loc}}^{\pm 1} &= 
    Q_{\bar{x}\bar{z}}  \pm i  Q_{\bar{y}\bar{z}} 
    = \frac{1}{\sqrt{2}}(S^\pm S^{\bar{z}} + S^{\bar{z}} S^\pm)
    \nonumber
    \\
    Q_{\text{loc}}^{\pm 2} &= Q_{\bar{x}^2-\bar{y}^2} \pm i Q_{\bar{x}\bar{y}} = \frac{1}{\sqrt{2}}(S^{\pm})^2,
    \label{eq:local_basis_raising_lowering}
\end{align}
where $S^\pm \equiv S^{\bar{x}} \pm i S^{\bar{y}}$.
Then the Hamiltonian can be written in this frame as
\begin{align}
	H_{ij} 
	=& 
    \sum_{m\in\{1,2\}} 
	\Big[
    - \,
    K_{\pm}^m \,(Q_{i}^{+m} Q_{j}^{-m} + Q_{i}^{-m} Q_{j}^{+m})
 	\nonumber 
    \\
    &\qquad\qquad
    +\,
    K_{\pm\pm}^m \,\gamma_{ij}^* (Q_{i}^{+m} Q_{j}^{+m} + Q_{i}^{-m} Q_{j}^{-m})
    \nonumber
    \\
    &\qquad\qquad
    +\,
    K_{z\pm}^m\, \left(\zeta_{ij}^* (Q_{i}^{+m} Q_j^{0} + Q_i^{0} Q_{j}^{+m}) + \text{h.c.}\right)
	\Big] 
 	\nonumber 
    \\
	&\quad
    +\,
	K_{zz}\, Q_{i}^{0} Q_{j}^{0}
    \nonumber
    \\
    &\quad
    +\, 
    K_{\pm}'\, (Q_{i}^{+1} Q_{j}^{-2} + Q_{i}^{+2} Q_{j}^{-1} + \text{h.c.})
    \nonumber
    \\
    &\quad
    +\,
    K_{\pm\pm}'\, \Big(\gamma_{ij}^*\, (Q_{i}^{+1} Q_{j}^{+2} + Q_{i}^{-1} Q_{j}^{-2}) + \text{h.c.}\Big)
    ,
	\label{eq:H_local_Kpm}
\end{align}
which closely parallels the form of \cref{eq:H_local_Jpm}.
Each basis may be useful for identifying interesting symmetric combinations of parameters.
In practice, we can always map back to the global basis \cref{eq:interaction_matrix_Jglo} for computational purposes, where there are no complex phase factors appearing the Hamiltonian. 
The mappings are given in \cref{apx:parameter_maps}.

\section{Irreps, Multipoles, and Symmetry Broken Phases}
\label{sec:irrep_multipoles}

In this section we study the Hamiltonian \cref{eq:H_local_Jpm} in more detail, by identifying linear combinations of quadrupole operators on each tetrahedron which transform irreducibly under the point group symmetry of a tetrahedron, $T_d$. 
This is equivalent to identifying the set of possible translationally-symmetric symmetry breaking patterns and their order parameters.
An alternative statement is that these irreducible multiplets define a set of symmetric normal modes for fluctuations on each tetrahedron. 
This method has been used to great success in the study of the dipolar Hamiltonian and its phase diagram~\cite{yanTheoryMultiplephaseCompetition2017,chungMappingPhaseDiagram2024,noculakClassicalQuantumPhases2023,lozano-gomezAtlasClassicalPyrochlore2024}.
Given the remarkable structure of the interaction matrix in the local frame, \cref{eq:H_local_Jpm}, one may guess that the phase structure looks much like that of the dipolar case, but with all phases involving the local-$xy$ components ``doubled''. 
Indeed we will see that this is the case.

\begin{table}[t]
    \centering
    \begin{booktabs}{
        width = 8.4cm, 
        colspec = {X[1,l] X[2.5,l] X[3,l] X[2.3,l] X[1.2,l]}
    }
         \toprule
         \Bstrut Irrep & Order Par. & Moment & $t_{2g}$ & $e_g$ \\
         \midrule
         $A_1$ 
         & $\Tr[\bar{\mathcal{O}}]$ 
         & Octupole 
         & AIAO 
         & \hspace{2mm} --- 
         \\
         \midrule
         $E$   
         & $\mathcal{Q}^{\alpha\alpha}$ 
         & Quadrupole 
         & \hspace{2mm} --- 
         & $\psi_2/\psi_3$ 
         \\
         $E$   
         & $\mathcal{R}^{\alpha\alpha}$ 
         & Toroid. Quad. 
         & $\psi_2/\psi_3$ 
         & \hspace{2mm} --- 
         \\
         \midrule
         $T_1$   
         & \hspace{.8mm}$ \epsilon_{\alpha\beta\gamma} \hspace{1.4mm} \bar{\mathcal{O}}^{\beta\gamma}$ 
         & Octupole 
         & PC 
         & $-$PC 
         \\
         $T_1$   
         & $\vert\epsilon_{\alpha\beta\gamma}\vert\mathcal{R}^{\beta\gamma}$ 
         & Toroid. Quad. 
         & PC 
         & $+$PC \\
         \midrule
         $T_2$   
         & $\vert\epsilon_{\alpha\beta\gamma}\vert\mathcal{Q}^{\beta\gamma}$ 
         & Quadrupole 
         & col-FM 
         & \hspace{2mm} --- 
         \\
         $T_2$   
         & $\vert\epsilon_{\alpha\beta\gamma}\hspace{0.6mm}\vert\bar{\mathcal{O}}^{\beta\gamma}$ 
         & Octupole 
         & biax-AFM 
         & $-$SFM \\
         $T_2$   
         & \Bstrut \hspace{.8mm} $\epsilon_{\alpha\beta\gamma} \hspace{1.0mm} \mathcal{R}^{\beta\gamma}$
         & Toroid. Quad. 
         & biax-AFM
         & $+$SFM 
         \\
         \bottomrule
    \end{booktabs}
    \caption{Irreducible symmetry multiplets of quadrupolar operators on a single tetrahedron obtained from the multipole construction. The first column lists the irreps from the decomposition \cref{eq:decomp}. The second lists the (unnormalized) operator multiplets in terms of multipole tensors and the third column lists the corresponding multipole moment of the tetrahedron from which this multiplet descended. The fourth and fifth list the decomposition in the $t_{2g}$ and $e_g$ sectors using \cref{eq:t2g_eg_vector_mapping} in terms of the analogous modes found in the dipolar model~\cite{yanTheoryMultiplephaseCompetition2017,rauFrustratedQuantumRareEarth2019,chungMappingPhaseDiagram2024}. Here we use the shorthand AIAO for all-in-all-out; PC for Palmer-Chalker; col-FM for collinear ferromagnet (called $T_{1\parallel}$ in \cite{chungMappingPhaseDiagram2024}); biax-AFM for biaxial anti-ferromagnet (called $T_{1\perp}$ in \cite{noculakClassicalQuantumPhases2023,chungMappingPhaseDiagram2024}); and SFM for easy-plane splayed ferromagnet (called $T_{1,\text{planar}}$ in \cite{yanTheoryMultiplephaseCompetition2017,chungMappingPhaseDiagram2024}).
    }
    \label{tab:irreps_multipoles}
\end{table}

\subsection{Irreducible Representation Multiplets}

For each tetrahedron there are twenty linearly independent quadrupole operators, indexed as $Q_\mu^a$. 
They transform in a 20-dimensional reducible representation of $T_d$.
By decomposing this into its irreducible components, we can identify multiplets of operators which are equivalent under symmetry and whose eigenstates have the same energies. 
In other words, this decomposition block-diagonalizes the interaction matrix, with each block labeled by an irrep.
Each block may then be further diagonalized, and the multiplet with the lowest energy gives the zero-temperature semi-classical ground states on a single tetrahedron.

The 20-dimensional representation is a tensor product of the permutations of the sublattice index and the rotation of the five quadrupole operators given by the Wigner D-matrix,
\begin{equation}
    Q_\mu^a \mapsto \sum_{\nu,b} \pi_{\mu\nu} D^{ab} Q_\nu^b
\end{equation}
Direct calculation gives the reduction\footnote{
    We use lowercase $t_{2g}$ and $e_g$ to denote the decomposition of a single quadrupole into irreps, whereas we use the capital letter names to distinguish the collective decomposition of all four quadrupoles in a tetrahedron.
}
\begin{align}
    \pi\otimes(t_{2g} \oplus e_g) 
    &= (A_1 \oplus E \oplus T_1 \oplus 2 T_2)_{t_{2g}} \oplus (E \oplus T_1 \oplus T_2)_{e_g}
    \nonumber
    \\
    &= A_1 \oplus 2E \oplus 2T_1 \oplus 3 T_2
    \label{eq:decomp}
\end{align}
where $\pi$ is a 4-dimensional permutation matrix. 
From this decomposition we can already see the ``doubling'' of the easy-plane modes relative to the dipolar case.
To see this, first note that the quadrupoles operators are even under parity, whereas dipoles are odd (that is, dipoles transform as $t_{1g}$).
Therefore the decomposition for dipoles is $\pi\otimes t_{1g} = A_2 \oplus E \oplus T_2 \oplus 2T_1$~\cite{mcclartyEnergeticSelectionOrdered2009,yanTheoryMultiplephaseCompetition2017,chungMappingPhaseDiagram2024}.
To compare them we must exchange $A_1 \leftrightarrow A_2$ and $T_1 \leftrightarrow T_2$. 
In the dipolar case, the $E$, $T_2$, and one of the $T_1$ irreps involve only the local-$xy$ components, and indeed we have one extra copy of each of these irreps.

\begin{table*}[ht]
    \centering
    \includegraphics[width=\textwidth]{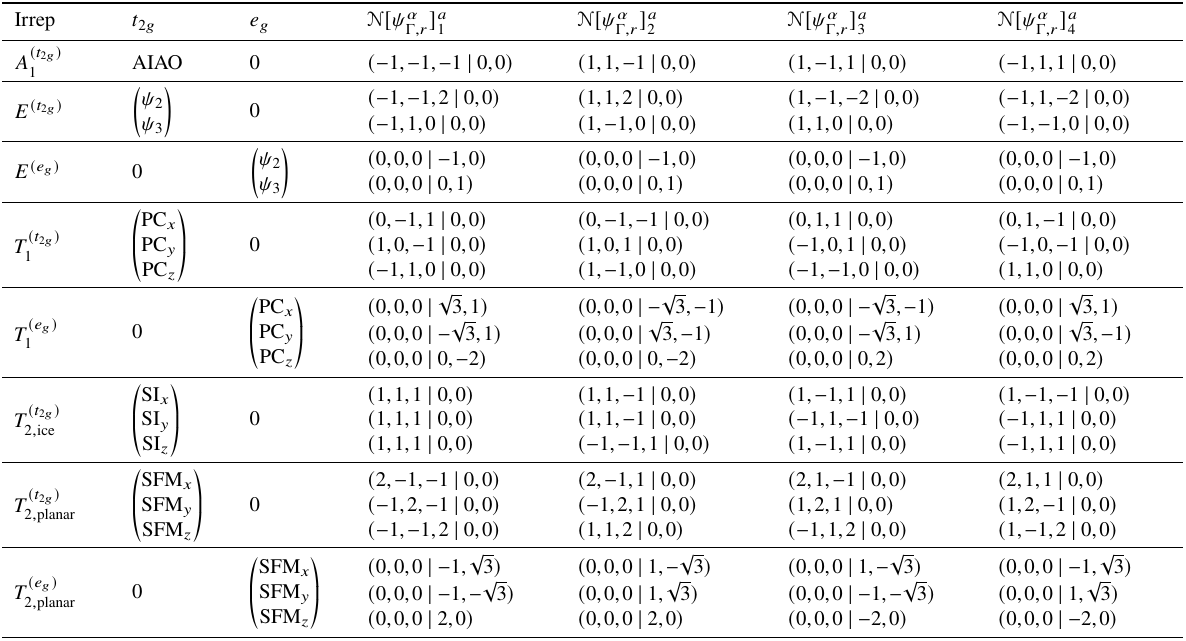}
    \caption{
    Choice of orthonormal basis for the irreducible representations of $T_d$ under which the four quadrupoles in a tetrahedron transform, where we have separated the $t_{2g}$ and $e_g$ components.
    The basis obtained in the multipole decomposition defined by the order parameters in \cref{tab:irreps_multipoles} contains linear combinations of $t_{2g}$ and $e_g$ components.  
    Here we use the abbreviations for comparison with the corresponding degrees of freedom in the dipolar case~\cite{yanTheoryMultiplephaseCompetition2017,chungMappingPhaseDiagram2024}: AIAO$=$all-in-all-out; PC$=$Palmer-Chalker; SI=spin ice; SFM=splayed ferromagnet.
    Irreps from the $t_{2g}$ sector reproduce those of the dipolar model, while ground states in the $e_g$ sector duplicate the easy-plane irreps. 
    These are mixed by the couplings $J_{\pm}'$ and $J_{\pm\pm}'$ in \cref{eq:interaction_matrix_Jloc}. 
    The 20-component vectors are separated by sublattice, where each resulting 5-component vector is given in the basis of \cref{eq:quadrupole_5_vector}, with the $t_{2g}$ and $e_g$ contributions separated by a vertical line. $\mathcal{N}$ is a normalization such that $\psi_{I,r}^\alpha$ is a unit vector. An analogous basis in the spherical harmonics local frame of \cref{sec:local_basis} is given in \cref{apx:irrepslocal}.
    }
    \label{tab:irrep_table_t2g_eg_basis}
\end{table*}

\begin{figure*}[t]
    \centering
    \includegraphics[width=.96\textwidth]{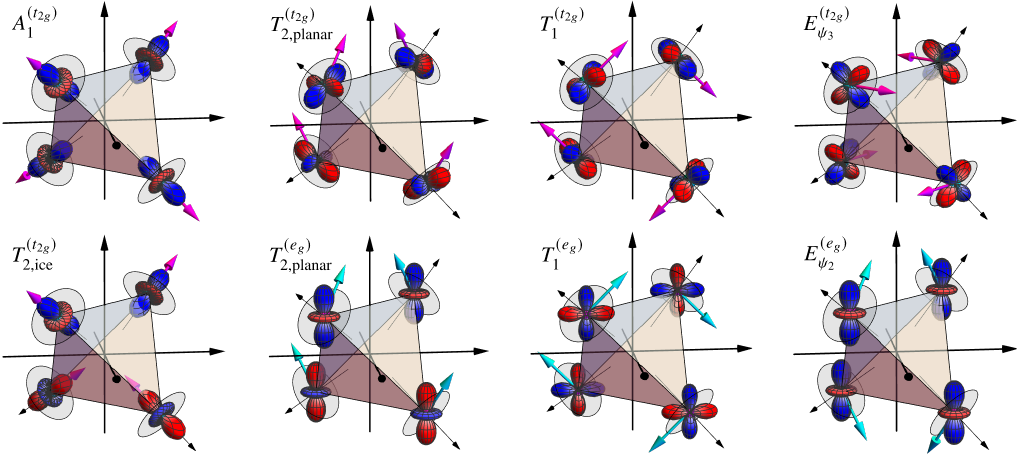}
    \caption{Illustrations of irreducible symmetry multiplets in the basis defined in \cref{tab:irrep_table_t2g_eg_basis}. Quadrupole operators on each site of form $\sum_{a=1}^5 [\psi_{\Gamma,r}^\alpha]_\mu^a Q_\mu^a$ are depicted by mapping the five coefficients to a symmetric trace-free tensor (using \cref{eq:vector_to_tensor}) and plotting the corresponding orbitals as in \cref{fig:orbitals}. Magenta (cyan) arrows depict the corresponding decomposition of the coefficients into a $t_{2g}$ and $e_g$ vector using \cref{eq:t2g_eg_vector_mapping}, illustrating how these correspond to analogous modes in the dipolar Hamiltonian (where dipoles are represented as vectors)~\cite{yanTheoryMultiplephaseCompetition2017,rauFrustratedQuantumRareEarth2019,chungMappingPhaseDiagram2024}.}
    \label{fig:ground_states}
\end{figure*}

\subsection{Multipole Decomposition}

Following Ref.~\cite{chungMappingPhaseDiagram2024} we can explicitly construct the irreducible multiplets by decomposing the multipole moments of a single tetrahedron into irreducible pieces---multipoles are irreducible representations of O(3) which become reducible when restricting to the $T_d$ subgroup. 
Here we will show that the 20-dimensional representation cleanly separates into the 5-component net quadrupole moment, the 7-dimensional octupole moment, and a residual 8-component ``toroidal quadrupole moment''.
These further break down under the point group symmetries to yield the decomposition in \cref{eq:decomp}.

\subsubsection{Quadrupole Moment}
First, we construct the net quadrupole moment tensor, 
\begin{equation}
    \mathcal{Q}^{\alpha\beta}_t = \sum_{i \in t} Q_i^{\alpha\beta}.
    \label{eq:net_quad_moment}
\end{equation}
This naturally decomposes as $E \oplus T_2$ since a single quadrupole decomposes as such using \cref{eq:quadrupole_5_vector}, corresponding to trace-free diagonal and symmetric off-diagonal components, respectively, accounting for five of twenty components.

\subsubsection{Octupole Moment}

Next, we construct a rank-3 moment tensor, which will contain the remaining irrep components,
\begin{equation}
    \mathcal{M}^{\alpha\beta\gamma}_t 
    = \sum_{i \in t} \hat{r}_i^\alpha Q_{i}^{\beta\gamma}
    = \sum_{i \in t} \hat{z}_i^\alpha Q_{i}^{\beta\gamma},
\end{equation}
where $\bm{r}_i$ is the displacement of corner $i$ from the center of the tetrahedron.
This is symmetric trace-free in the second two indices, so contains $3\times 5 = 15$ independent components. 
We can extract the remaining irreps by decomposing $\mathcal{M}$ into irreducible components. 
The decomposition of a general rank-3 tensor into irreducible representations of O(3) is given in Ref.~\cite{urruMagneticOctupoleTensor2022}.
The obvious component is the octupole tensor, which is obtained by symmetrizing and trace-subtracting $\mathcal{M}$. 
We fully symmetrize all the indices, defining the tensor 
\begin{equation}
    \mathcal{S}^{\alpha\beta\gamma}_t = \mathcal{M}^{(\alpha\beta\gamma)}_t,
\end{equation}
where parentheses indicate averaging over all permutations of the contained indices.
Next, we obtain the its vector trace,
\begin{equation}
    \chi^\alpha_t = S^{\alpha\beta\beta}_t,
    \label{eq:trace}
\end{equation}
with summation of the repeated index, from which the octupole moment is~\cite{urruMagneticOctupoleTensor2022}
\begin{equation}
    \mathcal{O}^{\alpha\beta\gamma}_t = \mathcal{S}^{\alpha\beta\gamma}_t - \frac{1}{5}(\chi_{t}^\alpha \delta^{\beta\gamma} + \chi_{t}^\beta \delta^{\gamma\delta} + \chi_{t}^\gamma \delta^{\alpha\beta}),
\end{equation}
where $\delta^{\alpha\beta}$ is the Kronecker delta. 
The octupole tensor transforms in a 7-dimensional irreducible representation of O(3), and is further reducible when restricted to the subgroup ${T_d \subseteq \mathrm{O}(3)}$.
To obtain the irreducible components under $T_d$ we define the following rank-2 tensor,
\begin{equation}
    \bar{\mathcal{O}}^{\alpha\beta}_t = \vert \epsilon_{\alpha\rho\gamma}\vert
    \mathcal{O}^{\rho\gamma\beta}_t,
\end{equation}
where $\epsilon$ is the fully-anti-symmetric Levi-Civita symbol, so that $\vert\epsilon\vert$ is fully symmetric and off-diagonal.\footnote{
    One can check that $\vert \epsilon_{\alpha\beta\gamma}\vert $ is an invariant symbol of $T_d$, meaning that 
    \[
        \vert \epsilon_{\alpha\beta\gamma}\vert = \vert \epsilon_{\rho\delta\sigma}\vert R(g)_{\rho\alpha} R(g)_{\delta \beta} R(g)_{\sigma\gamma}
    \]
    for all $g \in T_d$, where $R(g)$ is the corresponding O(3) rotation. 
}
One can check that all diagonal elements $\bar{\mathcal{O}}^{\alpha\alpha}$ are equal, so it has only seven independent degrees of freedom. Its trace transforms in the $A_1$ irrep, while its off-diagonal symmetric (anti-symmetric) components transform in the $T_2$ ($T_1$) irrep.
This accounts for all seven of the octupole moment components. 

\subsubsection{Toroidal Quadrupolar Moment}

The remaining eight degrees of freedom are contained in the so-called residue tensor,\footnote{
    The tensor $\mathcal{R}$ is one of three rank-2 ``residue tensors'' defined by pairwise contraction of the anti-symmetric Levi-Civita symbol with a general rank-3 tensor~\cite{urruMagneticOctupoleTensor2022}. In our case two are identical and the third is zero, because $\mathcal{M}^{\alpha\beta\gamma}=\mathcal{M}^{\alpha\gamma\beta}$ following from the symmetry of the quadrupole tensors~$Q$.
}
\begin{equation}
    \mathcal{R}^{\alpha\beta}_t = 
    \epsilon_{\alpha\rho\gamma}\,
    \mathcal{M}^{\rho\gamma\beta}_t. 
\end{equation}
Physically, this tensor can be described as a \emph{toroidal quadrupole moment}.
The analogous tensor in the dipolar case is the toroidal dipole moment, $\sum_i \bm{r}_i \times \bm{S}_i$, which is saturated in the chiral Palmer-Chalker ground states~\cite{chungMappingPhaseDiagram2024}. 
The tensor $\mathcal{R}$ may be written schematically as ${\mathcal{R}} = \sum_i \bm{r}_i \times {Q}_i$, where the cross product is taken with the first component of the quadrupole tensor ${Q}$. 
This tensor has zero trace.
Its two independent diagonal elements transform in the 2D $E$ irrep.
Its off-diagonal symmetric components transform in the $T_1$ irrep and its anti-symmetric components transform in the $T_2$ irrep,\footnote{
    In fact its anti-symmetric components are proportional to the trace vector $\chi^\alpha_t$, \cref{eq:trace}.
} thus accounting for the remaining eight degrees of freedom.

This exhausts all of the irreps and provides a canonical global frame orthonormal basis for them, listed in \cref{tab:irreps_multipoles}. 
Each irreducible multiplet derived this way is a linear combination of the twenty independent quadrupole operators on a single tetrahedron, of the form 
\begin{equation}
    [\mathscr{O}_{\Gamma,r}^{\alpha}]_t = \sum_{i\in t} \sum_{a=1}^5 [\psi_{\Gamma,r}^{\alpha}]_{i}^{a} Q_i^a,
    \label{eq:irrep_multiplet_operators}
\end{equation}
where $\Gamma$ labels an irreducible representation of $T_d$, $r$ labels the copies of each irrep, and $\alpha \in \{1\cdots d_{\Gamma}\}$ labels its components, where $d_\Gamma$ is the dimension of the irrep.
The coefficients $\psi_{\Gamma,r}^{\alpha}$ form a 20-component unit vector.
Note that the redundant third component of each $E$ irrep can be removed by taking the the inner product of these 3-component vector multiplets with $\hat{\bm{x}}_1$ and $\hat{\bm{y}}_1$ from \cref{eq:local_xyz_basis} to extract the two non-trivial components, since the inner product with $[111]$ computes the trace of the corresponding tensor, which is zero.
One can check that the twenty components from this decomposition indeed form an orthogonal basis.

In \cref{tab:irreps_multipoles} we have given a description of the projections of each $\psi_{\Gamma,r}^{\alpha}$ into the $t_{2g}$ and $e_g$ sectors in terms of normal modes of the dipolar case. 
The resulting configurations are described in the last two columns of \cref{tab:irreps_multipoles} in terms of their equivalent configurations in the dipole case~\cite{chungMappingPhaseDiagram2024}.
We know from the dipole case that a local symmetry-adapted frame is most well-suited to understand the ways that the irreps mix.
In \cref{tab:irrep_table_t2g_eg_basis} we give a basis for the irreducible multiplets appropriate for the mixed local-global frame introduced in \cref{sec:local_basis}. 
This table makes clear that the $e_g$ components of the quadrupoles duplicate all of the easy-plane components of the $t_{2g}$ sector. 
These multiplets are illustrated in \cref{fig:ground_states}, in which the vectors $[\bm{\psi}_{\Gamma,r}]_\mu$ are converted to symmetric trace-free tensors and plotted as spherical harmonics using the same method as in \cref{fig:orbitals}.
The $t_{2g}$ and $e_g$ components are mapped to vectors according to \cref{eq:quadrupole_5_vector}, which are shown in magenta and cyan, respectively, which can be compared directly to corresponding vector configurations in Ref.~\cite{chungMappingPhaseDiagram2024}.

\subsection{Decoupled Multiplets and Canting}
\label{sec:canting}

Symmetry forbids linear couplings between multiplets in different irreps.
However, multiple copies of a single irrep may couple, leading to a mixing of multiplets within each irrep which varies continuously as the parameters in the Hamiltonian are varied. 
Given an orthonormal choice of irrep multiplets $\bm{\mathscr{O}}_{\Gamma,r}$, the Hamiltonian can be put in a block-diagonalized form in terms of matrices $\mathcal{J}_\Gamma$,
\begin{equation}
    H = \frac{1}{2} \sum_t \sum_{\Gamma} \sum_{r,r'} [\mathcal{J}_\Gamma]_{r,r'}\bm{\mathscr{O}}_{\Gamma,r}\cdot \bm{\mathscr{O}}_{\Gamma,r'} .
\end{equation}
If there is more than one multiplet of a given irrep, symmetry allows for a coupling between them, i.e. between $r \neq r'$. 
This coupling is removed by a parameter-dependent orthogonal transform $U$ which diagonalizes the symmetric real matrix $\mathcal{J}_\Gamma$, defining new multiplets
\begin{equation}
    \tilde{\mathscr{O}}_{\Gamma,\tilde{r}} = \sum_{r} [U(J_1, \cdots, J_9)]_{\tilde{r},r} \mathscr{O}_{\Gamma,r}
\end{equation}
In the dipolar case there is only one repeated irrep, which appears twice. 
Its coupling is removed by a parameter-dependent SO(2) rotation, causing the dipoles to cant continuously as the parameters are tuned~\cite{chungMappingPhaseDiagram2024,yanTheoryMultiplephaseCompetition2017,rauFrustratedQuantumRareEarth2019}.
In the quadrupolar case the $T_1$ and $E$ irreps both appear twice, and require an SO(2) rotation to decouple, while the $T_2$ irrep appears three times, requiring an SO(3) rotation to decouple.
Once we have rotated to the proper irreducible multiplets $\tilde{\bm{\mathcal{O}}}_{\Gamma,r}$ the Hamiltonian can be expressed as a sum of decoupled quadratic multiplets,
\begin{equation}
    H = \frac{1}{2} \sum_t \sum_{\Gamma,\tilde{r}} J_{\Gamma,\tilde{r}} \vert \tilde{\bm{\mathcal{O}}}_{\Gamma,\tilde{r}}\vert^2.
    \label{eq:Hamiltonain_diagonalized}
\end{equation}
This form of the Hamiltonian allows one to read off the ground states at the semi-classical level from the relative $J_{\Gamma,\tilde{r}}$ when the quadrupolar operators are normalized and when the quadrupoles are not overly constrained (discussed further in Section~\ref{sec:quantum_quadrupoles}).
For the unique 1-dimensional $A_1$ irrep no decoupling is required and we have simply
\begin{equation}
    \mathcal{J}_{A_1} = 3J_{zz}
\end{equation}
For the remaining irreps we work in the basis given in \cref{tab:irrep_table_t2g_eg_basis} and use the corresponding local frame defined in \cref{sec:mixed_frame}. 

For the $E$ irrep, we have in terms of the local frame parameters given in \cref{eq:interaction_matrix_Jloc} 
\begin{equation}
    [\mathcal{J}_E]_{r,r'} = 
    -\frac{1}{6} \begin{pmatrix}
        J_{\pm}^{t_{2g}} & J_{\pm}' \\[2ex]
        J_{\pm}' & J_{\pm}^{e_g}
    \end{pmatrix},
    \label{eq:irrep_matrix_E}
\end{equation}
which can be diagonalized to give the energies of the two decoupled $E$ irreps, 
\begin{equation}
    J_{E\pm} 
    = 
    -\frac{1}{3}
    (J_{\pm}^{t_{2g}}
    +J_{\pm}^{e_{g}}) 
    \pm 
    \frac{1}{3}
    \sqrt{(J_{\pm}^{t_{2g}}-J_{\pm}^{e_{g}})^2 + 4 (J_\pm')^2}\,,
\end{equation}
where the SO(2) canting angle is given by 
\begin{equation}
    \tan(2\theta_E) = \frac{2J_{\pm}'}{J_{\pm}^{e_g} - J_{\pm}^{t_{2g}} }\,.
\end{equation}
For the $T_1$ irrep, the coupling is given by
\begin{equation}
    [\mathcal{J}_{T_1}]_{r,r'} = 
    \frac{1}{2}
    \begin{pmatrix}
        J_{\pm}^{t_{2g}} - 2 J_{\pm\pm}^{t_{2g}} & -(J_{\pm}' + 2 J_{\pm\pm}') 
        \\[2ex]
        -(J_{\pm}' + 2 J_{\pm\pm}') & J_{\pm}^{e_g} - 2 J_{\pm\pm}^{e_g}
    \end{pmatrix},
    \label{eq:irrep_matrix_T1}
\end{equation}
and the eigenvalues are
\begin{align}
    J_{T_{1}\pm} &= (J_{\pm}^{t_{2g}}  - 2J_{\pm\pm}^{t_{2g}})+(J_{\pm}^{e_g}-2J_{\pm\pm}^{e_g}) 
    \nonumber
    \\
    &
    \pm
    \sqrt{[(J_{\pm}^{t_{2g}} - 2J_{\pm\pm}^{t_{2g}}) -(J_{\pm}^{e_g} -2 J_{\pm\pm}^{e_g}) ]^2 + 4(J_\pm'+2J_{\pm\pm}')^2}\,.
\end{align}
The SO(2) canting angle which decouples them is 
\begin{equation}
    \tan(2\theta_{T_1}) = \frac{2(J_{\pm}'+2J_{\pm\pm}')}{(J_{\pm}^{t_{2g}} - 2J_{\pm\pm}^{t_{2g}})- (J_{\pm}^{e_g} - 2J_{\pm\pm}^{e_g})}.
\end{equation}
For the triple $T_2$ irrep, we have
\begin{equation}
    [\mathcal{J}_{T_2}]_{r,r'} = \begin{pmatrix}
        J_{\pm}^{t_{2g}} + 2 J_{\pm\pm}^{t_{2g}} & 2 J_{z\pm}^{t_{2g}} & 2J_{\pm\pm}'-J_{\pm}' 
        \\[2ex]
        2J_{z\pm}^{t_{2g}} & -J_{zz}/2 & 2 J_{z\pm}^{e_g}
        \\[2ex]
        2J_{\pm\pm}'-J_{\pm}'  & 2 J_{z\pm}^{e_g} & J_{\pm}^{e_g} + 2 J_{\pm\pm}^{e_g}
    \end{pmatrix}.
    \label{eq:irrep_matrix_T2}
\end{equation}
Unfortunately there is no straightforward way to express analytically the eigenvalues of such a $3\times 3$ matrix, or to express the formula for the SO(3) rotation which diagonalizes it.
This prevents us from performing the type of analysis done in Ref.~\cite{chungMappingPhaseDiagram2024} in which the parameters were re-expressed as functions of the eigenvalues, which greatly aided in parameterizing the phase diagram and all degenerate intersections of phases.  
Nevertheless, there is no issue in numerically diagonalizing the matrix. 
It is useful, however, to define a canting angle between the two easy-plane $T_2$ modes,
\begin{equation}
    \tan(2\theta_{T_{2,\text{planar}}}) = \frac{2(2J_{\pm\pm}'-J_{\pm}')}{(J_{\pm}^{e_g} + 2J_{\pm\pm}^{e_g}) - (J_{\pm}^{t_{2g}} + 2J_{\pm\pm}^{t_{2g}})}
\end{equation}
which decouples them when both $J_{z\pm}$ couplings are zero. 
It makes sense in general to first determine each of the three above-defined canting angles between the easy-plane modes and rotate into the corresponding basis which decouples them. 
This produces new values for the two $z$-$\pm$ couplings, and the remaining difficulty is diagonalizing the $T_2$ sector, which is in general not possible to do analytically unless one of the resulting $z$-$\pm$ couplings is zero.

\subsection{Symmetry Breaking Phases and Order Parameters}

At the level of irreps we expect there to be nominally four phases, one for each unique irrep in the decomposition. 
The exception is the $E$ irrep, where in the dipolar case it is known that a fluctuation-driven order-by-disorder mechanism splits this irrep into multiple symmetry-inequivalent phases.
We find a similar situation for quadrupoles, discussed further in \cref{sec:obd}.

Since all interactions are within a single unit cell, and due to the corner-sharing nature of the pyrochlore lattice, it is only possible to stabilize zero momentum ground states.
To define order parameters for the symmetry breaking patterns, let
\begin{equation}
    \mathscr{Q}_{\mu}^a = \frac{1}{N_S}\sum_{t \in \text{A}} Q_{t,\mu}^a 
\end{equation}
be the net quadrupole moment of the entire system resolved by sublattice (written as a 5-component vector via \cref{eq:quadrupole_5_vector}), where each site is indexed by an A tetrahedron~$t$ and a sublattice~$\mu$. 
Summing over all A tetrahedra is equivalent to summing over all primitive FCC unit cells, thus each site is included once. 
We have added a normalization factor~$N_S$ such that the maximum normalized quadrupole moment per site is unity, which depends on the spin quantum number~$S$ and is given later in \cref{eq:quadrupole_normalization} and discussed in more detail there.

For each irrep we define a projector into the corresponding subspace of the 20-dimensional space of single-tetrahedron degrees of freedom,
\begin{equation}
    P^{(\Gamma)} = \sum_{r} \sum_{\alpha} \bm{\psi}_{\Gamma,r}^\alpha \otimes [\bm{\psi}_{\Gamma,r}^\alpha]^{\dagger},
\end{equation}
in terms of an orthonormal basis such as the one given in \cref{tab:irrep_table_t2g_eg_basis} (the projector itself is independent of the basis). 
We extract the projections of the total quadrupole moment to each irrep,
\begin{equation}
    [\phi_{\Gamma}]_{\mu}^{a}= \frac{1}{N} \sum_{\nu=1}^4 \sum_{b=1}^5 P^{(\Gamma)}_{\mu,a;\nu,b} \mathscr{Q}_{\nu}^b
\end{equation}
where $N$ is the number of lattice sites, and define scalar order parameters for each irrep,
\begin{equation}
    \Phi_{\Gamma} = \vert\langle \bm{\phi}_{\Gamma} \rangle\vert \in [0,1],
\end{equation}
where the angle brackets indicate the quantum expectation value at fixed temperature.

\begin{figure}[t]
    \centering
    \begin{overpic}[width=\linewidth]{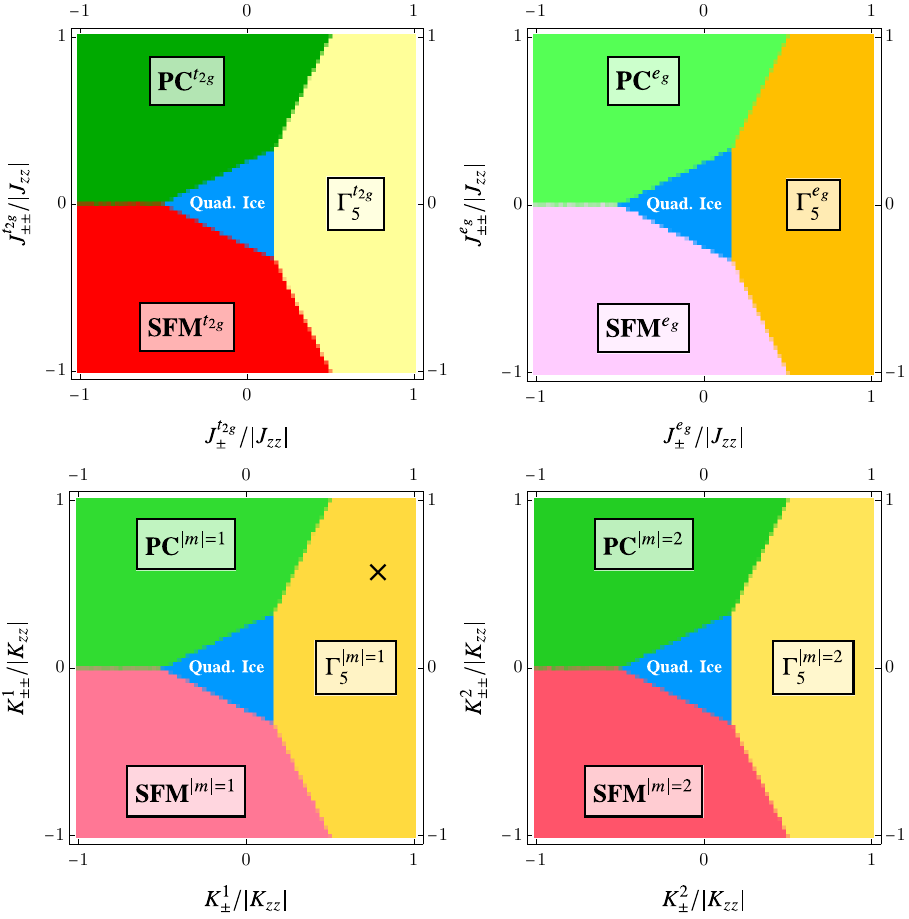}
        \put(00,95){(a)}
        \put(50,95){(b)}
        \put(00,45){(c)}
        \put(50,45){(d)}
    \end{overpic}
    \includegraphics[width=.8\linewidth]{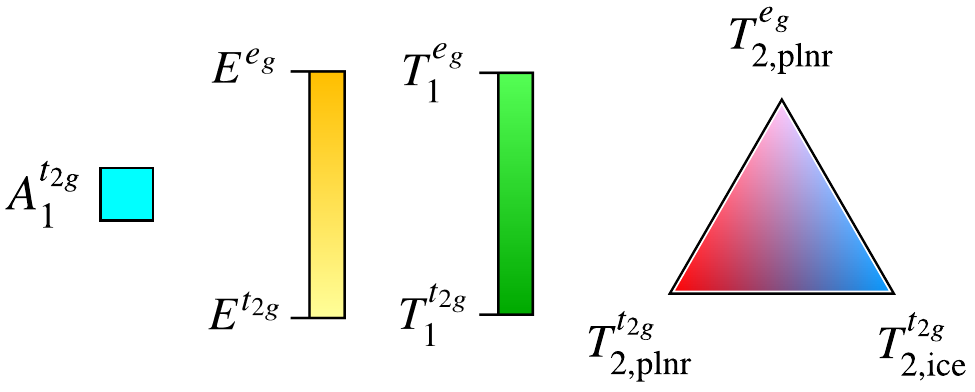}
    \caption{
    Luttinger-Tisza phase diagrams with $z$-$\pm$ couplings in either \cref{eq:H_local_Jpm} or \cref{eq:H_local_Kpm} set to zero and $J_{zz} > 0$, which reproduce the analogous ones of the dipolar model (unspecified parameters are zero). 
    We have labeled the phases using the dipolar conventional names---Palmer-Chalker (PC, $T_1$), Splayed Ferromagnet (SFM, $T_2$), and $\Gamma_5$ ($E$). The analog of the spin ice phase is labeled as quadrupolar ice. 
    (a) Phase diagram where the $t_{2g}$ sector in \cref{eq:H_local_Jpm}  minimal energy, and (b) case where the $e_g$ sector has minimal energy. (c) Phase diagram where the $\vert m \vert = 1$ sector in \cref{eq:H_local_Kpm} has minimal energy, and (d) case where the $\vert m\vert = 2$ sector has minimal energy. 
    The phases are colored according to the canting angle of the ground state irrep as illustrated in the color legends below. 
    The cross in (c) corresponds to the simulation parameters for \cref{fig:monte_carlo_OBD}.
    }
    \label{fig:dipolar_phase_diagrams}
\end{figure}

\section{Structure of the Semi-Classical Phase Diagram}
\label{sec:phase-diagram}

We now turn to an exploration of the basic structure of the phase diagram.
To do so, we investigate how the energies of the single-tetrahedron irreducible symmetry multiplets vary as functions of the nine couplings parameters. 
The single-tetrahedron mean field ground states are the maximal weight eigenstates of the lowest-energy multiplet operators.
These then form zero-momentum ground states of the entire system by translating them to all $A$ tetrahedra, which is a ground state on all of the $B$ tetrahedra as well. 
This is analogous to the Luttinger-Tisza type of analysis applied to dipolar Hamiltonians~\cite{luttingerTheoryDipoleInteraction1946,litvinLuttingerTiszaMethod1974}, and we will refer to it as such. 
Later, in \cref{sec:quantum_quadrupoles}, we will discuss how constraints arising from the quantum spin Hilbert space and the behavior of semi-classical quadrupoles modify these pictures.
In particular, while we do expect that the minimal irrep captures the correct symmetry breaking in the ground state, it may occur that additional irreps are also active.
In \cref{sec:obd} we will discuss how the $E$ irrep ground state is split into multiple distinct phases by quantum effects and fluctuations not captured by mean field theory. 
While we do not present a complete picture of the entire phase diagram. we highlight (at the level of the minimal energy irreps) some of its important subspaces and intersection points.

With nine parameters to vary, the phase diagram is very difficult to visualize. 
Furthermore, due to the presence of three copies of the $T_2$ irrep it is not possible to use the method of Ref.~\cite{chungMappingPhaseDiagram2024} to parameterize all of the intersections of phases. 
Nevertheless, we can gain significant intuition for the general structure by using the fact that the Hamiltonian can be put in forms which directly parallel the dipolar one, \cref{eq:H_local_Jpm,eq:H_local_Kpm}.
For example, if we turn off all couplings except those in the $t_{2g}$ sector in \cref{eq:H_local_Jpm}, then it is clear that the phase diagram (again at the level of irrep energies) must be the same as that of the dipolar case, for which a comprehensive picture is available in Ref.~\cite{chungMappingPhaseDiagram2024}.
It also must be the case that turning off the easy-plane $t_{2g}$ couplings, leaving only $Q_{t_{2g}}^{z_{\text{loc}}}$ and the two $e_g$ degrees of freedom, the Hamiltonian again has the same for as the dipolar one, so this limit of the phase diagram is also the same at the level of irrep energies.
On the other hand, we can do the same in \cref{eq:H_local_Kpm} by turning off either the $m = \pm 1$ or the $m= \pm 2$ couplings, and again the Hamiltonian reduces to a form identical to the dipolar case and will again show the same phase diagram. 
This is illustrated in \cref{fig:dipolar_phase_diagrams}. 
We have plotted phase diagrams in which the coupling between easy-axis and easy-plane modes is set to zero, and either the (a) $t_{2g}$, (b) $e_g$, (c) $\vert m \vert = 1$, or (d) $\vert m \vert = 2$ sectors are the ground states.
This reproduces the analogous phase diagram found in the dipolar Hamiltonian~\cite{rauFrustratedQuantumRareEarth2019,chungMappingPhaseDiagram2024}, with phases labeled accordingly, but with different mixing of the easy-plane $E$, $T_1$, and $T_2$ irreps as indicated by the varying colors. 

We expect that in fact one can continuously rotate the parameterization in such a way that the Hamiltonian always reduces to the form of of~\cref{eq:H_local_Jpm,eq:H_local_Kpm} and the phase diagram looks the same, but with a continuous rotation of the canting angles of the irreps.
That is, rotating into the basis defined by the easy-plane canting angles given in \cref{sec:canting}, the Hamiltonian may in general be rewritten in the form of \cref{eq:H_local_Jpm,eq:H_local_Kpm} with new couplings $\widetilde{J}_{\pm}^{\tilde{r}}$, $\widetilde{J}_{\pm\pm}^{\tilde{r}}$, etc., but \emph{without} any direct coupling between the two sets of easy-plane modes, i.e. with $\widetilde{J}_{\pm}'=\widetilde{J}_{\pm\pm}' = 0$. 
For example, if $J_{\pm}'=J_{\pm\pm}'=0$ in \cref{eq:H_local_Jpm}, then $K_{\pm}'$ and $K_{\pm\pm}'$ are non-zero in \cref{eq:H_local_Kpm}.

\begin{figure}
    \centering
    \includegraphics[width=\linewidth]{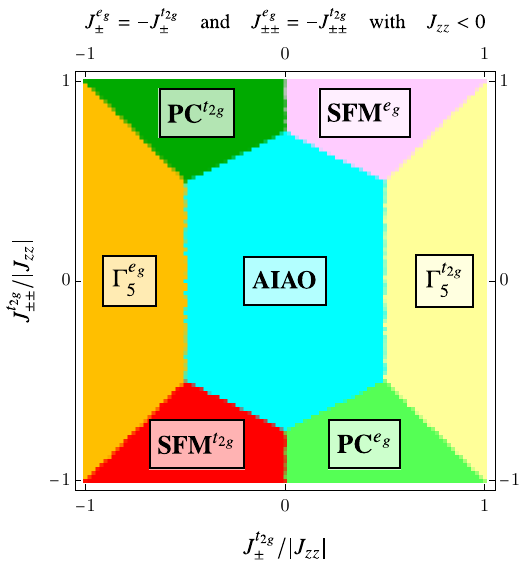}
    \caption{A Luttinger-Tisza phase diagram cut which exhibits all easy-plane phases surrounding the quadrupolar all-in-all-out phase.}
    \label{fig:phase_diagram_diamond}
\end{figure}

To clearly exhibit the doubling of all easy-plane phases, we show in \cref{fig:phase_diagram_diamond} a cut of the phase diagram which exhibits all of the easy-plane phases bordering the all-in-all-out phase (an analogous phase diagram with quadrupolar spin ice in the middle is obtained by switching the sign of $J_{zz}$). 
This is achieved by setting $J_{\pm}^r$ to have opposite signs in the two sectors, and similarly for $J_{\pm\pm}^r$.

\begin{figure*}[t]
    \centering
    \begin{overpic}[width=\textwidth]{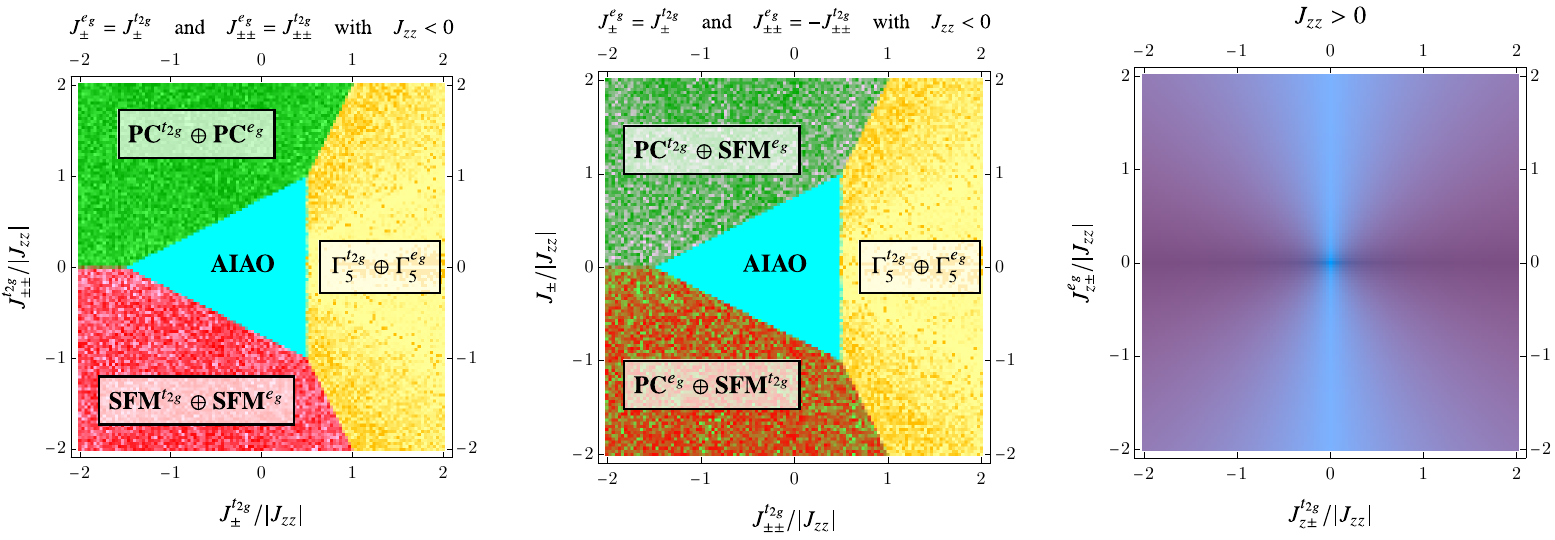}
        \put(00.0,32){(a)}
        \put(31.5,32){(b)}
        \put(66.3,32){(c)}
    \end{overpic}
    \caption{
    Luttinger-Tisza phase diagram cuts illustrating different dualities of the Hamiltonian. (a) A phase diagram cut which is self-dual under the anti-diagonal duality which exchanges the two easy-plane sectors, (b) An ``anti-self-dual'' cut which combines the anti-diagonal duality with a duality switching the sign of one $J_{\pm\pm}$ couplings (when $J_{z\pm} =0$).
    (c) A cut which demonstrates the duality under reversing the signs of either of the $J_{z\pm}$.
    The color legend is the same as in \cref{fig:dipolar_phase_diagrams}. 
    Unspecified parameters are zero.
    }
    \label{fig:phase_diagrams_duality}
\end{figure*}

\subsection{Dualities}

In the dipolar Hamiltonian it is known that there are two dualities---first, switching the sign of $J_{z\pm}$ is equivalent to to a $\pi$ pseudo-spin rotation about the easy-axis, which maps each irrep to itself. 
Second, when $J_{z\pm} = 0$, switching the sign of $J_{\pm\pm}$ is equivalent to a $\pi/2$ rotation about the easy-axis, which switches the Palmer-Chalker and splayed ferromagnetic phases (i.e. switches the easy-plane $T_1$ and $T_2$ irreps). 
Clearly these dualities are also present in the Hamiltonian \cref{eq:H_local_Jpm} when the couplings terms between the two easy-plane sectors, $J_{\pm}'$ and $J_{\pm\pm}'$, are zero.
When they are non-zero, we can can make a rotation between the two easy-plane sectors to remove them, then both dualities are present with respect to the decoupled basis.

Additionally, \cref{eq:interaction_matrix_Jloc} reveals a remarkable duality structure of the Hamiltonian, which \emph{exchanges} the two easy-plane sectors. 
We refer to this as anti-diagonal duality, since it transposes the matrix \cref{eq:interaction_matrix_Jloc} across the anti-diagonal. 
It is interesting to consider phase diagram cuts which are self-dual under this anti-diagonal duality. 
We show examples of such cuts in \cref{fig:phase_diagrams_duality}(a,b). 
\Cref{fig:phase_diagrams_duality}(a) shows precisely such a cut with both $J_{z\pm}^{r}$ couplings set to zero. 
This corresponds to making both easy-plane sectors from \cref{fig:dipolar_phase_diagrams}(a) and (b) degenerate, so all canting angles in each phase are degenerate.

Since both $J_{z\pm}^{r}$ are zero in \cref{fig:phase_diagrams_duality}(a), we can also combine the anti-diagonal duality with the two $J_{\pm\pm}^{r}$ dualities, switching the sign of one of the $J_{\pm\pm}^{r}$ couplings, in order to obtain an ``anti-self-dual'' phase diagram shown in \cref{fig:phase_diagrams_duality}(b).
This phase diagrams has the same form, but instead of having $T_1$ degenerate with $T_1$ and $T_2$ degenerate with $T_2$ in the two sectors, we have $T_1$ degenerate with $T_2$ and vice-versa in the two sectors.

Lastly, \cref{fig:phase_diagrams_duality}(c) demonstrates the dualities under switching the signs of either $J_{z\pm}^{r}$ coupling. 
Here we take $J_{zz} > 0$ to put the quadrupolar ice phase in the center, and vary $\smash{J_{z\pm}^{t_{2g}}}$ and $\smash{J_{z\pm}^{e_g}}$, which continuously mixes the three $T_2$ irreps. 
This cut shows symmetry under flipping the sign of either of the $J_{z\pm}$ couplings.

\subsection{Diabolical Loci}

\begin{figure*}
    \centering
    \begin{overpic}[width=\textwidth]{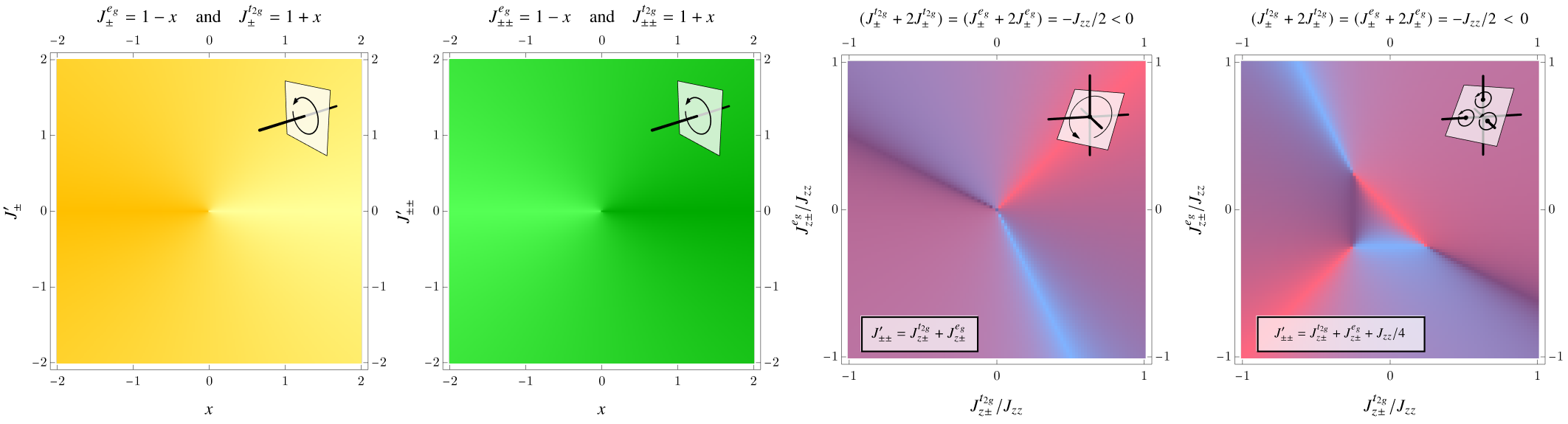}
        \put(00.0,25.5){(a)}
        \put(25.0,25.5){(b)}
        \put(50.5,25.5){(c)}
        \put(75.5,25.5){(d)}
    \end{overpic}
    \caption{Examples of Luttinger-Tisza phase diagram cuts which intersect the diabolical loci where multiplets in the same irrep become degenerate with each other. Insets illustrate the geometry of the cuts relative to the loci that they intersect. (a) Cutting the $E$ locus. (b) Cutting the $T_1$ locus. (c) Cutting through the triply degenerate $T_2$ locus, where all three $T_2$ irreps are degenerate. (d) The triple $T_2$ locus sits at the intersection of three doubly $T_2$ loci, and here we make a cut of the phase diagram which intersects all three, showing three singular points. The color legend is the same used in \cref{fig:dipolar_phase_diagrams}. Unspecified couplings are zero.
    }
    \label{fig:diabolical_loci}
\end{figure*}
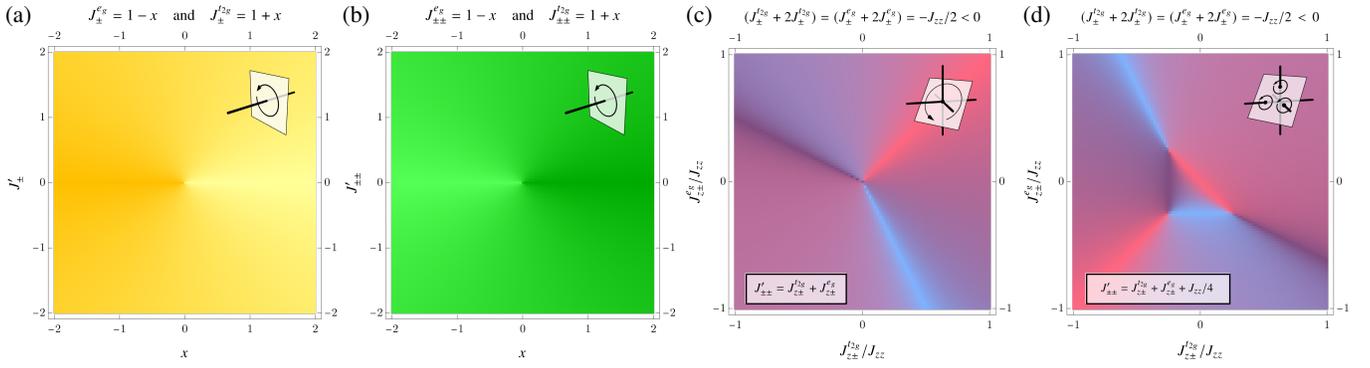

Ref.~\cite{chungMappingPhaseDiagram2024} demonstrated that in the phase diagram of the dipolar Hamiltonian there is a locus along which the doubly repeated irrep becomes degenerate with itself.
Whereas a phase boundary occurs in codimension-1, meaning that a single parameter must be tuned to cross it, the locus on which an irrep can become degenerate with another copy of itself codimension-2. 
This is a manifestation of level repulsion of matrix eigenvalues, due to the fact that symmetry does not forbid a coupling between copies of the same irrep. 
For example, using \cref{eq:irrep_matrix_E,eq:irrep_matrix_T1}, in order to make the two $E$ irreps or the two $T_1$ irreps degenerate, it is clear that we must set the two diagonal elements of the matrix to be the same and the off-diagonal matrix element to be zero, thus requiring two parameters to be tuned to find this locus.
As such, these loci are analogs of diabolical points (conical intersections) in quantum mechanics, or diabolical loci in quantum field theory~\cite{hsinBerryPhaseQuantum2020}.
They may be viewed as sources of Berry curvature in the space of couplings, in the sense that if the system is adiabatically transported around a path linking the locus, the corresponding ground state canting angle must wind non-trivially.
\Cref{fig:diabolical_loci}(a) and (b) show examples of phase diagram cuts which intersect the diabolical loci of the $E$ and $T_1$ phases.

One particularly interesting aspect of the pyrochlore quadrupole Hamiltonian is that it contains three copies of the $T_2$ irrep. 
Making any two of them degenerate again requires tuning two parameters, but it is also possible for all three to be degenerate. 
This occurs in codimension-5, because we must set the three off-diagonal elements to be zero and set two pairs of diagonal entries to be the same. 
Clearly all three of the double-degenerate $T_2$ loci must intersect at the triple $T_2$ locus. 
\Cref{fig:diabolical_loci}(c) shows a cut through the triple-degenerate locus, showing a singularity at the center around which variation in all three $T_2$ irrep components can be seen. 
\Cref{fig:diabolical_loci}(d) shows a cut where all three double-degenerate $T_2$ singularities are visible.

\section{Quantum Quadrupoles}
\label{sec:quantum_quadrupoles}

For a dipolar Hamiltonian, involving only bilinears of spin operators, the standard semi-classical approximation is to replace the spin operators by classical unit-length vectors. 
The situation for quadrupoles is much more subtle, and a semi-classical treatment requires a careful analysis. 
For example, while it is true that the quadrupole operators are bilinears of spin operators, if we simply replace those spin operators with vectors then we would be restricted to only describe uniaxial quadrupole configurations, and would fail to capture the biaxial configurations which appear as ground states of the single-tetrahedron Hamiltonian illustrated in \cref{fig:ground_states}. 
In this section we study the problem of the semi-classical treatment of quadrupoles in detail. 
We briefly review for juxtaposition the case of dipoles, where the semi-classical configuration space is the unit sphere in three-dimensions.

\subsection{Dipolar Coherent States and Unit Vectors}

We begin by briefly reviewing the standard approach for dipoles, in order to contrast it with what we find for quadrupoles. 
Consider the problem of identifying the mean-field ground state of a bilinear dipole Hamiltonian of the form~\cref{eq:Hamiltonian-S-Q} without the quadrupolar terms. 
At a single site, the mean-field Hamiltonian has the form
\begin{equation}
    H_{S,j}^{\text{MF}} = \sum_{j} \langle S_i^\alpha \rangle [\mathcal{J}_S]_{ij}^{\alpha\beta} S_j^\beta \equiv -\vert \bm{h}_j\vert \hat{\bm{h}}_j \cdot \bm{S}_j
\end{equation}
where $\bm{h}_j$ is the local molecular field due to the neighboring spin configurations. 
The mean field ground state, which neglects entanglement between different spins, must be the maximum weight eigenstate of the operator $\hat{\bm{h}}\cdot \bm{S}$. 
Such a state is a coherent state, 
\begin{align}
    \ket{\hat{\bm{n}}} &= \text{max. positive weight eigenstate of } \hat{\bm{n}}\cdot \bm{S} 
    \nonumber
    \\
    &= R(\theta,\phi) \ket{S},   
    \label{eq:dipolar_coherent_state}
\end{align}
where $\hat{\bm{n}}$ is the direction of the dipole moment and $R(\theta,\phi)$ the rotation operator from the global $z$-axis to $\hat{n}$.
The states $\ket{\hat{\bm{n}}}$ maximize the dipole moment $\vert \langle \bm{S} \rangle \vert$, such that it remains to minimize the interaction energy by adjusting the angular directions $\hat{\bm{n}}$. 
Thus the semi-classical description of a dipolar Hamiltonian is in terms of unit length 3-component vectors, whose configuration space is a 2-sphere, regardless of the spin quantum number. 
Semi-classical finite-temperature Monte Carlo simulations can be performed with respect to the classical energy functional
\begin{align}
    H^{\text{MC}}_S 
    &=  
    \sum_{ij} 
    \sum_{\alpha\beta} 
    \bra{{\hat{\bm{n}}_i}}{S_i^\alpha}\ket{\hat{\bm{n}}_i}
    \,
    [\mathcal{J}_S]_{ij}^{\alpha\beta} 
    \,
    \bra{\hat{\bm{n}}_j} S_j^\beta \ket{\hat{\bm{n}}_j},
    \nonumber
    \\
    &= 
    S^2 
    \sum_{ij} 
    \sum_{\alpha\beta} 
    \hat{n}_i^\alpha
    [\mathcal{J}_S]_{ij}^{\alpha\beta} 
    \hat{n}_j^\beta
    \label{eq:dipolar_monte_carlo_coherent_states}
\end{align}
where the unit vectors $\hat{\bm{n}}_i$ are varied, interpreted as varying the angular direction of the local expectation value of the dipole moment. 
This is equivalent to using a variational product state ansatz for the many-body wavefunction, $\ket{\psi} = \bigotimes_i \ket{\hat{\bm{n}}_i}$.
This reproduces the mean-field zero-temperature ground states and agrees with the lowest order of a high-temperature series expansion of the partition function~\cite{stoudenmire2009}.
The weighting of angular configurations on the surface of the 2-sphere are uniform, because all of the spin operators $\hat{\bm{n}}\cdot \bm{S}$ are unitarily related.

\subsection{Quadrupolar Coherent States and Biaxial Nematics}

Now consider the same problem for a quadrupolar Hamiltonian. The mean-field Hamiltonian can be written 
\begin{align}
    H_{Q,j}^{\text{MF}} &= \sum_{j} \langle Q_i^{\alpha'\beta'} \rangle [\mathcal{J}_Q]_{ij}^{\alpha'\beta',\alpha\beta} Q_j^{\alpha\beta}
    \nonumber
    \\
    &\equiv 
    -\Tr[{q}^2_j]^{\frac{1}{2}} \Tr[\hat{{q}}_j  Q_j]
\end{align}
where $q$ is a symmetric trace-free tensor, $\Tr[q^2]^{1/2}$ is its Frobenius norm, and $\hat{q}$ is the normalized form of the tensor with unit norm, describing the ``angular configuration'' analogous to $\hat{\bm{n}}$. 
The ground states of such a Hamiltonian are maximal weight eigenstates of quadrupolar operators $\Tr[\hat{q}\, Q]$, which we will refer to as ``quadrupolar coherent states'',
\begin{equation}
    \ket{\hat{q}} = \text{max. positive weight eigenstate of } \Tr[\hat{q} \, Q]
    \label{eq:quad_coherent_state}
\end{equation}
Such states are not generally constructed by simple rotations of a fixed state as in the dipolar coherent states \cref{eq:dipolar_coherent_state}.
Only uniaxial quadrupole states can be written in this way, but the space of quadrupolar configurations also contains biaxial states. 
Rather, the tensor $\hat{q}$ lives on the unit 4-sphere in the five-dimensional space of symmetric trace-free matrices, which parameterizes the set of quadrupolar coherent states in the same way that the 2-sphere parameterizes the set of dipolar coherent states. 
The map from 5-component vectors to symmetric tensors is given by 
\begin{equation}
    \begin{pmatrix}
        q_x \\
        q_y \\ 
        q_z \\
        q_{3z^2 - r^2}\\
        q_{x^2 - y^2}
    \end{pmatrix}
    \mapsto
    \begin{pmatrix}
        -\frac{q_{3z^2 - r^2}}{\sqrt{6}}+\frac{q_{x^2 - y^2}}{\sqrt{2}}
        & 
        \frac{q_x}{\sqrt{2}}
        & 
        \frac{q_y}{\sqrt{2}} 
        \\[1ex]
        \frac{q_x}{\sqrt{2}} 
        & 
        -\frac{q_{3z^2 - r^2}}{\sqrt{6}}-\frac{q_{x^2 - y^2}}{\sqrt{2}} 
        & 
        \frac{q_z}{\sqrt{2}} 
        \\[1ex]
        \frac{q_y}{\sqrt{2}} 
        & 
        \frac{q_z}{\sqrt{2}} 
        &
        \frac{2q_{3z^2 - r^2}}{\sqrt{6}}
    \end{pmatrix}\!.
    \label{eq:vector_to_tensor}
\end{equation}
We can interpret this 4-sphere as the configuration space of a biaxial nematic with fixed volume. 
This is naively the natural candidate for the semi-classical configuration space of quantum quadrupoles, akin to the unit vectors for dipoles.
A given biaxial nematic configuration can be expressed in terms of an orthonormal frame $\hat{\bm{e}}_\alpha$ as 
\begin{equation}
    q_{\text{diag}} = \frac{\mathcal{S}}{\sqrt{6}}\left( 3\hat{\bm{e}}_z \otimes \hat{\bm{e}}_z -  \mathds{1}\right) + \frac{\mathcal{T}}{\sqrt{2}} \left(\hat{\bm{e}}_x \otimes \hat{\bm{e}}_x - \hat{\bm{e}}_y \otimes \hat{\bm{e}}_y \right),
    \label{eq:biaxial_tensor}
\end{equation}
where $\mathcal{T} = 0$ is the pure uniaxial case and $\mathcal{S}=0$ is the pure biaxial case, normalized so that $\Tr[q_{\text{diag}}^2] = \mathcal{S}^2 + \mathcal{T}^2$.
A general biaxial nematic configuration can then be described by performing an SO(3) rotation of this tensor and varying the $\mathcal{S}$ and $\mathcal{T}$ parameters,\footnote{
    This description is redundant and covers the 4-sphere multiple times, for example $q(0,\pi/2,0,\mathcal{S},-\mathcal{T}) = q(0,0,0,\mathcal{S},\mathcal{T})$. In practice one can simply work with a unit five-component vector and reconstruct the symmetric trace-free matrix without any redundancy. 
}
\begin{equation}
    q(\theta,\phi,\psi,\mathcal{S},\mathcal{T}) = R(\theta,\phi,\psi) q_{\text{diag}}(\mathcal{S},\mathcal{T}) R(\theta,\phi,\psi)^{-1}.
    \label{eq:general_biaxial_rotated}
\end{equation}
Any symmetric trace-free tensor, such as the expected quadrupole moment $\langle Q^{\alpha\beta}\rangle$, can be written in the form of \cref{eq:biaxial_tensor} in its eigenframe.
This provides an intuitive picture of semi-classical quadrupole configurations, analogous to the unit vectors for semi-classical dipole configurations, but it does not address the issue of how different classical quadrupole configurations ought to be weighted.

\begin{figure*}[t]
    \centering
    \begin{overpic}[width=.99\textwidth]{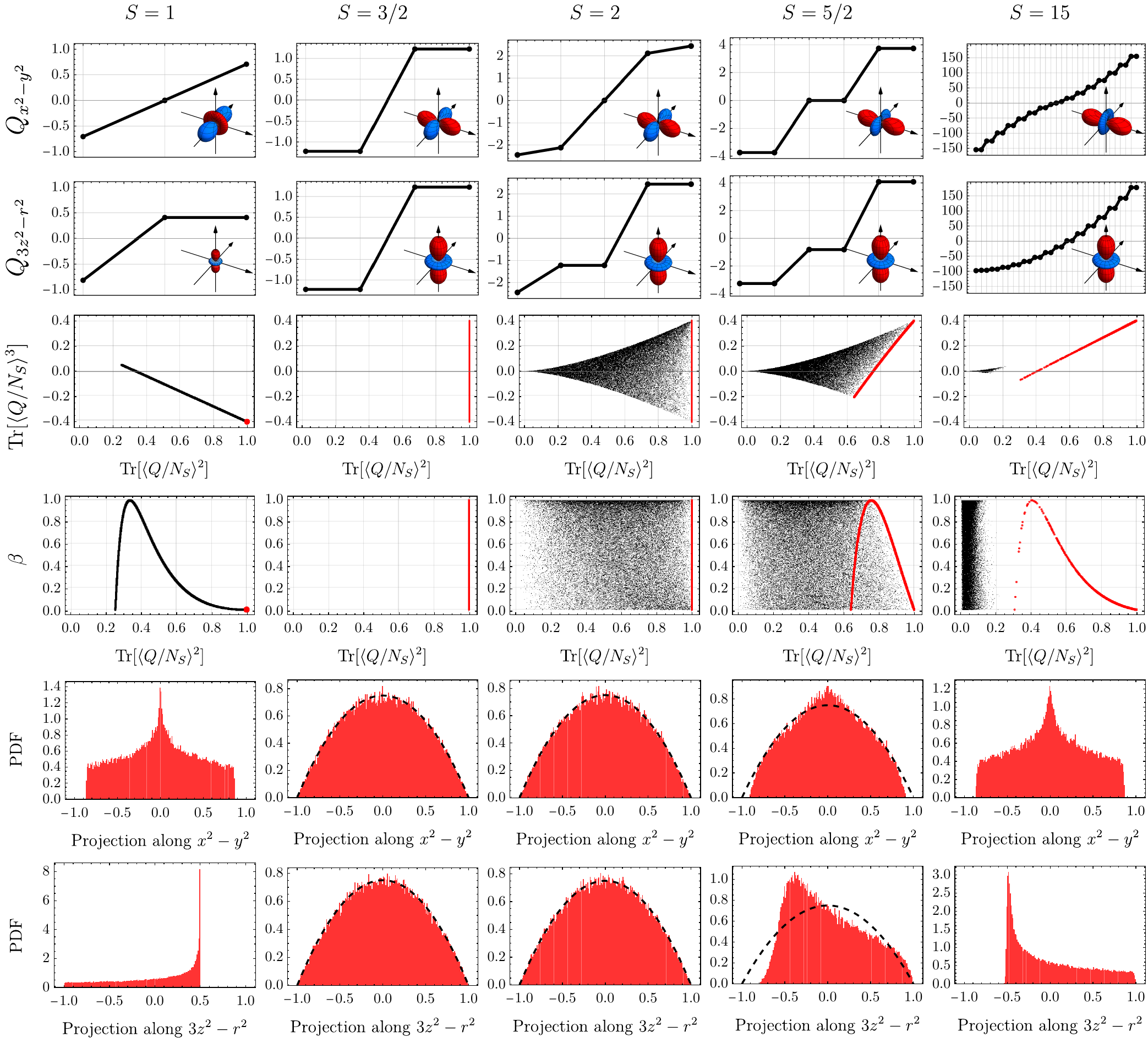}
        \put(0.5,87.0){(a)}
        \put(0.5,75.0){(b)}
        \put(0.5,62.5){(c)}
        \put(0.5,47.0){(d)}
        \put(0.5,30.0){(e)}
        \put(0.5,14.0){(f)}
        \put(9.0,12.0){\footnotesize disk-like}
        \put(9.2,10.0){\footnotesize uniaxial}
        \put(91.0,12.0){\footnotesize rod-like}
        \put(91.0,10.0){\footnotesize uniaxial}
    \end{overpic}
    \caption{Plots showing the $S$-dependence of the expected quadrupole moment. (a) Spectrum of $\smash{Q_{x^2 - y^2} = [(S^x)^2 - (S^y)^2]/\sqrt{2}}$. The operators $Q_{xy}$, $Q_{yz}$, and $Q_{xz}$ have the same spectrum. Inset: the shape of the expected quadrupole moment $\langle Q^{\alpha\beta}\rangle$ evaluated in the maximal weight eigenstate of $Q_{x^2 - y^2}$. 
    For $S=3/2$ and $S=2$ the shape matches that of the corresponding orbital shown in \cref{fig:orbitals}; for $S=1$ it is a disk-like uniaxial with director along $y$; for $S>2$ it tends towards a rod-like uniaxial moment oriented along $x$. 
    (b) Spectrum of $\smash{Q_{3z^2 -r^2}}$. The spectrum matches that of the other four operators only in the case $S=3/2$. For $S=2$ the maximum eigenvalues are the same, while for $S>2$ the maximum eigenvalue of $\smash{Q_{3z^2 -r^2}}$ is larger. 
    (c) Plots of $\smash{\Tr[\langle Q/N_S\rangle^3]}$ versus $\smash{\Tr[\langle Q/N_S\rangle^2]}$ for randomly sampled spin-$S$ states (black dots) and randomly sampled quadrupolar coherent states (red).
    (d) Variation of biaxial parameter $\beta$ versus $\Tr[\langle Q/N_S\rangle^2]$. Note that the endpoints of the red curves at $\beta = 0$ correspond to the states $\ket{S}+\ket{-S}$ (with maximum value of $\Tr[\langle Q \rangle^2]$) and either $\ket{0}$ for integer spin or $\ket{1/2} + \ket{- 1/2}$ for half-integer spin (with minimal value of $\Tr[\langle Q \rangle^2]$). In the large-$S$ limit the red curve approaches the same distribution as the $S=1$ case. 
    (e) Probability density function (PDF) of the $x^2 - y^2$ component of $\bra{\hat{q}} Q^{\alpha\beta}\ket{\hat{q}}$ evaluated for random coherent states with uniform distribution of $\hat{q}$ on the unit 4-sphere in the space of symmetric trace-free tensors.
    Dashed lines show the corresponding distribution of uniformly distributed points on the 4-sphere. 
    (f) Distribution of $3z^2 - r^2$ components of coherent states. For $S=3/2$ and $S=2$ the distribution appears uniform on the 4-sphere, but for other values of $S$ the $3z^2 - r^2$ components shows significant anisotropy and cannot take all values in $[-1,1]$.
    }
    \label{fig:quantum-spin-S-dependence}
\end{figure*}
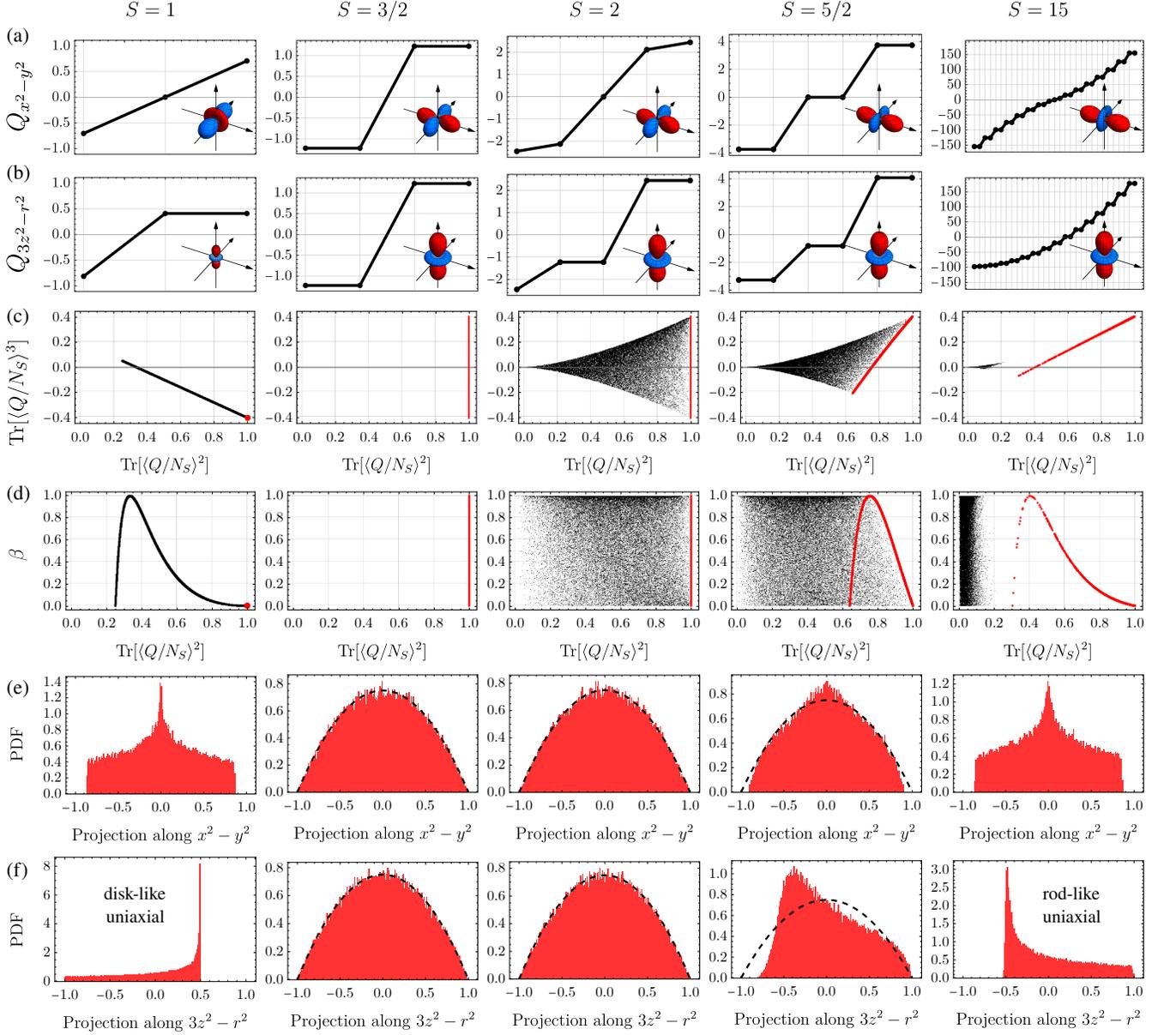

\subsection{Dependence on Spin Quantum Number}
\label{sec:S_dependence}

In fact, the answer to the question of how to appropriately weight the quadrupolar states is strongly dependent on the spin quantum number $S$.  
A key reason for this, which highlights the distinction of quadrupoles compared to dipoles, is that, while there are five linearly independent quadrupole operators transforming irreducibly under the action of SO(3), the operator $Q_{3z^2 - r^2}$ is \emph{not unitarily equivalent} to $Q_{xy}$, $Q_{yz}$, $Q_{xz}$, or $Q_{x^2 - y^2}$, which are all unitarily equivalent to each other by a rotation. 
This is clearly seen in the corresponding orbital structures shown in \cref{fig:orbitals}---four of them are just rotated versions of each other while the fifth is not related by a rotation but rather a change in the $\mathcal{S}$ and $\mathcal{T}$ parameters. 
In terms of \cref{eq:biaxial_tensor}, $Q_{3z^2 - r^2}$ corresponds to a $\mathcal{T}=0$ uniaxial tensor, while the other four are all $\mathcal{S}=0$ biaxial tensors related by rotations. 
Indeed, the spectrum of $Q_{3z^2 - r^2}$ is different from the others, as shown in \cref{fig:quantum-spin-S-dependence}(a,b). 
For $S>2$ its maximal eigenvalue is larger than the others', and it grows more quickly in the $S\to \infty$ limit. 
As a result, uniaxial states tend to dominate at large spin, and there is a tradeoff between maximizing the total magnitude of the quadrupole moment, as measured by $\Tr[\langle Q\rangle^2]$, and satisfying the interaction energy between neighboring quadrupoles. 
Meanwhile, the behavior for $S\leq 2$ is strongly constrained.
This does not occur for dipoles, where all three operators are unitarily related and the semi-classical description is given by uniform weighting of the 2-sphere for any spin quantum number.

There are two invariants for a symmetric trace-free tensor, $\Tr[q^2]$ and $\Tr[q^3]$. 
For uniaxial configurations, the sign of $\Tr[q^3]$ indicates whether it is rod-like (positive) or disk-like (negative).\footnote{
    Note that for unaxial tensors with $T=0$ in \cref{eq:biaxial_tensor}, $\Tr[q^3]=S^3/\sqrt{6}$, so the sign of $\Tr[q^3]$ is equivalent to the sign of $S$.
}
From them we can define a parameter quantifying the degree of biaxiality
\begin{equation}
    \beta = 1 - 6 \frac{\Tr[q^3]^2}{\Tr[q^2]^3} \in [0,1]
\end{equation}
which is zero for purely uniaxial configurations and unity for purely biaxial configurations. 
For a given spin quantum number $S$, the maximum of $\Tr[\langle Q \rangle^2]$ occurs for the states 
\begin{equation}
    \ket{\psi_{\text{uni}}} = \frac{1}{\sqrt{2}}\big(\ket{S} + \ket{-S}\big),
\end{equation}
which has maximal uniaxial quadrupole moment.
The maximum values are given by\footnote{
    For $S>1$ it is sufficient to evaluate $\bra{S} Q^{\alpha\beta} \ket{S}$, but for $S=1$ there is also a non-trivial cross term coming from $\bra{-1} Q^{\alpha\beta} \ket{+1}$. 
}
\begin{equation}
    N_S^2 = \Tr[\bra{\psi_{\text{uni}}} Q \ket{\psi_{\text{uni}}}^2] = \begin{cases}
        S^2 (1-2S)^2 / 6 &\quad S>1, \\
        2/3 &\quad S=1.
    \end{cases}
    \label{eq:quadrupole_normalization}
\end{equation}
With these definitions we are prepared to explore the variation of the behavior of quantum and semi-classical quadrupoles for different spin quantum number. 
Normalizing the quadrupole operator by the square root of this quantity, in \cref{fig:quantum-spin-S-dependence}(c,d) we plot the variation of $\Tr[\langle Q/N_S \rangle^2]$ versus $\Tr[\langle Q/N_S \rangle^3]$ and $\beta$ for a uniform sampling of random states $\ket{\psi}$ in the spin-$S$ Hilbert space, and a uniform sampling of random quadrupole coherent states $\ket{\hat{q}}$, \cref{eq:quad_coherent_state}. 
\cref{fig:quantum-spin-S-dependence}(e,f) show the distributions of the uniaxial and biaxial components of randomly generated quadrupolar coherent states. 
The results are strikingly different for $S=1$, $3/2$, $2$ and $S> 2$.

\subsubsection{\texorpdfstring{$S=1$}{S=1}}

Referring to \cref{fig:quantum-spin-S-dependence}(c,d), for $S=1$ there is a strong constraint on the possible values of the two invariants\footnote{
    We thank Karlo Penc for pointing out that such a constraint exists for $S=1$.
}
\begin{equation}
    \Tr[\langle Q \rangle^3] + \frac{1}{2} \Tr[\langle Q \rangle^2]= \frac{1}{9}  \qquad (S=1)
\end{equation}
and the moment is bounded, ${\Tr[\langle Q \rangle^2] \in [1/6,2/3]}$.
This means that are no states which have zero quadrupole moment (i.e. $\Tr[\langle Q \rangle^2] = 0$), and the biaxiality of a state is determined entirely by the magnitude of the quadrupole moment. 
The only states which saturate the quadrupole moment, i.e. with $\Tr[\langle Q \rangle^2] = 2/3$, are disk-like uniaxial ones, with $\beta = 0$ and $\Tr[\langle Q\rangle^3] < 0$. 
All of the quadrupolar coherent states are such maximal quadrupole moment disk-like uniaxial ones, seen as a red point in \cref{fig:quantum-spin-S-dependence}(c,d). 
For this case an appropriate semi-classical description is therefore in terms of a uniaxial disk-like nematic.
This can only explore a subspace of the 4-sphere isomorphic to the real projective plane $\mathds{RP}^2$, i.e. a 2-sphere with antipodal points identified. 
To see this more clearly, in \cref{fig:quantum-spin-S-dependence}(e,f) we have plotted the distribution of the biaxial and uniaxial components, respectively, of the 5-component vectors (\cref{eq:t2g_eg_vector_mapping}) corresponding to the expected quadrupole tensor in a quadrupolar coherent state. 
This distribution exactly matches that of a disk-like uniaxial tensor, i.e. one of the form
\begin{equation}
    q_{\text{disk-uniaxial} } = -\frac{1}{\sqrt{6}}\left(3\,\hat{n}^\alpha \,\hat{n}^\beta - \delta^{\alpha\beta}\right)
    \label{eq:disk_uniaxial_tensor}
\end{equation}
for a random 3-component unit vector $\hat{\bm{n}}$.

\subsubsection{\texorpdfstring{$S=3/2$}{S=3/2} and \texorpdfstring{$S=2$}{S=2}}

The case $S=3/2$ is also highly constrained, in that \emph{every} state has maximal magnitude of the quadrupole moment. 
In fact, $S=3/2$ is the only case where all five quadrupole operators have the same spectrum, as seen in in \cref{fig:quantum-spin-S-dependence}(a,b). 
In the case $S=2$, the spectra of the quadrupole operators are not identical, but the maximum eigenvalues are. 
Indeed one can check numerically that the quadrupole coherent states satisfy
\begin{equation}
    \bra{\hat{q}} Q^{\alpha\beta}\ket{\hat{q}} \propto \hat{q}^{\alpha\beta},
    \label{eq:coherent_4-sphere_distribution}
\end{equation}
for $S=3/2$ and $S=2$, directly parallel to the equivalent statement for dipolar coherent states (which holds for arbitrarily spin quantum number), but not for other values of $S$. 
Thus for both $S=3/2$ and $S=2$ we expect the semi-classical configuration space, corresponding to the space of expected coherent-state quadrupole moments $\bra{\hat{q}}Q^{\alpha\beta} \ket{\hat{q}}$, to be the 4-sphere of unit-norm symmetric trace-free tensors.
For $S=3/2$ the states are uniformly weighted on this sphere because all quadrupole operators are unitarily equivalent. This has the important implication that spin-$3/2$ quadrupoles behave like spins on the 4-sphere (i.e. unit-length 5-component vectors) with an effective orientation given by the sign of the quadrupole and the bilinear quadrupolar model built from these has an effective time reversal symmetry that is broken by ordered states.
For $S=2$ the quadrupolar expectation values $\langle Q^{\alpha\beta}\rangle$ can take arbitrary values within the unit sphere for a general (non-coherent) state, reflected in the ``cloud'' of black points in \cref{fig:quantum-spin-S-dependence}(c,d), and the coherent states indeed explore all of the maximal quadrupolar moment configurations with arbitrary biaxiality. 
Furthermore, \cref{fig:quantum-spin-S-dependence}(e,f) demonstrate that the distributions are the same for the uniaxial and biaxial components of the coherent state quadrupole moments, as implied from \cref{eq:coherent_4-sphere_distribution}.

\subsubsection{\texorpdfstring{$S>2$}{S>2}}
\label{sec:S_gt_2_quantum_quadrupoles}

All spins $S > 2$ are qualitatively similar to each other, uniformly approaching the large-$S$ limit. 
In all of these cases the maximum eigenvalue of the uniaxial operator $Q_{3z^2 - r^2}$ is larger than that of the four biaxial operators, 
approaching the limiting ratio
\begin{equation}
    \lim_{S\to\infty} \frac{\text{max. eigenvalue of } Q_{\text{uniaxial}}}{\text{max. eigenvalue of }Q_{\text{biaxial}}} = \frac{\sqrt{3}}{2}.
\end{equation}
They share the property that only rod-like uniaxial states ($\Tr[\langle Q \rangle^3] > 0)$ maximize the quadrupole moment.
This is clearest in the allowed values of the biaxial parameter $\beta$---the only states which have maximum quadrupolar moment are uniaxial ones with $\beta = 0$, and biaxial states with large $\beta$ necessarily have suppressed quadrupole moment. 
This effect becomes more pronounced for larger spin. 
Therefore for $S>2$ we expect a competition between maximizing the quadrupole moment and satisfying the interaction energy between quadrupoles, which may prefer biaxial configurations. 
Interestingly, the limiting curve for traced by the coherent states in \cref{fig:quantum-spin-S-dependence}(c,d) appears to be the same as the curve for random $S=1$ states, but with the opposite sign of $\Tr[\langle Q\rangle^3]$.
This is further emphasized by \cref{fig:quantum-spin-S-dependence}(e,f), which shows that the distribution tends towards that of a rod-like uniaxial tensor, i.e. one of the form \cref{eq:disk_uniaxial_tensor} but with opposite sign.
This partially justifies the replacement of spin operators with unit vectors in the definition of the quadrupole tensor, \cref{eq:quadrupole_operator}, resulting in only rod-like uniaxial states. 
Importantly, however, the quadrupolar coherent states are still able to explore the full range of biaxial configurations, despite being heavily weighted towards the uniaxial ones, so such an approximation is ultimately not fully appropriate as it cannot capture biaxial states, which are present in the quantum Hilbert space.

\begin{figure*}[th]
    \centering
    \begin{overpic}[width=.9\textwidth]{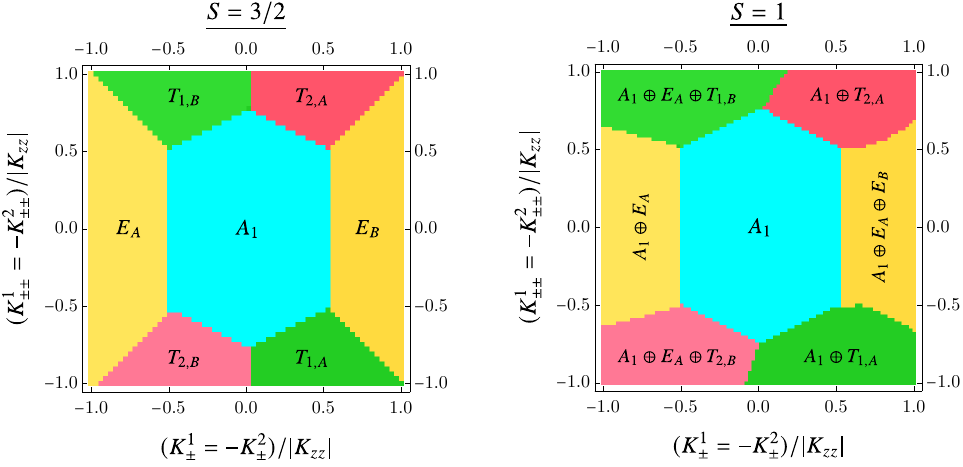}
        \put(01,45){(a)}
        \put(54,45){(b)}
    \end{overpic}
    \caption{Mean field phase diagrams equivalent to \cref{fig:phase_diagram_diamond} but in the spherical harmonics basis of \cref{eq:H_local_Kpm}, for (a) $S=3/2$ and (b) $S=1$, showing the spin dependence of the ground states and phase boundaries. 
    The $S=3/2$ phase boundaries agree with Luttinger Tisza, while for $S=1$ the strong constraints force uniaxial $A_1$ order to coexist with the biaxial orders. 
    An $E$ irrep is also switched on, since one the $E$ irreps include global uniaxial ferro-quadrupolar order (c.f. \cref{fig:ground_states} bottom right panel labeled $\smash{E_{\,\smash{\psi_2}}^{\smash{(e_g)}}}$). 
    The labeling of the irreps refers to the local frame states listed in Appendix~\ref{apx:irrepslocal}, where a subscript $A$ ($B$) refers to $\vert m \vert = 2$ ($1$). 
    For these phase diagrams $K_{zz} < 0$ and $K_{\pm}' = K_{\pm\pm}' = 0$.
    The color scheme of (a) is the same as in\cref{fig:dipolar_phase_diagrams}, while in (b) we keep the same color scheme for the phases without adding additional colors for the extra switched on irreps. 
    }
    \label{fig:mean_field_phase_diagram_S1_vs_S32}
\end{figure*}

\subsection{Semi-Classical Monte Carlo Simulation of Quadrupoles}
\label{sec:monte_carlo}

To perform classical Monte Carlo simulations for a quadrupolar Hamiltonian of the form of \cref{eq:Hamiltonian-Q-5-generic}, one could evaluate it in the quadrupolar coherent state basis, directly analogous to \cref{eq:dipolar_monte_carlo_coherent_states},
\begin{align}
    H^{\text{MC}}_Q 
    &=  
    \sum_{ij} 
    \sum_{a,b=1}^5 
    \bra{{\hat{q}_i}} Q_i^{a} \ket{\hat{q}_i}
    \,
    \mathcal{J}_{ij}^{ab} 
    \,
    \bra{\hat{q}_j} Q_j^{b} \ket{\hat{q}_j},
    \nonumber
    \\
    &= 
    N_S^2
    \sum_{ij} 
    \sum_{a,b=1}^5 
    \tilde{q}^a_i \mathcal{J}_{ij}^{ab} \tilde{q}^b_j,
    \label{eq:H_MC_quadrupolar_coherent}
\end{align}
where $\tilde{\bm{q}}_i$ are classical 5-component vectors with $\vert \tilde{\bm{q}}_i\vert \leq 1$ defined by the sequence of maps
\begin{equation}
    \hat{q} \in S^4 \mapsto \ket{\hat{q}} \in \mathds{CP}^{2S} \mapsto \tilde{\bm{q}} = \sum_{a=1}^5\bra{\hat{q}} Q^{a} \ket{\hat{q}}\,\tilde{\bm{\epsilon}}_a \in\mathds{R}^5,
    \label{eq:coherent_4sphere_mapping}
\end{equation}
In the first map we take a point on the 4-sphere (the space of quadrupole operators up to scale) and obtain the corresponding quadrupolar coherent state in the complex projective space of quantum states via \cref{eq:quad_coherent_state}. 
In the second we compute the quadrupole moment expectation value, mapped to a 5-component vector via \cref{eq:t2g_eg_vector_mapping}, where $\tilde{\bm{\epsilon}}_a$ are orthonormal basis vectors in the space of symmetric trace-free matrices. 
This is a well-defined Monte Carlo protocol, which correctly captures the number of degrees of freedom of a classical quadrupole, but involves diagonalizing the operator $\hat{q}_i^a Q^a$ for each update, and so is quite inefficient. 
It would be preferable if we could directly identify the classical configuration space, which depends on $S$ and is the image of the mapping $S^4 \to \mathds{R}^5$ given in \cref{eq:coherent_4sphere_mapping}, and sample it directly. 
However, it is unclear how to analytically identify it or the corresponding distribution. 
For $S=3/2$ and $S=2$, the image is apparently just the 4-sphere again (e.g. due to \cref{eq:coherent_4-sphere_distribution}), but the distribution on the 4-sphere may not be uniform (though numerically, according to \cref{fig:quantum-spin-S-dependence}(e,f), it does appear to be).

To sidestep these concerns, we take a somewhat simpler approach to semi-classical simulation. 
We treat the quantum problem exactly at the level of a single site, by taking the internal configuration space to be that of unit length ${(2S+1)}$-component complex vectors representing a general spin-$S$ product state
\begin{equation}
    \ket{\psi_{\text{sc}}} = \bigotimes_i \ket{\psi}_i , \qquad \ket{\psi}_i = 
    \sum_{m=-S}^S \psi_i^m \ket{m}_i 
\end{equation}
where $\ket{m}_i$ are the local eigenstates of $S_i^z$. 
This method was introduced in Ref.~\cite{stoudenmire2009} for $S=1$, where it was called ``semi-classical SU(3)'' since it simulates 3-component complex unit vectors. 
Its direct generalization may naturally be referred to as semi-classical SU($2S+1$). 
We then perform finite-temperature classical Monte Carlo simulations wherein the product state is varied according to a Metropolis update with respect to the variational energy given by 
\begin{align}
    E_{\text{sc}}(\psi) &= \bra{\psi_{\text{sc}}}H \ket{\psi_{\text{sc}}} 
    \\
    &= \sum_{ij} \sum_{\substack{\alpha\beta \\ \alpha'\beta'}} \bra{\psi_i} Q_i^{\alpha\beta}\ket{\psi_i}\, \mathcal{J}_{ij}^{\alpha\beta,\alpha'\beta'} \bra{\psi_j} Q_j^{\alpha'\beta'} \ket{\psi_j},
    \nonumber
\end{align}
As argued in Ref.~\cite{stoudenmire2009}, this agrees at zero temperature with the mean-field ground state and corresponds to the first-order term in a high-temperature series expansion of the partition function. 
It allows us to capture single-site quantum fluctuations and the correct distribution of quadrupole moments without the need for identifying the precise semi-classical configuration space.

The downside to this method is that it does not correctly capture the number of independent degrees of freedom, because the number of fluctuating modes increases linearly with the dimension of the Hilbert space, $2S+1$. 
For example, ferro-quadrupolar order corresponds to the ordering of the 5-component quadrupolar order parameter, leading to four quadratic modes and a low-temperature specific heat of $2k_B$ per spin, while this method would predict $4S$ quadratic modes, arising from fluctuations of a $(2S+1)$ component complex unit vector minus one ordered real component and one real gauge degree of freedom in the overall phase of the wavefunction. 
Notably, this gives the correct count for $S=1/2$ dipolar order and $S=1$ quadrupolar order, but not for $S>1$ in either case.
Nevertheless, for the purposes of obtaining symmetry-broken ground states and order parameters, identifying phase transitions, and studying order-by-disorder effects, it should be a reliable method which sidesteps the issue of identifying the semi-classical configuration space, and is much more efficient to implement than having to compute the quadrupolar coherent state in \cref{eq:H_MC_quadrupolar_coherent} with each update.

\subsection{Spin dependence of Mean Field Ground States}

The foregoing discussion demonstrated that $S=3/2$ and $S=2$ are relatively isotropic in the space of quadrupoles, while all other spins host uniaxial quadrupoles when the quadrupolar moment is saturated. 
In general this fact has ramifications for the mean field ground state phase diagram. 
\Cref{fig:mean_field_phase_diagram_S1_vs_S32} gives one example showing spin dependence of the phase diagram.
It shows the analogous phase diagram cut to \cref{fig:phase_diagram_diamond} but in the local spherical harmonics frame defined in \cref{sec:local_basis}.
\Cref{fig:mean_field_phase_diagram_S1_vs_S32}(a) shows the $S=3/2$ phase diagram. 
In accordance with the isotropic of the configuration space, the ground states are pure irreps and the phases and phase boundaries agree with the results obtained from the Luttinger-Tisza analysis. 
The labeling of the irreps refers to the local frame given in Appendix~\ref{apx:irrepslocal}, where a subscript $A$ ($B$) refers to $\vert m \vert = 2$ (${\vert m \vert = 1}$).


For spin one, however, the constraints on the quadrupolar length and biaxiality alter the nature of the phases and the phase boundaries, as seen in the bottom panel. Although the ground state irrep observed from Luttinger-Tisza appears in the corresponding region of the $S=1$ phase diagram, a non-vanishing $A_{1}$ component is switched on simultaneously. In addition, for three of the seven regions we find, in addition, that an additional $E$ irrep is switched on. Furthermore, the left and right regions composed of $A_1\oplus E_A$ and $A_1\oplus E_A\oplus E_B$, respectively, are suppressed compared to the corresponding regions in the $S=3/2$ phase diagram.  

In general, we expect the Luttinger-Tisza calculation to be captured well for spins $3/2$ and $2$ and for ground states to have mixed irreps for all other spins.

\section{Order By Disorder Selection}
\label{sec:obd}

In the dipolar case it is well-known that the $E$ irrep ground state has an accidental semi-classical U(1) symmetry in the ground state manifold which is lifted by thermal or quantum fluctuations to $6$-fold degenerate discrete states of two possible types denoted $\psi_2$ (the $n\pi/3$ states in the local easy-plane as measured from the $+\hat{\bm{x}}_i$ axis) and $\psi_3$ (the $n\pi/3+ \pi/6$ states). This physics is directly relevant to the material \ce{Er2Ti2O7}~($\psi_2$)~\cite{championEr2Ti2O7EvidenceQuantum2003,savaryOrderQuantumDisorder2012}, and, most likely, also \ce{Yb2Ge2O7}~($\psi_3$)~\cite{dun2015}.

In the quadrupolar case, the two and three dimensional irreps are potentially continuously degenerate. However, in the $T_1$ and $T_2$ cases, only a discrete set of states can satisfy the constant quadrupolar length constraint. This leaves the $E$ sector. Unlike the dipolar case, the quadrupolar problem has two $E$ irreps. In each, the quadrupolar length constraint is satisfied by a continuous set of states implying a $U(1)$ degeneracy that may be broken by fluctuations as in the dipolar case. In this section, we study the fate of these states in the two $E$ irreps finding a rich set of order-by-disorder phenomena involving a complex interplay between the two $E$ irreps and the size of the spin. In general, six-fold selection is possible but fine-tuned, with three-fold selection being the typical outcome.

\subsection{Time Reversal Symmetry, Landau Theory, and 3-Fold Selection}

To set the scene, we discuss Landau theory within the $E$ sector where we see three-fold selection arising by symmetry. 
For this discussion it is most convenient to use the spherical harmonics basis from \cref{sec:local_basis}, whose behavior under rotations is straightforward and for which the $E$ irreps can be separated into the simple forms
\begin{align}
    \bm{\Phi}_{E_A} &= \begin{pmatrix}
        \Phi_{E_{A}}^{\psi_2}
        \\[1ex]
        \Phi_{E_{A}}^{\psi_3}
    \end{pmatrix}
    = \sum_{\mu}
    \begin{pmatrix}
        Q^{\bar{x}^2 - \bar{y}^2}_\mu
        \\[1ex]
        Q^{\bar{x}\bar{y}}_\mu
    \end{pmatrix}
    \qquad 
    (\vert m \vert = 2)
    \nonumber
    \\[2ex]
    \bm{\Phi}_{E_B} &= \begin{pmatrix}
        \Phi_{E_{B}}^{\psi_2}
        \\[1ex]
        \Phi_{E_{B}}^{\psi_3}
    \end{pmatrix}
    = \sum_{\mu}
    \begin{pmatrix}
        Q^{\bar{x}\bar{z}}_\mu
        \\[1ex]
        Q^{\bar{y}\bar{z}}_\mu
    \end{pmatrix}
    \qquad\quad \,
    (\vert m \vert = 1)
\label{eq:E_irreps_local}
\end{align}
We have observed that these appear at the quadratic level through
\begin{equation}
\eta_1 \vert \boldsymbol{\Phi}_{E_{A}} \vert^2 + \eta_2 \vert \boldsymbol{\Phi}_{E_{B}} \vert^2 + \eta_3 \boldsymbol{\Phi}_{E_{A}} \cdot  \boldsymbol{\Phi}_{E_{B}} 
\end{equation}
We then introduce
\begin{align}
    \Phi_{E_A}^\pm \equiv \Phi_{E_A}^{\psi_2} \pm i \Phi_{E_A}^{\psi_3},
    \nonumber
    \\
    \Phi_{E_B}^\pm \equiv \Phi_{E_B}^{\psi_2} \pm i \Phi_{E_B}^{\psi_3}.
\end{align}
In this basis, transformations under the point group $O_h$ are are simple as possible. In summary, the two-fold rotations about the cubic axes act as the identity in this frame. The eight $C_3$ rotations $(n=1,2,3)$ act as local frame rotations such as:
\begin{align}
\Phi_{E_A}^+ & \rightarrow e^{4\pi i n / 3} \Phi_{E_A}^+ \nonumber
\\
\Phi_{E_B}^+ & \rightarrow e^{2\pi i n / 3} \Phi_{E_B}^+ 
\end{align}
The remaining elements are two-fold rotations about axes in the local easy plane ($m=-1,0,1$)
\begin{align}
\Phi_{E_A}^+ & \rightarrow e^{ 2\pi i m / 3} \Phi_{E_A}^- \nonumber
\\
\Phi_{E_B}^+ & \rightarrow e^{ 4\pi i m / 3} \Phi_{E_B}^-. 
\end{align}
Time reversal symmetry does not forbid terms cubic in the quadrupoles as it does for dipoles. Symmetry allows: 
\begin{align}
\left( \Phi_{E_A}^+ \right)^3 + \left( \Phi_{E_A}^- \right)^3 \nonumber\\
\left( \Phi_{E_B}^+ \right)^3 + \left( \Phi_{E_B}^- \right)^3 
\end{align}
It also allows for cubic terms that combine irreps A and B 
\begin{align}
\left( \Phi_{E_A}^+ \right)^2  \Phi_{E_B}^- + {\rm c.c.}
\end{align} 

The presence of cubic terms in the Landau theory implies that the $U(1)$ degeneracy present at the quadratic order in $E_A$ and $E_B$ is generally broken by fluctuations. This order-by-disorder takes place in such a way that three states either at angles $2 n \pi  /3$ or  $\pi/3 + 2 n \pi  /3$ are selected. This is in contrast to the dipolar case where six-fold selection into angles $n \pi  /3$ ($\psi_2$) or $\pi/6 + n \pi  /3$ ($\psi_3$) is typical and three-fold selection is forbidden by time-reversal symmetry~\cite{wongGroundStatePhase2013,javanparastOrderbydisorderCriticalityXY2015,rauFrustratedQuantumRareEarth2019}. 

The cubic term coupling the two irreps \emph{locks} the angles in the two $E$ sectors. A short calculation reveals that $\phi_B = -\phi_A$ or $\phi_B = \pi - \phi_A$. It follows that four distinct cubic selection sectors are possible. 

We further note that six-fold selection terms are also allowed by symmetry as they are in the dipolar case. Some degree of fine-tuning is necessary to bring the three-fold selection terms to zero so that six-fold section becomes the primary degeneracy breaking mechanism. 

In the dipolar case, it is possible to show that $\psi_2$ states and not $\psi_3$ states are susceptible to uniform canting in the local frame. 
More precisely, there is a coupling between the $A_2$ irrep and the $E$ irrep of the form $m_z \cos3\phi$~\cite{javanparastOrderbydisorderCriticalityXY2015}. For the quadrupole, the $A$ irrep couples to quadrupoles through $\vert \Psi_E \vert^2 \Psi_{A_{2}}$. In other words, an $A$ irrep component tends to be present whenever there is ordering in either of the $E$ irreps or a mixture of the two. It is also straightforward to show that there are couplings living within the product $E \otimes T \otimes T$. 

\subsection{Flavor Wave Selection and Order by Disorder}

{
\begin{figure}[t]
    \centering
    \includegraphics[width=\linewidth]{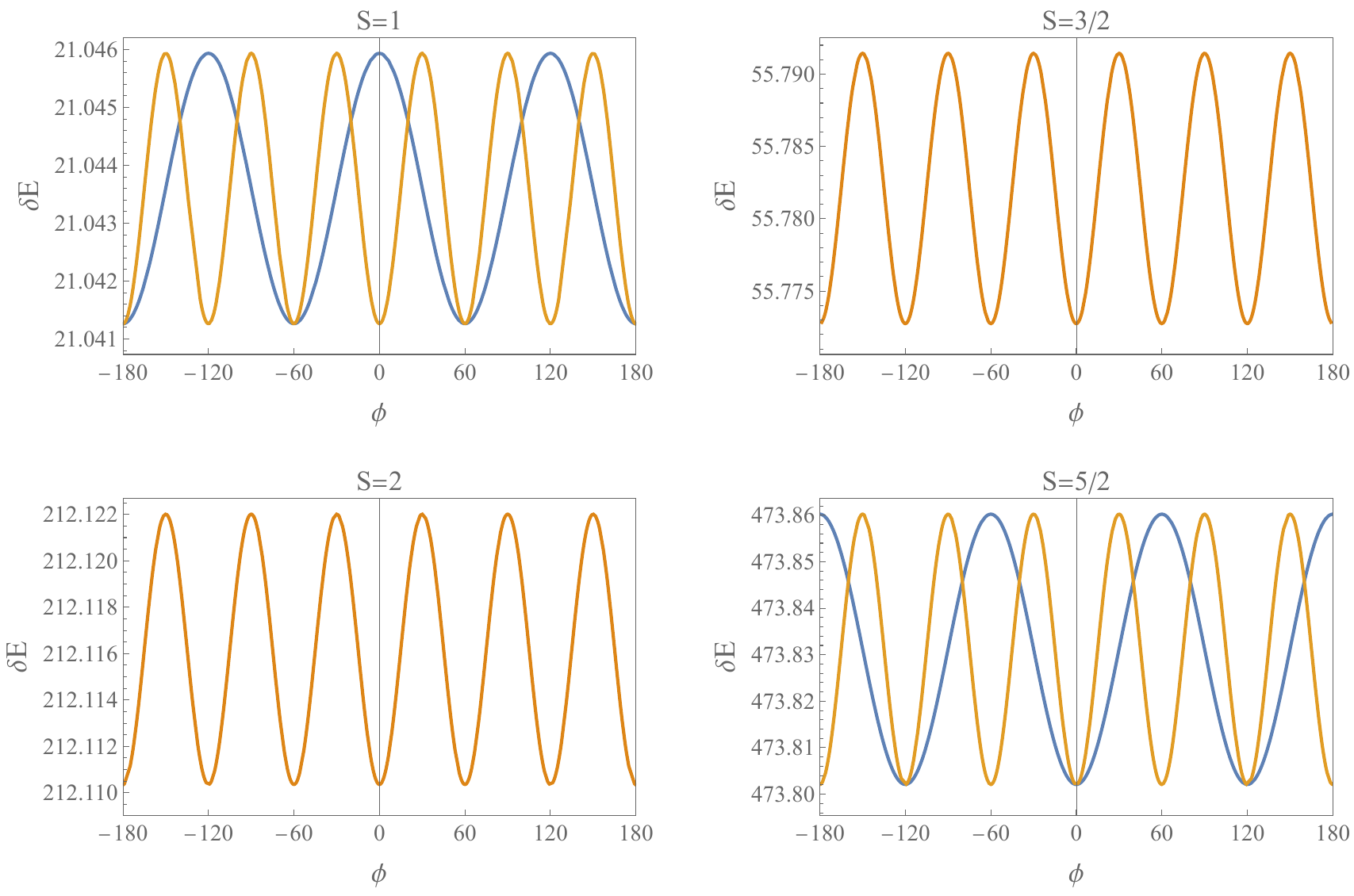}
    \caption{Zero point energy computed from flavor wave theory for model with $K_1=-K_2=-1$ coupling only the $E_B$ sector. Each panel has the energy on the vertical axis and the horizontal axis is an angle $\phi$ that parametrizes the states. While there is, in principle, a separate angle for each sector and we overlay both on the same axes with blue for $E_A$ and orange for $E_B$. The four panels are for spins $1$, $3/2$, $2$, $5/2$ from left to right. There is six-fold selection in the $E_B$ sector for all spins. For $S=1$ and $5/2$ the constraints on the quadrupoles are such that the $E_A$ acquires a nonzero expectation value with three fold selection---the $E_A$ angle being doubled relative to the $E_B$ angle. }
    \label{fig:ObDeg}
\end{figure}
\begin{figure}[h!]
    \centering
    \includegraphics[width=.91\columnwidth]{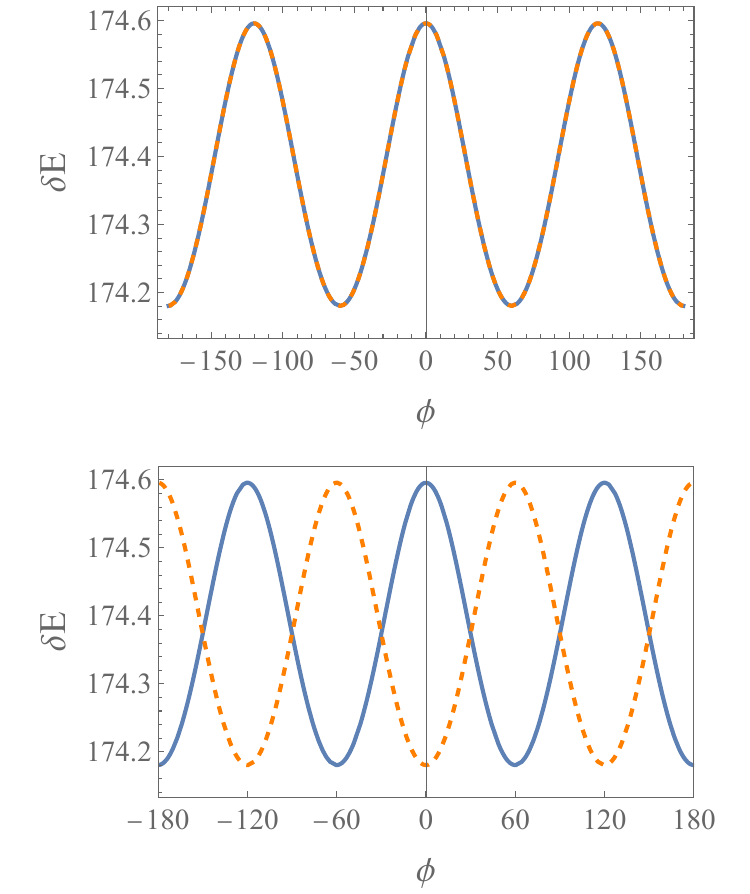}
    \hspace{3ex}
    \caption{Zero point energy computed from flavor wave theory for models coupling irreps $E_A$ and $E_B$: $K_1=-1$, $K_2=\pm 0.5$, $K_3=\pm 0.1$ for $S=2$. Each plot shows the ordering in the $E_A$ (solid) and $E_B$ (dashed) sectors. The $E_A$ sector orders always into configurations with $\phi=60,180,300$ degrees. The $E_B$ sector orders with the same angles as $E_A$ for $K_2=\pm 0.5$ (with $\phi_A=-\phi_B$ locking) and $K_3=+0.1$ and with angles $0,120,240$ (with $\phi_A=\pi-\phi_B$ locking) for $K_3=-0.1$. }
    \label{fig:ObD3F}
\end{figure}
}

In this section, we begin our exploration of the semi-classical dynamics of the quadrupoles. We do this through a multi-boson expansion (otherwise known as flavor waves) of the quadrupolar model for underlying spin $S$. Our starting point is the following Hamiltonian:
\begin{equation}
H = \frac{1}{2}\sum_{i,j} J_{ij}^{\alpha\beta} \hat{O}^\alpha_i \hat{O}^\beta_j - \sum_{i,\alpha} B_i^\alpha  \hat{O}^\alpha_i
\end{equation}
where the $ \hat{O}^\alpha_i$ are some kind of on-site operators---later assumed to be quadrupolar operators. 

We carry out local mean field theory on this model to obtain the spectrum on each magnetically distinct site $\vert i,\mu\rangle$ and we introduce boson operators that create these states on top of the vacuum $\vert i,\mu\rangle \equiv A^\dagger_{i,\mu} \vert {\rm VAC} \rangle$ for $\mu=0,\ldots,2S$. There is a constraint that the boson number is equal to one per site $\sum_\mu A^\dagger_{i\mu}A_{i\mu}=1$. For the purposes of formulating a spin wave expansion we take
\begin{equation}
\sum_\mu A^\dagger_{i\mu}A_{i\mu}=M
\end{equation}
for fixed $M$. The ground state is a condensate of bosons so that
\begin{equation}
A_{i0} = A^\dagger_{i0} = \sqrt{ M - \sum_{\mu=1}^{2S} A^\dagger_{i\mu}A_{i\mu}  }.
\end{equation}
We now write the operators in the Hamiltonian in the on-site basis
\begin{equation}
\hat{O}^\alpha_i = \sum_{\mu,\nu} \vert i\mu \rangle \langle i\mu \vert  \hat{O}^\alpha_i \vert i\nu \rangle \langle i\nu \vert \rightarrow \sum_{\mu,\nu} \langle i\mu \vert  \hat{O}^\alpha_i \vert i\nu \rangle A^\dagger_{i\mu}A_{i\nu}.
\end{equation}
We expand the Hamiltonian around the mean field ground state and redefine $B_{i}^\alpha = M \tilde{B}_{i}^\alpha$.  We find
\begin{equation}
H = M^2 H^{(0)} + M^{3/2} H^{(1)} + M H^{(2)} + \ldots 
\end{equation}
where 
\begin{equation}
H^{(0)} =  \frac{1}{2}\sum_{i,j} J_{ij}^{\alpha\beta} [\hat{O}^\alpha_i]_{00} [ \hat{O}^\beta_j ]_{00} - \sum_{i,\alpha} B_i^\alpha [ \hat{O}^\alpha_i ]_{00}
\end{equation}
using notation $ \langle i\mu \vert  \hat{O}^\alpha_i \vert i\nu \rangle\equiv [\hat{O}^\alpha_i]_{\mu\nu}$.  

The terms linear in the bosonic operators should vanish in any mean field local minimum. About such a minimum we compute the quadratic Hamiltonian 
\begin{equation}
H^{(2)} = \frac{1}{2} \sum_{\mathbf{k}} \boldsymbol{\Upsilon}_{\mathbf{k}}^\dagger \left( \begin{array}{cc} 
\boldsymbol{A}(\mathbf{k}) &  \boldsymbol{B}(\mathbf{k})  \\
\boldsymbol{B}^\star(-\mathbf{k}) &  \boldsymbol{A}^*(-\mathbf{k})  
 \end{array} \right) \boldsymbol{\Upsilon}_{\mathbf{k}}
 \label{eq:H2}
\end{equation}
with $\boldsymbol{\Upsilon}_{\mathbf{k}} = (A_{\mathbf{k}1}, \cdot , A_{\mathbf{k}2S}, A^\dagger_{-\mathbf{k}1}, \cdot , A^\dagger_{-\mathbf{k}2S})^T$. We find
\begin{widetext}
\begin{align}
A_{ab}^{\mu\nu}(\mathbf{k}) & = \sum_{\alpha\beta} J_{ab}^{\alpha\beta}(\mathbf{k}) [ O_a^\alpha ]_{\mu 0} [ O_b^\beta ]_{0\nu} + \delta_{ab} \sum_c J_{ac}^{\alpha\beta}(\mathbf{k}=0) \left[   [ O_a^\alpha ]_{\mu \nu} [ O_c^\beta ]_{0 0} -  \delta_{\mu\nu} [ O_a^\alpha ]_{0 0} [ O_c^\beta ]_{00}  \right] \nonumber \\
& - \sum_{\alpha} \tilde{B}_a^\alpha \left(  [ O_a^\alpha ]_{\mu\nu} - \delta_{\mu\nu} [  O_a^\alpha ]_{00} \right)
\end{align}
\end{widetext}
and
\begin{align}
B_{ab}^{\mu\nu}(\mathbf{k}) & = \sum_{\alpha\beta} J_{ab}^{\alpha\beta}(\mathbf{k}) [ O_a^\alpha ]_{\mu 0} [ O_b^\beta ]_{\nu 0} 
\end{align}

We carry out a bosonic Bogoliubov transformation on Eq.~\ref{eq:H2} to obtain the spectrum of states with energies $\omega_{\mathbf{k}m}$ for $m=1,\ldots,8S$.

\begin{figure*}
    \centering
    \phantom{a}
    \begin{overpic}[width=.945\linewidth]{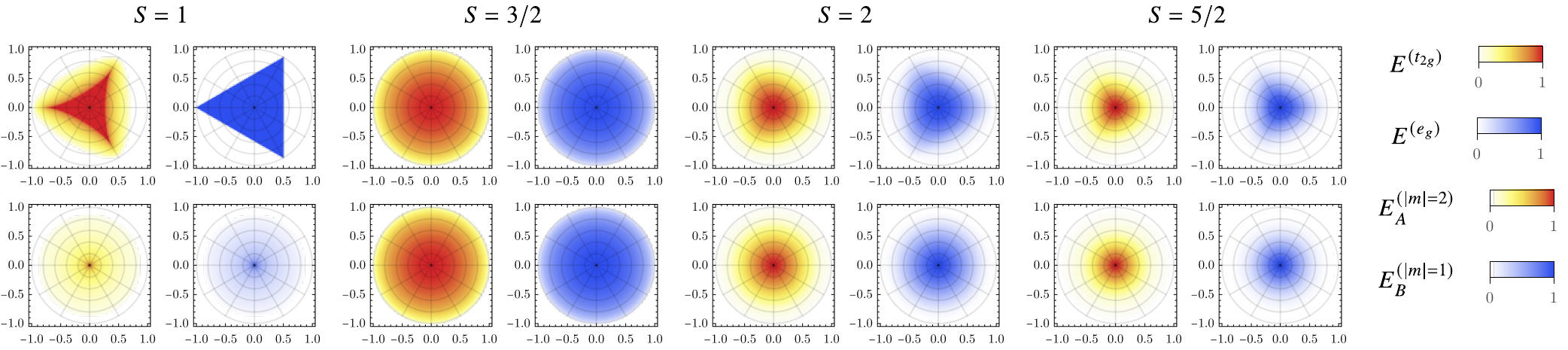}
        \put(01,21){(a)}
        \put(23,21){(b)}
        \put(45,21){(c)}
        \put(67,21){(d)}
    \end{overpic}
    \hfill 
    \phantom{a}
    \\[3ex]
    \begin{overpic}[width=\linewidth]{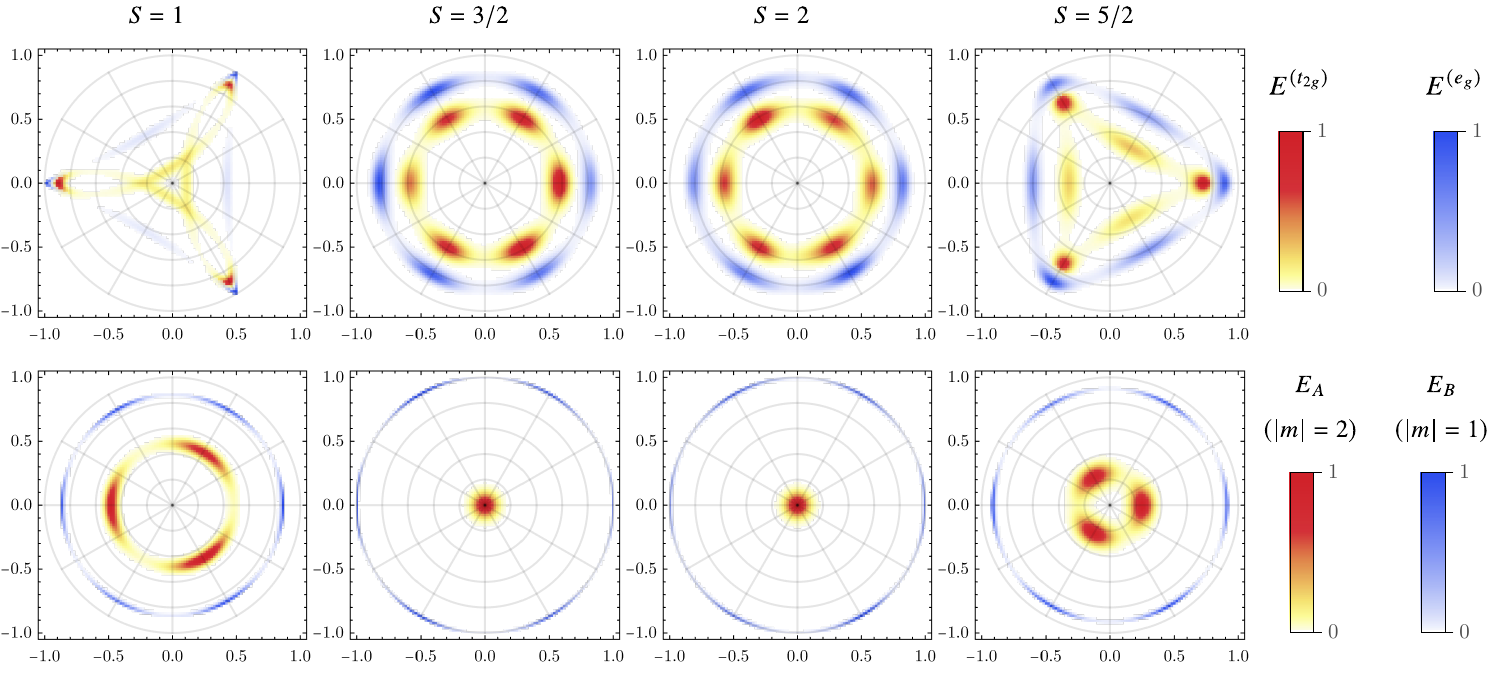}
        \put(02,43.5){(e)}
        \put(23,43.5){(f)}
        \put(44,43.5){(g)}
        \put(65,43.5){(h)}
    \end{overpic}
    \caption{
    Monte Carlo results demonstrating the order by disorder selection in the two $E$ irreps for parameters $(K_1,K_2,K_9)=K(-1,1,1)$ and all other $K_n = 0$, which corresponds to the point marked ``$\bm{\times}$'' in \cref{fig:dipolar_phase_diagrams}(c). 
    Plots show histograms of the projection of the 20-component quadrupolar configuration on each tetrahedron $\smash{Q_{t,\mu}^a}$ into two orthogonal $E$ subspaces, decomposed as either $E^{(t_{2g})}\oplus E^{(e_g)}$ (c.f. \cref{tab:irrep_table_t2g_eg_basis}) or $E_A\oplus E_B$ (\cref{eq:E_irreps_local}).
    The ground state irrep is $E_B$. 
    (a-d): High temperature $T/K = 100$ deep in the paramagnetic phase, demonstrating how the spin quantum number affects the distribution of quadrupolar configurations in the $E$ sector. This is clear in the $t_{2g}\oplus e_g$ decomposition: for $S=1$ not all possible $E$ configurations are accessible; for $S=3/2$ all configurations are accessible and exhibit a uniform distribution; for $S\geq 2$ all configurations are accessible, but a three-fold anisotropy is clearly visible in the distribution. 
    (e-h): Low-temperature $T/K = 0.03$ deep in the ordered phase, where $E_A$ is the ground state irrep, showing three-fold and six-fold selection patterns. Data is taken for an $L=8$ system ($4L^3$ sites), thermalized over $10^7$ sweeps starting from a random state and annealed by continuously lowering the temperature incrementally to the target temperature, averaged over $10^7$ sweeps, and further averaged over 64 independent runs. 
    Results are consistent with quantum flavor wave calculations. 
    }
    \label{fig:monte_carlo_OBD}
\end{figure*}
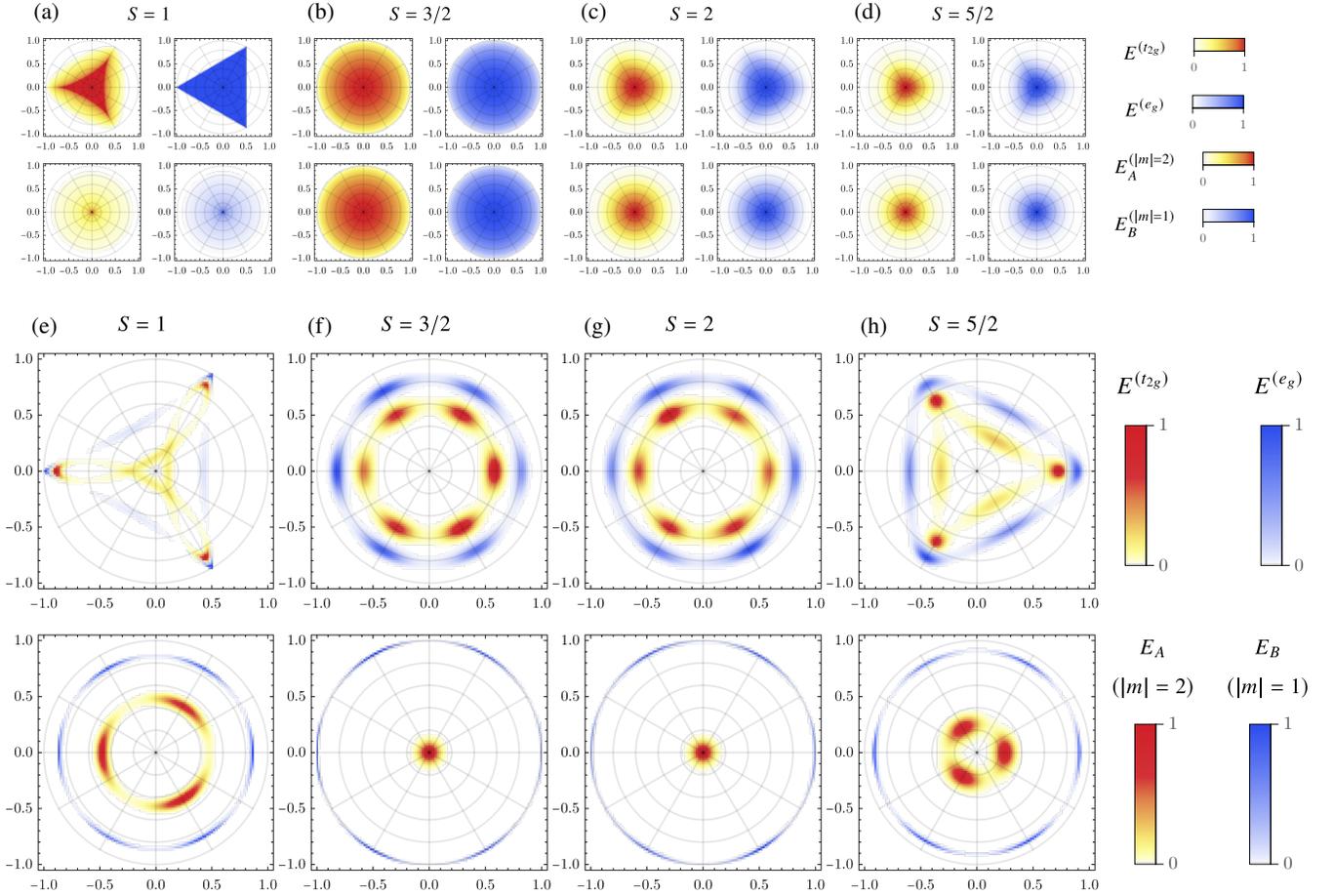

We are mainly interested in cases where there is an accidental $U(1)$ degeneracy in the ground state as happens under certain circumstances within the $E$ irrep. In these cases, the fluctuations computed from spin wave theory will tend to lift the degeneracy. This so-called {\it order-by-disorder} mechanism~\cite{sobralOrderDisorderPyrochlore1997,henleyOrderingDueDisorder1989} can often be captured through the leading quantum correction to the energy
\begin{equation}
\delta E = \frac{1}{4 N} \frac{1}{2S} \sum_{\mathbf{k},n} \omega_{\mathbf{k}n}. 
\label{eq:zpe}
\end{equation}

In the dipolar model, when the single $E$ irrep is the lowest energy mean field state there is always an accidental degeneracy and this linear spin wave correction to the energy leads to one of two classes of six-fold degenerate states denoted $\psi_2$ and $\psi_3$. For the quadrupolar model, one might expect greater complexity coming from the appearance of a pair of $E$ irreps. This is indeed the case yet the results are also highly dependent on the choice of spin $S$. 

We begin by discussing the case where only one of the irreps $E_A$ or $E_B$, defined in \cref{eq:E_irreps_local}, is coupled. We shall see that this turns out to be a special basis. Concretely, we couple only components of $E_B$, $Q^{\bar{x}\bar{z}}$ and $Q^{\bar{y}\bar{z}}$, by switching on $K_1$ and $K_2$. With only $K_1$, for any spin, there is a U$(1)$ degeneracy that is not lifted once corrections from \cref{eq:zpe} are included. Including $K_2$ leads to order-by-disorder as shown in Fig.~\ref{fig:ObDeg}. For all spins, the irrep that is directly coupled orders into a discrete set of states with six-fold degeneracy. The angles match those of $\psi_2$ order familiar from the dipolar case. Spins $3/2$ and $2$ exhibit this selection alone with expectation values of the other components vanishing. However, other values of $S$ have constraints that switch on expectation values for all components. We have discussed these constraints in some detail in \cref{sec:quantum_quadrupoles}. In summary, spin $1$ and $S\geq 5/2$ are biased towards uniaxial nematic states. We find that, for spin $1$ and $5/2$, $E_A$ expectation values are non-vanishing and the U$(1)$ angle parameterizing this irrep is locked to double the $E_B$ angle resulting in $3$-fold selection in $E_A$. From a Landau theory perspective, all cubic terms vanish in the $S=3/2$ and $2$ cases reflecting the fine-tuning of microscopic parameters. 

The generic case where both irreps $E_A$ and $E_B$ are coupled also exhibits a clear dependence on the spin length. The most strongly constrained cases---$S=1$ and $S\geq 5/2$---cease to exhibit order-by-disorder altogether. Instead, there is three-fold selection of states within both $E_A$ and $E_B$ at the mean field level. Two states appear depending on the couplings $\phi_A = -\phi_B$ and $\phi_A = \pi - \phi_B$. Spins $3/2$ and $2$ do generally have a U$(1)$ degeneracy at the mean field level and there is fluctuation selection. The configuration space of spin $3/2$ is remarkably isotropic such that the selection for this spin is mimicked by Monte Carlo on the four-sphere with an effective time reversal symmetry. For this reason, in general, $S=3/2$ has six-fold selection. In contrast, $S=2$ with coupled $E_A$ and $E_B$ sectors orders into states with a three-fold degeneracy. Figure~\ref{fig:ObD3F} shows generic features of the order-by-disorder selection for $S=2$.

\subsection{Finite-Temperature Monte Carlo}

We now turn to exploring the selection due to thermal fluctuations, using the Monte Carlo method described in \cref{sec:monte_carlo} for various spin quantum number. 
To do so, we perform simulations at finite temperature for parameters in the $E$ phase, with parameters $-K_1 = K_2 = K_9 = K >0$ and other $K_n = 0$ in terms of \cref{eq:interaction_matrix_Kloc}, corresponding to the point marked $\bm{\times}$ in \cref{fig:dipolar_phase_diagrams}(c). 
This choice keeps the $E$ irreps decoupled in the spherical harmonics basis $(K_{\pm}' = K_{\pm\pm}'=0$), and the ground state irrep is $E_B$ ($\vert m \vert = 1$). 
To visualize the order-by-disorder selection we measure the quadrupole configuration on each tetrahedron and project it into the 4D $E \oplus E$ space, and further separate this into two 2D orthogonal $E$ subspaces.
We choose two different separations---as $E_{t_{2g}}\oplus E_{e_g}$ (defined in \cref{tab:irrep_table_t2g_eg_basis}), and the local frame $E_A \oplus E_B$ (defined in \cref{eq:E_irreps_local}).
We then produce density histograms within each 2D subspace, the results of which are shown in \cref{fig:monte_carlo_OBD}.

First, as a control to differentiate the effects of the spin quantum number, we take high-temperature data in the paramagnetic phase at $T/K=100$, shown in \cref{fig:monte_carlo_OBD}(a-d).
At this temperature each spin is practically independent and assumes nearly random states in its local Hilbert space, such as represented by the clouds of black points in \cref{fig:quantum-spin-S-dependence}(c,d). 
The strong constraints on $S=1$ spins are starkly visible in \cref{fig:monte_carlo_OBD}(a), particularly in the $e_g$ subspace, where the allowed quadrupolar configurations all lie inside of the blue triangle. 
The $S=3/2$ case appears completely isotropic, with no difference between the two different cuts of the $E\oplus E$ space. 
The $S=2$ and $S=5/2$ cases exhibit a three-fold triangular anisotropy in the distributions in the $e_g$ and $t_{2g}$ subspaces. 
These three-fold anisotropies for $S \neq 3/2$ are not visible in the $\vert m \vert = 1$ and $\vert m \vert = 2$ subspaces. 

Next we thermally anneal the system down to a low temperature well in the ordered phase, at $T/K = 0.03$, shown in \cref{fig:monte_carlo_OBD}(e-h). 
Focusing first on the isotropic $S=3/2$ case, we can clearly see a six-fold selection of states out of a continuous U(1) manifold of $E_B$ ground states.
The same appears to be the case for $S=2$, despite the high-temperature data showing a three-fold anisotropy. 
This is likely because at low temperatures the system prefers to maximize the quadrupole moment, selecting locally quadrupolar coherent states, which from \cref{fig:quantum-spin-S-dependence}(e,f) have a uniform distribution. 
This is consistent with the fact that the uniaxial and biaxial operators do not have the same spectrum for S=2, but they do have the same maximum eigenvalue.
The cases $S=1$ and $S = 5/2$ are different: the $E_A$ components are switched on despite costing more energy, with the $E_B$ components no longer saturating.
This is a manifestation of the previously-mentioned competition between maximizing the local quadrupole moment, which is only possible for uniaxial configurations, and minimizing the exchange energy, which may prefer biaxial ground states (\cref{sec:S_gt_2_quantum_quadrupoles}). 
The $E_A$ components exhibit three-fold selection while the $E_B$ components exhibit six-fold selection, which is consistent with the flavor wave calculations. 
This effect becomes far more pronounced when looking at the $t_{2g}\oplus e_g$ cross sections---while $S=2$ and $S=3/2$ show 6-fold selection out of an isotropic distribution, $S=1$ and $S=5/2$ show that states at the corners of the triangular distributions seen in \cref{fig:monte_carlo_OBD}(a) and (d) are selected, which are maximally uniaxial ones (for example $\smash{E^{(e_g)}_{\psi_2}}$ in \cref{fig:ground_states}). 
This implies that for $S \neq 3/2$ or $2$ there is a direct selection of three discrete ground states, rather than a fluctuation-driven selection from a continuous ground state manifold.

\section{Quadrupolar Spin Liquids}
\label{sec:spin_liquids}

We now turn to consider generalized spin liquids in the quadrupolar model. 
First, there is obviously the quadrupolar ice spin liquid, which occurs when the ground state irrep is $T_{2,\text{ice}}^{(t_{2g})}$. 
Its physics is identical at the semi-classical level to that of ordinary classical spin ice, i.e. it has the usual set of 2-in-2-out ground states, where ``in'' means rod-like uniaxial configurations along the easy-axes and ``out'' means disk-like uniaxial. 
More interesting for our purposes are the quadrupolar spin liquids which occur when multiple irrep multiplets are degenerate in the ground state.

In dipolar spin liquids, a basic classification for bilinear Hamiltonians is given by examining the band structure of the interaction matrix, with classical spin liquids corresponding to a set of flat bands at the base of the spectrum~\cite{yanClassificationClassicalSpin2024,yanClassificationClassicalSpin2024a,fangClassificationClassicalSpin2024,henleyPowerlawSpinCorrelations2005}.
The flat bands provide an extensive set of degenerate zero-energy modes (i.e. an extensive number of Luttinger-Tisza ground states), and topologically non-trivial touching between the flat bands and higher-energy dispersive bands encode Gauss law constraints of a coarse-grained gauge field theory. 
In dipolar models, it is common to use the so-called self-consistent Gaussian approximation (SCGA) as a starting point for analysis, wherein the semi-classical spin length constraint is replaced by a Lagrange multiplier, and in which the zero-temperature two-point correlation function is proportional to the projector to the flat bands~\cite{henleyPowerlawSpinCorrelations2005,chungMappingPhaseDiagram2024}.

\begin{figure*}
    \centering
    \begin{overpic}[width=.9\linewidth]{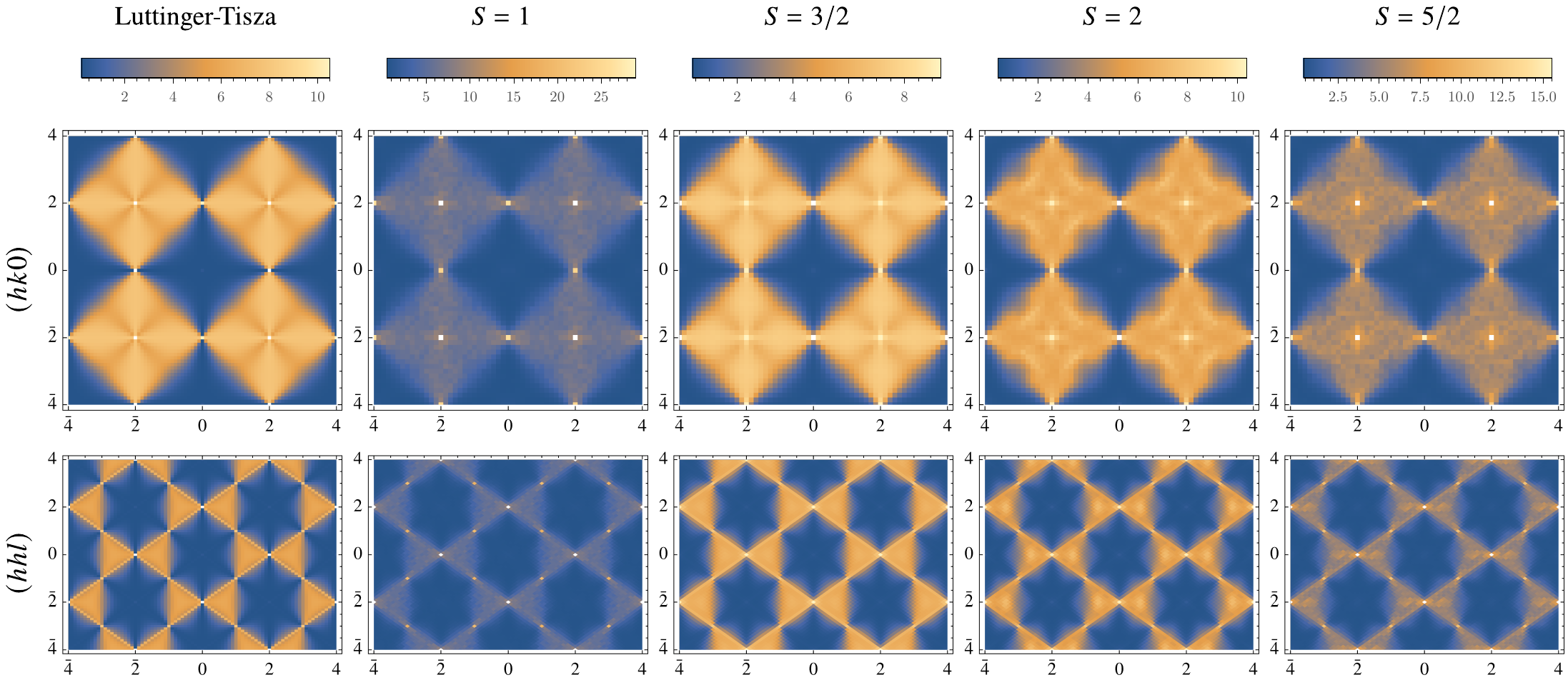}
        \put(0, 40){(a)}
        \put(0,-05){(b)}
    \end{overpic}
    \\
    \includegraphics[width=.9\linewidth]{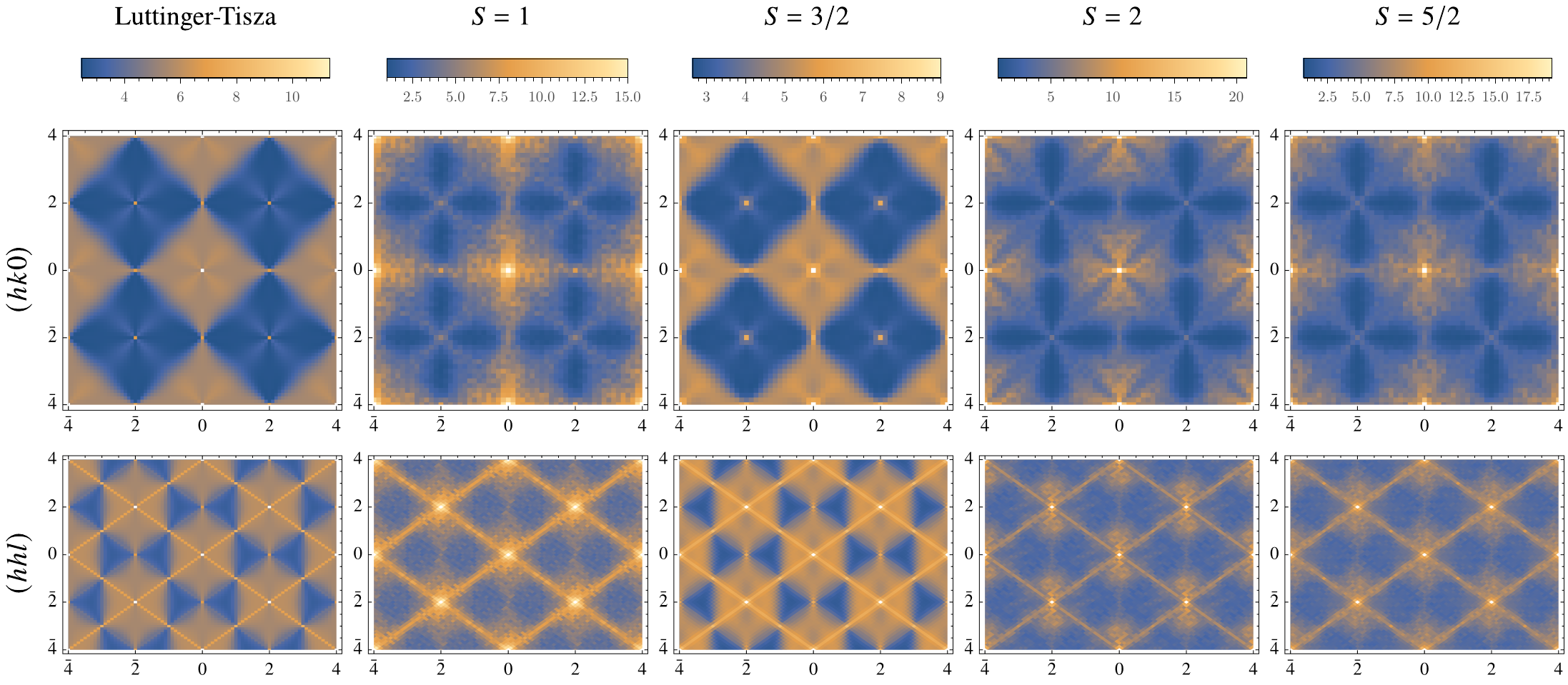}
    \caption{(a) Example of a spin liquid which arises from setting the $e_g$ couplings to zero and tuning the $t_{2g}$ couplings to those of a rank-2 tensor spin liquid in the dipolar Hamiltonian, corresponding to a point on the triple intersection of the Palmer-Chalker, ferromagnetic, and $\Gamma_5$ phases in the $\smash{t_{2g}}$ sector. The ground state irreps are $E \oplus T_1 \oplus T_2$. Here we show the quadrupolar structure factor \cref{eq:quadrupolar_sf} computed from the projector to the flat bands, \cref{eq:gaussian_projector_sf}. (b) The anti-diagonal dual of the spin liquid from (a), with switched easy-plane sectors. Simulations are performed for a system with $\smash{4L^3}$ spins at $L=14$, thermalized for $10^5$ sweeps and averaged over $10^6$ sample configurations and further averaged over $64$ independent simulations. }
    \label{fig:A2_SL}
\end{figure*}

For a Hamiltonian containing dipolar and quadrupolar bilinears in the form of \cref{eq:Hamiltonian-S-Q}, this general approach remains a viable starting point for identifying potential spin liquids with quadrupolar degrees of freedom.
In particular, the Hamiltonian can be put in the form
\begin{equation}
    H = \frac{1}{2} \sum_{ij} \sum_{\alpha,\beta,a,b} 
    \begin{pmatrix}
        S^\alpha_i \\ Q^a_i
    \end{pmatrix}
    \begin{pmatrix}
        [\mathcal{J}_S]_{ij}^{\alpha\beta} & 0 \\
        0 & [\mathcal{J}_Q]_{ij}^{ab}
    \end{pmatrix}
    \begin{pmatrix}
        S^\beta_j \\ Q^b_j
    \end{pmatrix}
\end{equation}
and this block-diagonal interaction matrix can be Fourier transformed and its band structure analyzed. 
The dipolar interaction matrix $\mathcal{J}_S$ has been extensively studied and all of its degenerate irrep combinations cataloged for nearest-neighbor interactions~\cite{chungMappingPhaseDiagram2024}, and the stability of the corresponding classical spin liquids in the semi-classical limit analyzed~\cite{lozano-gomezAtlasClassicalPyrochlore2024}. 
Unfortunately a complete classification and parameterization of all degenerate combinations of irreps as was done in Ref.~\cite{chungMappingPhaseDiagram2024} is not possible for the quadrupolar interaction matrix due to the triply degenerate $T_2$ irrep, which cannot be diagonalized analytically. 
A detailed analysis could be carried out in the case that two of the off-diagonal terms in \cref{eq:irrep_matrix_T2} are turned off, but we will not pursue it here. 
Instead we will highlight a number of readily identifiable interesting cases. 

One can in principle apply a Gaussian-type approximation to the bilinear quadrupolar Hamiltonian, by treating the quadrupoles as 5-component vectors with Gaussian length constraint.
Based on our discussions in \cref{sec:quantum_quadrupoles}, this is likely not a very accurate characterization except possibly in the cases $S=3/2$ and $S=2$.
Nevertheless, the low-energy band structure of the interaction matrix---its flat bands and band touchings---should accurately reflect the emergent gauge structure of these semi-classical quadrupolar spin liquids. 
The projector to the flat bands also provides a useful baseline point of comparison for low-temperature correlation functions derived from Monte Carlo, i.e. 
\begin{equation}
    \langle Q_\mu^a(-\bm{q}) Q_\nu^b(\bm{q})\rangle_{\text{Gauss. $T=0$}} \propto [P_{\text{flat}}]_{\mu\nu}^{ab}(\bm{q}),
    \label{eq:gaussian_projector_sf}
\end{equation}
where $\bm{q}$ is the wavevector and $\mu,\nu$ index fcc sublattices. 
For our purposes we will only show the isotropic quadrupole-quadrupole correlator, 
\begin{equation}
    S_Q(\bq) = \frac{1}{4L^3} \sum_{i,j} \langle \Tr[Q_i Q_j] \rangle e^{-i\bq\cdot(\bm{r}_j - \bm{r}_i)}
    \label{eq:quadrupolar_sf}
\end{equation}
where we use the four-site basis fcc primitive unit cell with $L^3$ unit cells and periodic boundaries.

\subsection{Dipolar-like Liquids and Evidence for Fragmentation}

As discussed in \cref{sec:canting}, we can in general make a change of basis to remove the couplings $J_{\pm}'$ and $J_{\pm\pm}'$ which couple the two easy-plane sectors from the Hamiltonian, then if we further turn off one of the resulting $J_{z\pm}$ couplings the decoupled irrep multiplets behave like those of the dipolar model. 
It follows then that all of the flat band cases of the dipolar model, categorized in Ref.~\cite{chungMappingPhaseDiagram2024}, are reproduced in the phase diagram of the quadrupolar model, once in each easy-plane sector, and as an extended manifold as the $J_{\pm}'$ and $J_{\pm\pm}'$ couplings are varied (along with the $J_{z\pm}$ couplings to keep one of them zero in the rotated basis).
The actual ground state behavior, however, depends on the parameters and the spin quantum number, even if the flat band content is the same.

Here we demonstrate with an example. 
The dipolar model has a continuous 1-dimensional family of spin liquids at the intersection of the Palmer-Chalker, splayed-ferromagnet, and $\Gamma_5$ phases which is a rank-2 tensor spin liquid with pinch line singularities along the high-symmetry $(hhh)$ directions of reciprocal space~\cite{chung2FormU1Spin2025,lozano-gomezAtlasClassicalPyrochlore2024,bentonSpinliquidPinchlineSingularities2016}.
Ref.~\cite{chungMappingPhaseDiagram2024} gives a parameterization of the family in terms of a parameter $x \in [-1,1]$.
By taking those couplings and using them for the $t_{2g}$ couplings in \cref{eq:interaction_matrix_Jloc} with all other couplings zero, we reproduce the same form of the Hamiltonian and obtain the same set of flat bands and band touchings in the interaction matrix spectrum, and find the ground state irreps are $E \oplus T_1 \oplus T_2$ as expected from the dipolar model. 
In particular, this family of Hamiltonians has 1-dimensional lines in reciprocal space where a quadratically dispersing band touches the flat bands, giving rise to pinch line features in the structure factors. 

\begin{figure}[t]
    \centering
    \includegraphics[width=.85\linewidth]{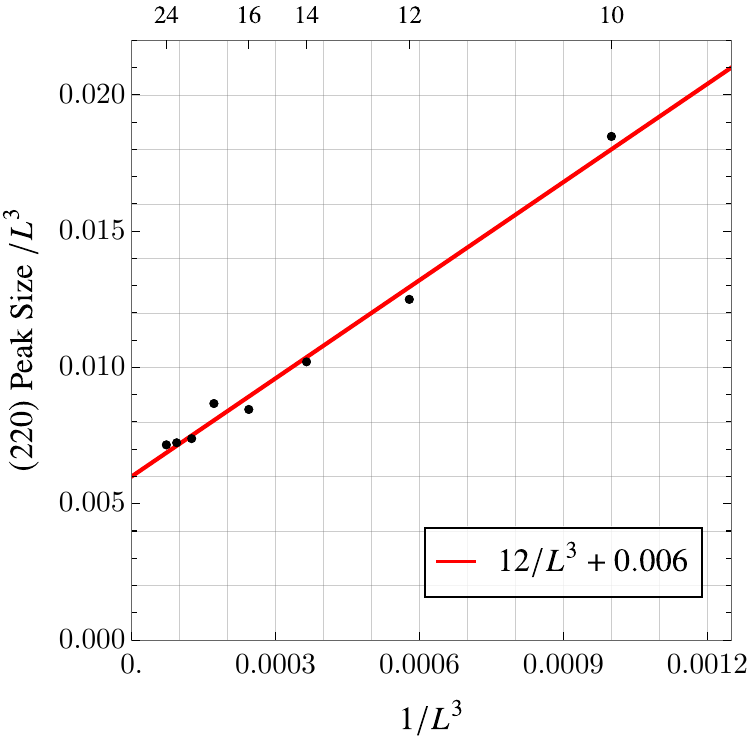}
    \hfill\phantom{a}
    \caption{Finite-Size scaling of the (220) peak in the dipolar-like spin liquid from \cref{fig:A2_SL}(a) for $S=1$. The peak height increases with system size, but per unit cell it approaches a non-zero constant in the thermodynamic limit. This suggests that this classical quadrupolar spin liquid is fragmented for $S=1$. }
    \label{fig:SL_A2_Peak_Scaling}
\end{figure}

\Cref{fig:A2_SL}(a) shows the corresponding quadrupolar structure factor computed from the projector to the flat bands as well as from semi-classical Monte Carlo simulations for spin quantum numbers ranging from $S = 1$ to $S = 5/2$. 
The $S=3/2$ simulations agree closely with the projector calculation reflecting the isotropic behavior of the quadrupole operators, while other spin quantum numbers show some deviations. 
In particular, $S=1$ shows a relatively strong peak at some zone centers compared to the other cases, although the background intensity is qualitatively consistent and at the same intensity scale as the Luttinger-Tisza. 
We have seen that in the $S=1$ case it may occur that multiple irreps become active simultaneously when Luttinger-Tisza predicts only a single irrep ground state, such as illustrated in \cref{fig:mean_field_phase_diagram_S1_vs_S32}, raising the possibility that the peak may indicate the activation of another irrep and coexistence of a spin liquid with long range quadrupolar order, i.e. the spin liquid may be fragmented~\cite{lefrancoisFragmentationSpinIce2017}, somewhat like the case reported in Refs.~\cite{chung2FormU1Spin2025,rougemailleAmperePhaseFrustrated2025} where a spin liquid coexists with a very weak partially saturated long-range order.
To investigate this we check the finite system size scaling of the peak height, shown in \cref{fig:SL_A2_Peak_Scaling}. 
This data is consistent with a finite Bragg peak consisting of only a small finite fraction of the total spectral weight. This is situation is commonly referred to as spin liquid fragmentation~\cite{lefrancoisFragmentationSpinIce2017}.

We can obtain another family of spin liquids by applying the anti-diagonal duality, switching the easy-plane $t_{2g}$ and $e_g$ components. 
\Cref{fig:A2_SL}(b) shows the resulting structure factors, again from the projector to the flat bands and from semi-classical Monte Carlo. 
Again we see good agreement for $S=3/2$ and some deviations for other spin quantum number. 
The qualitative features in this case remain consistent, and in particular the four-fold pinch points at $(220)$ points remain well-resolved for all spin quantum number.

\subsection{Liquids Connected to the Biquadratic Point --- Planar Band Touchings}

In the dipolar Hamiltonian there is a special point in the phase diagram at which all four phases meet---the isotropic Heisenberg anti-ferromagnet~\cite{chungMappingPhaseDiagram2024,moessnerLowtemperaturePropertiesClassical1998}.
Correspondingly, at this point four of the five irreps are degenerate in the ground state, six of twelve bands are flat, and the system is described three decoupled copies of a U(1) gauge theory~\cite{conlonAbsentPinchPoints2010,isakovDipolarSpinCorrelations2004,henleyPowerlawSpinCorrelations2005,lozano-gomezAtlasClassicalPyrochlore2024,henleyCoulombPhaseFrustrated2010}.
By lifting the degeneracy of any one of the irreps it is possible to move along a lines where three irreps remain degenerate, which gives families of flat band models~\cite{chungMappingPhaseDiagram2024,lozano-gomezAtlasClassicalPyrochlore2024,franciniHigherrankSpinLiquids2025}.

\begin{figure*}
    \centering
    \includegraphics[width=.9\linewidth]{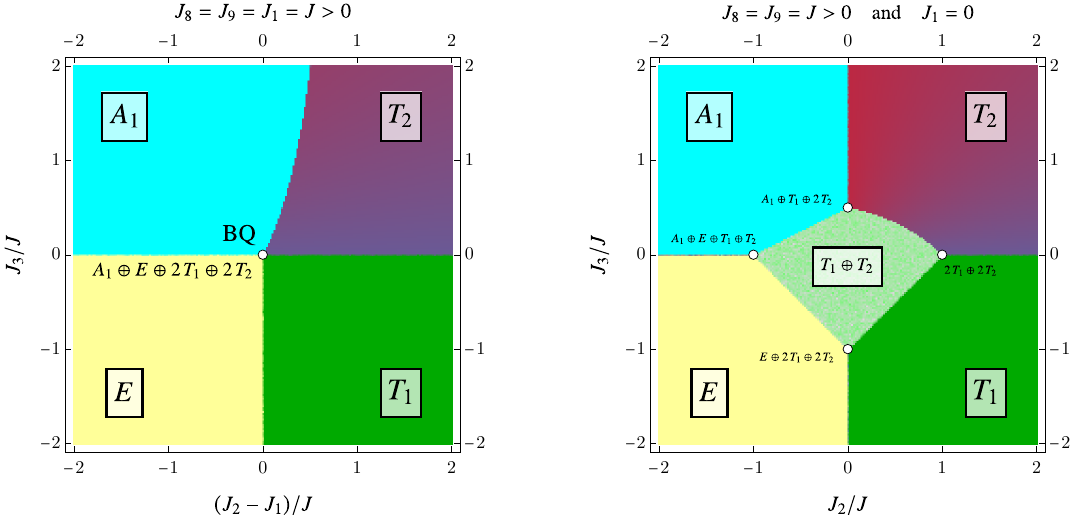}
    \hfill
    \phantom{a}
    \caption{(a) Phase diagram cutting through the biquadratic (BQ) point, which has maximal degeneracy of irreps in the ground state, with ten flat bands. Perturbing it to lift the degeneracy of some of the irreps one can move along loci in parameter space where many remain degenerate, giving many flat band models. (b) Another cut where one parameter has been perturbed, showing that the BQ point splits into multiple degenerate combinations, each of which has a number of flat bands. The color legend is the same as in \cref{fig:dipolar_phase_diagrams}. The four points labeled with white dots belong to families of Hamiltonians emerging from the BQ point containing flat bands with various flat line touchings. The two points with $J_3 =0$ have flat plane touchings, which do not occur in the dipolar case. Their structure factors are shown in \cref{fig:spin_liquids_global_frame}.}
    \label{fig:BQ_point_phase_diagram}
\end{figure*}

The most symmetric point is the pure biquadratic point, where the Hamiltonian can be written in the global frame as
\begin{align}
    H_{\text{BQ}} &= J \sum_{t} \sum_{\langle ij \rangle} \Tr[Q_i Q_j] 
    \nonumber
    \\
    &\propto \text{const.} + \frac{J}{2} \sum_t 
    \Tr\left[\left(\sum_{i \in t} Q_i \right)^2 \right].
    \label{eq:BQ}
\end{align}
Ground states therefore have zero quadrupole moment on each tetrahedron.
In terms of the global frame parameters~\cref{eq:interaction_matrix_Jglo}, it has $J_1 = J_2 = J_8 = J_9 > 1$ and all off-diagonal couplings zero. 
This is the analog of the Heisenberg AFM point in the dipolar case, where every tetrahedron has zero dipole moment in the ground state. 
It has ten flat bands (out of twenty) and, in a naive coarse-graining, is described by five decoupled copies of U(1) spin liquids (whereas the dipolar case is described by three~\cite{isakovDipolarSpinCorrelations2004,lozano-gomezAtlasClassicalPyrochlore2024,henleyCoulombPhaseFrustrated2010}). 
Monte Carlo simulation (c.f. \cref{sec:monte_carlo}) shows that it is stable for all spin quantum numbers and exhibits ordinary 2-fold pinch points in the quadrupolar structure factor identical to those seen in the dipolar HAFM.

\Cref{fig:BQ_point_phase_diagram}(a) shows a cut of the phase diagram in terms of global frame parameters in \cref{eq:interaction_matrix_Jglo} which intersects the BQ point, showing the meeting of all four phases. 
This point is zero-dimensional in the phase diagram (up to overall energy scale), and has six minimal energy irreps $A_1 \oplus E \oplus 2T_1 \oplus 2T_2$, with gapped irreps $\smash{E^{(e_g)}}$ and $\smash{T_2^{(t_{2g})}}$ corresponding to the net quadrupole moment of a single tetrahedron, \cref{eq:net_quad_moment}.
Perturbing it, there will be 1-dimensional lines of parameters emerging from this point where multiple irreps remain degenerate.
The precise number of irreps depends on how the irreps are split, since the number of parameters required to be tuned to make multiple copies of the same irrep degenerate is two, while the number of parameters to be tuned to make different irreps degenerate is one. 
An example is illustrated in \cref{fig:BQ_point_phase_diagram}(b), where $J_1$ has been set to zero, showing in the middle a diamond-shaped region where $T_1$ and $T_2$ irreps are degenerate, and at the corners of this region points where either four or five irreps are degenerate. 
Each of these high-degeneracy points lies along a line where the irreps remain degenerate in the ground state, giving rise to a family of Hamiltonians which potentially host spin liquids. 

Indeed all of these have four flat bands, meaning that Luttinger-Tisza predicts that they are spin liquids. 
Similar to spin liquids found in the dipolar model, these all host flat lines---there is at least one quadratically dispersing band which touches the flat bands along either the $(hhh)$ or $(00l)$ high-symmetry lines of reciprocal space (or both). 
Moreover, the two corner points in \cref{fig:BQ_point_phase_diagram}(b) with $J_3 = 0$ host \emph{flat planes}---a quadratically dispersing band touches the flat bands everywhere on the 2D $(hhl)$ plane of reciprocal space. 
To our knowledge, these are the first examples of flat plane touchings in classical spin liquids. 
The possibility was raised in Ref.~\cite{yanClassificationClassicalSpin2024a} but no examples were given. 
Additionally, two more quadratically dispersing bands touch the flat bands along the 1D $(hhh)$ and $(h00)$ lines.

\begin{figure*}
    \centering
    \begin{overpic}[width=0.9\linewidth]{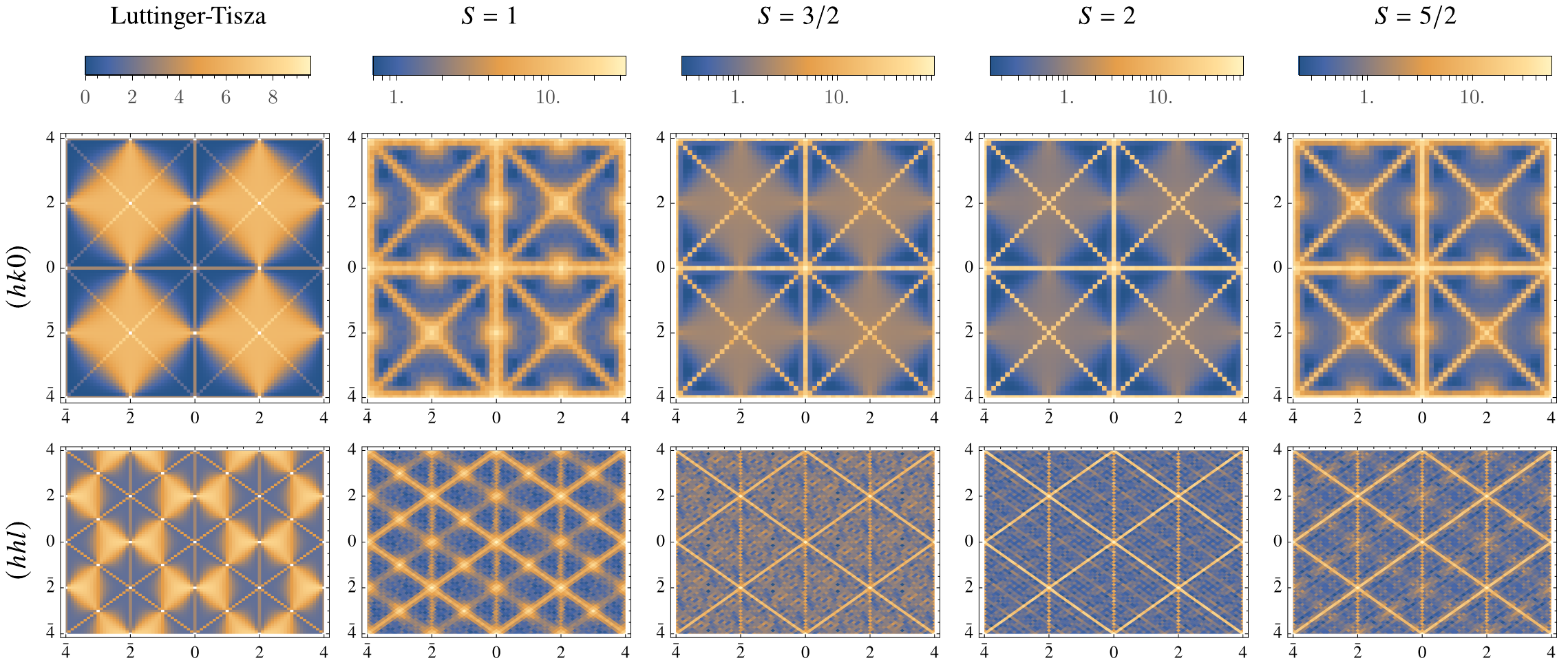}
        \put(0,40){(a)}
        \put(0,-5){(b)}
    \end{overpic}
    \\
    \includegraphics[width=0.9\linewidth]{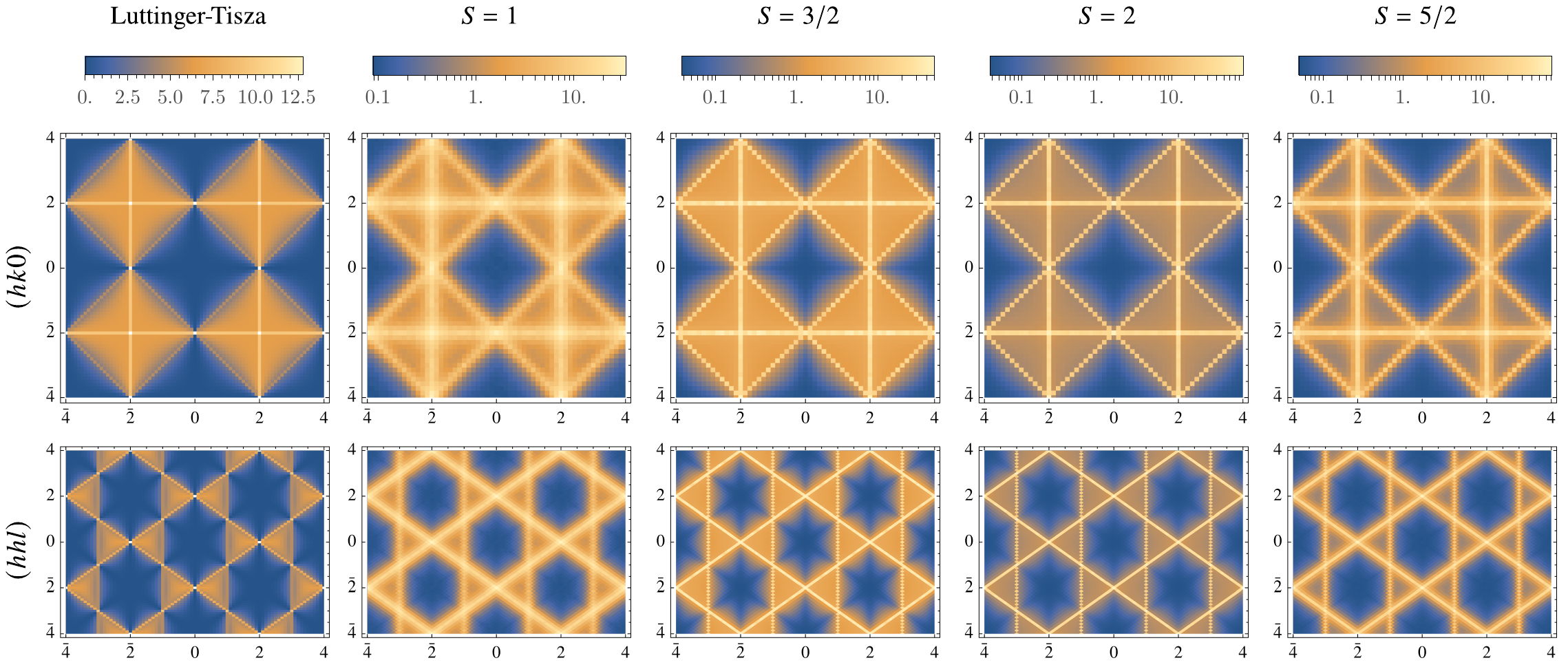}
    \caption{Examples of quadrupolar structure factors of semi-classical spin liquids emerging from perturbing the biquadratic model, corresponding to points in \cref{fig:BQ_point_phase_diagram}. (a) The $A_1 \oplus E \oplus T_1 \oplus T_2$ point ($J_2 < 0$). (b) The $2T_1 \oplus 2 T_2$ point ($J_2 > 0$). The first column shows the zero-temperature Luttinger-Tisza result, where the two-body correlation matrix is given by the projector to the flat bands. The remaining columns show Monte Carlo results for different spin quantum numbers. Both of these models have four flat bands and one \emph{flat plane} where a quadratically dispersing band touches the flat bands, along with two flat line band touchings along $(00h)$ and $(hhh)$, and three quadratic band touchings at the zone center.  }
    \label{fig:spin_liquids_global_frame}
\end{figure*}

In \cref{fig:spin_liquids_global_frame} we show quadrupolar structure factors computed from Luttinger-Tisza and Monte Carlo for different spin quantum numbers for these two points. 
Since we are taking 2D cuts of reciprocal space, the flat planes are intersected along 1D lines, which are clearly visible as lines along which the intensity is higher than the background. 
Note that \cref{fig:spin_liquids_global_frame}(b) is the anti-diagonal dual of the HAFM in the $t_{2g}$ sector.
It is unclear what types of pinch singularities can be associated to flat planes in a 3-dimensional system, as they cannot be linked by a closed manifold on which the dispersing band's eigenvectors can wind. 
It is also unclear what type of tensor gauge theory they can correspond to. 
We leave the detailed study of these curious band touchings for future research.

\subsection{Rank-3 Symmetric Tensor Spin Liquid}

The spin liquids arising from degenerate irreps discussed thus far, such as those in \cref{fig:A2_SL}, have complex pinch structures such as fourfold pinch points, which reflect a Gauss law of a tensor gauge theory~\cite{premPinchPointSingularities2018,yanRank2U1Spin2020,bentonSpinliquidPinchlineSingularities2016,hanRealizationFractonicQuantum2022}, which themselves arise in the description of fractonic phases of matter~\cite{pretkoFractonPhasesMatter2020,pretkoFractonGaugePrinciple2018,premEmergentPhasesFractonic2018,hanRealizationFractonicQuantum2022,williamsonFractonicMatterSymmetryenriched2019,vijayFractonTopologicalOrder2016,maFractonTopologicalOrder2018,nandkishoreFractons2019,bulmashHiggsMechanismHigherrank2018,pretkoGeneralizedElectromagnetismSubdimensional2017,pretkoSubdimensionalParticleStructure2017}.
For the dipolar Hamiltonian, all of the tensor spin liquids found in the nearest-neighbor phase diagram carry nodal lines, which give rise to their multi-fold pinch points~\cite{chungMappingPhaseDiagram2024}.
It is, however, possible to construct a ``clean'' rank-2 symmetric trace-free tensor spin liquid, which exhibits four-fold pinch points without pinch lines, if one allows for breathing anisotropy~\cite{yanRank2U1Spin2020}. 
The key is to recognize that the various irreps correspond to components of the multipole moments of a single tetrahedron. 
To obtain a theory whose low-energy limit is described by a symmetric rank-2 tensor field, the most natural place to consider is to suppress all multipole components except the net quadrupole moment on each tetrahedron. 
In the dipolar Hamiltonian this can be achieved by adding a Dzyaloshinskii-Moriya (DM) interaction to the Heisenberg antiferromagnet, which places the ground state on the phase boundary between the $\Gamma_5$ phase and the splayed ferromagnet phase, with the splay angle precisely tuned so that the allowed ground state configurations are quadrupole configurations. 
This is insufficient degeneracy to stabilize a spin liquid, but one can exploit the bipartiteness of the tetrahedra and relax some constraints on one set, by setting the DM interaction to zero on the ``B'' tetrahedra~\cite{yanRank2U1Spin2020}.
This leads to two flat bands with a quadratic band touching at the zone center which realizes the trace-free symmetric rank-2 U(1) spin liquid in the coarse-grained limit at the level of Luttinger-Tisza~\cite{yanRank2U1Spin2020}.
The system still orders at low temperature~\cite{sadouneHumanmachineCollaborationOrdering2024}, but there remains a regime above the ordering temperature in which four-fold pinch points can be seen.

Since the fundamental objects of the theory are already rank-2, it is natural to wonder whether we can construct a rank-3 spin liquid. 
Indeed, for a system of quadrupolar degrees of freedom we naturally have higher multipoles on each tetrahedron, as laid out in \cref{tab:irreps_multipoles}.
Following the intuition just described for the dipolar rank-2 spin liquid, such a spin liquid ought to occur when we tune the Hamiltonian parameters so that the seven components of the octupole moment are the ground state modes. 
We can work out the required fine-tuning by obtaining the equivalent of the matrices in ~\cref{eq:irrep_matrix_E,eq:irrep_matrix_T1,eq:irrep_matrix_T2} in the basis defined by \cref{tab:irreps_multipoles}.
We then require that the $T_1$ and $T_2$ components of the octupole moment do not cant, i.e. we restrict the off-diagonal couplings between them and the other copies of the same irreps to be zero, giving three constraints.
We then further require that the energies of the $A_1$, $T_1$, and $T_2$ components of the octupole moment be degenerate, giving another two constraints. 
In the global basis~\cref{eq:interaction_matrix_Jglo}, these constraints are satisfied if
\begin{align}
    J_5 &= J_1 - J_2 + 2J_3 \nonumber\\
    J_6 &= -J_1+J_2 - J_3 - 2J_4\nonumber\\
    J_7 &= J_3 - J_4 \nonumber\\
    3J_8 &= -4J_1 + 7J_2 -8J_3 \nonumber\\
    J_9 &= J_2
\end{align}
We then can choose $J_1$-$J_4$ such that the octupole moment irreps have the lowest energies. 
For example, $J_1 = J_2 = J> 0$, $J_3 = 0$ and $J_4 = 0.1J$ will do. 

While this gives the right ground state irreps, it does not yield a spin liquid---the interaction matrix has flat lines, but no flat bands.
In other words, the system is over-constrained, which is also what happened in the dipolar case. 
In order to relax some of the constraints, we can change the couplings on the B tetrahedra to be those of the biquadratic model (the analog of the HAFM), \cref{eq:BQ}, which contains many more irreps in its ground state and does not cant the octupole ones. 
Doing so, the resulting model yields two flat bands, with five quadratic bands touching the flat band at zero center.
Because the low-energy degrees of freedom on the A tetrahedra are given by the octupole moment---a rank-3 symmetric trace-free tensor---we expect the long-wavelength coarse-grained theory to be a rank-3 U(1) gauge theory with a single-derivative Gauss law (as the dispersion is quadratic),
\begin{equation}
    \partial_\alpha O^{\alpha\beta\gamma} = \rho^{\beta\gamma} = 0,
    \label{eq:rank-3-gauss}
\end{equation}
where $O$ is the coarse-grained equivalent of the local octupole moment on a single octahedron. 
Excitations of this tensor gauge theory are naturally quadrupoles, since violations of the Gauss law take the form of a symmetric trace-free rank-2 tensor charge density $\rho^{\beta\gamma}$. 
Note that this agrees with the band counting: for a long-wavelength expansion of the form $(q_\alpha O^{\alpha\beta\gamma})^2$, since the octupole tensor $O$ has seven degrees of freedom and the charge density $\rho^{\beta\gamma}$ has five degrees of freedom, the band structure should have five quadratically dispersing bands and two non-dispersing bands.   
Such a theory is expected to have a combination of fracton (immobile) and lineon (mobile along lines) excitations~\cite{pretkoSubdimensionalParticleStructure2017}.

\begin{figure}
    \centering
    \includegraphics[width=.95\linewidth]{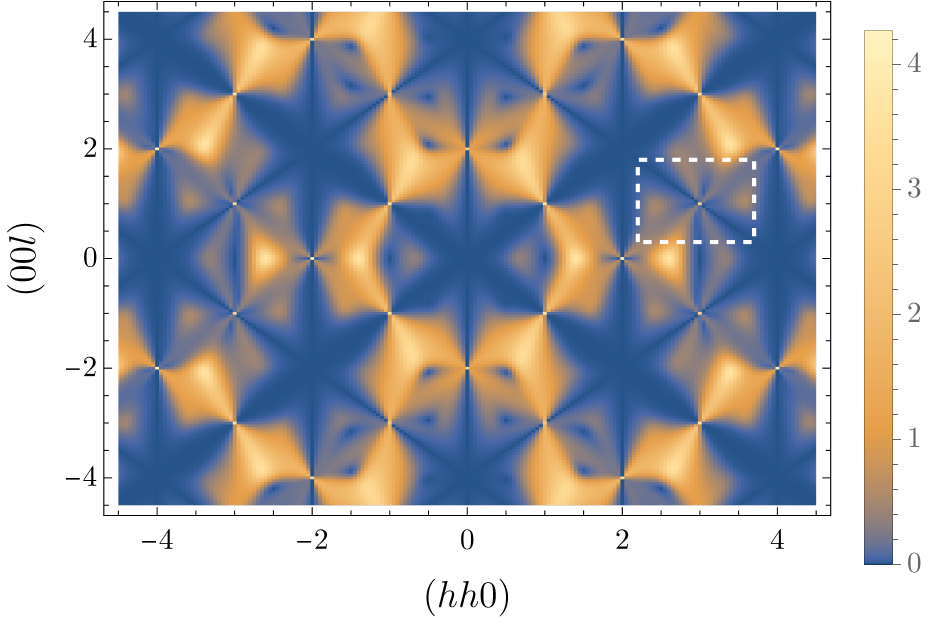}
    \hfill
    \phantom{a}
    \\
    \includegraphics[width=.97\linewidth]{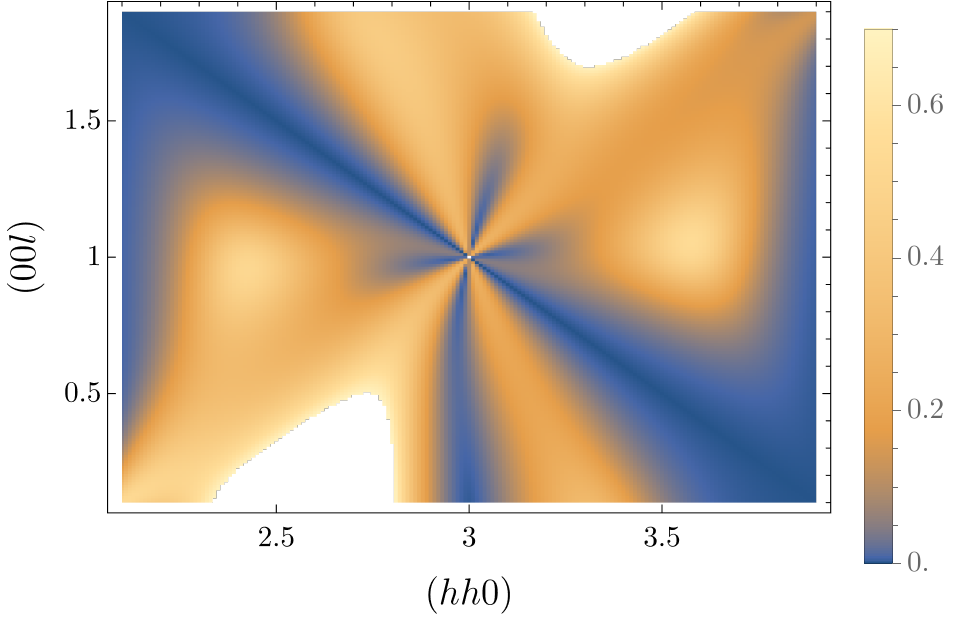}
    \hfill
    \phantom{a}
    \caption{(a) Longitudinal structure factor in the $t_{2g}$ channel in the $(hhl)$ plane computed from the projector to the flat bands for the rank-3 symmetric tensor spin liquid. The spin liquid is obtained by restricting the low-energy degrees of freedom to be those of the octupole moment on the A tetrahedra, and putting isotropic biquadratic interactions on the B tetrahedra. (b) Zoom in of a six-fold pinch point at the (331) point, confirming the rank-3 structure. }
    \label{fig:rank3_SL}
\end{figure}

The rank-3 theory is expected to exhibit 6-fold pinch points in correlation functions~\cite{premPinchPointSingularities2018}. 
This is because in the coarse-grained theory the zero-temperature two-point correlator of the rank-3 tensor field (carrying six total indices) must be zero when any index is contracted with the wave-vector, due to the Gauss law \cref{eq:rank-3-gauss}. 
As in the dipolar rank-2 spin liquid~\cite{yanRank2U1Spin2020}, the multifold pinch points do not appear in the isotropic quadrupolar structure factor. 
Instead it is necessary to study a structure factor in which the quadrupole moment indices are contracted with a wave-vector-dependent form factor. 
In \cref{fig:rank3_SL}(a) we show a structure factor for this model computed by the Luttinger-Tisza method (i.e. from the projector to the flat bands). 
The cross section shown is the longitudinal projection in the $t_{2g}$ sector,
\begin{equation}
    S_{t_{2g}}^{\text{long}}(\bq) = \sum_{\mu\nu} \sum_{a=1}^3 \hat{q}^a \langle Q_\mu^a(-\bq) Q_\nu^b(\bq)\rangle \hat{q}^b,
\end{equation}
where the wave-vector is contracted with the $t_{2g}$ components of the quadrupole (treated as a 3-component vector). 
This exhibits a complex pattern with clearly visible two-fold and four-fold pinch points. 
More importantly, however, is the six-fold pinch point faintly visible at the (331) point.
In \cref{fig:rank3_SL}(b) we have zoomed in and changed the color scale in order to highlight the six-fold singularity. 
This confirms this to be a rank-3 symmetric tensor spin liquid, to the best of our knowledge the first such identified in an anisotropic spin model.
Obviously this same construction can be generalized to higher multipoles to generate even higher rank spin liquids.
Furthermore, we expect that most other spin liquids in the quadrupolar model should also be described by rank-3 tensor gauge theories, albeit ones with pinch lines and planes, characterized by more complex Gauss laws.

\section{Putting Back The Dipoles}
\label{sec:dipoles}

So far in this paper, we have considered only quadrupolar degrees of freedom. But in materials, whether quadrupoles are important or not, there are certainly dipolar moments. Here, we consider the possible couplings between the quadrupolar order parameters and their dipolar counterparts. 

As a general rule, the minimal coupling between quadrupolar $\Phi^A_Q$ and  dipolar $\Phi^a_D$ order parameters takes the schematic form $\Phi^A_Q \Phi_D^a \Phi_D^b$ to respect time reversal symmetry. These order parameters may belong to different irreps. In such cases, one refers to $\Phi_Q$ as primary and $\Phi_D$ as secondary. In situations where there is a phase transition into a quadrupolar ordered phase, this may drive the on-set of dipolar order.  

We take a simple example specializing now to the pyrochlore lattice. We know that all-in/all-out order corresponds to the $A_{2g}$ irrep and the analog quadrupolar order transforms trivially (as $A_{1g}$). These may couple through $Q_{A_{1g}} D_{A_{2g}}^2$. Thus all-in/all-out dipolar order may naturally accompany the $A_{1g}$ quadrupolar order although they transform differently---the dipolar order breaking time reversal symmetry and switching sign under transformations that map ``up" to ``down" tetrahedra whereas the quadrupolar order breaks no lattice symmetries and preserves time reversal. 

A more interesting example is connected to the mixed dipolar-quadrupolar character of non-Kramers doublets at the single ion level. The dipolar matrix elements of such doublets may be interpreted as belonging to an Ising spin (transverse components being zero). This fixes possible dipolar long-range order to irreps $A_{2g}$ for AIAO order and $T_{1g}$ for spin ice states. If the dipolar irrep is $A_{2g}$ then cubic terms pin the multipolar order to $A_{1g}$. 

However, if the dipolar sector condenses in the ice manifold then, since $T_{1g}\otimes T_{1g} = A_{1g}\oplus E_{g} \oplus T_{1g} \oplus T_{2g}$, all quadrupolar phases may couple to these states. The $A_{1g}$ quadrupolar phase simply couples to the uniform $\boldsymbol{\Phi}_{D}^{T_{1g}}\cdot \boldsymbol{\Phi}_{D}^{T_{1g}}$. Considering next the $E_g$ quadrupoles, a calculation of the components gives 
\begin{align}
F_{E-{\rm Ice}} = & \frac{1}{2}\left( - \Phi^{E1}_Q - \sqrt{3} \Phi^{E2}_Q \right) \left( \Phi^{T_1 1}_D \right)^2 + \Phi^{E1}_Q \left( \Phi^{T_1 2}_D \right)^2 \nonumber \\ &  + \frac{1}{2}\left( - \Phi^{E1}_Q  + \sqrt{3} \Phi^{E2}_Q \right) \left( \Phi^{T_1 3}_D \right)^2 
\end{align}
which correlates one of three discrete quadrupolar $E$ states to the ice configuration on a single tetrahedron. Here the three components in the dipolar $T_{1g}$ basis correspond to three orthogonal ice states on a tetrahedron.  

Although a cubic coupling between $T_{1g}$ quadrupolar states and dipolar ice states is allowed by symmetry, the precise form of the term is antisymmetric in the dipolar terms and therefore vanishes. Finally, the coupling between $T_{2g}$ quadrupolar states and the ice states takes the following form
\begin{align}
F_{T_{2g}-{\rm Ice}} = \Phi^{T_{2g}1}_Q  \Phi^{T_{1}2}_D \Phi^{T_{1}3}_D + \Phi^{T_{2g}2}_Q  \Phi^{T_{1}1}_D \Phi^{T_{1}2}_D - \Phi^{T_{2g}3}_Q  \Phi^{T_{1}1}_D \Phi^{T_{1}3}_D.
\end{align}
This also vanishes within a uniform dipolar $T_{1g}$ state. 

The interpretation of these terms is subtle within a quasi-degenerate spin ice regime. But long-range quadrupolar order in any of the $A_{1g}$, $E_{g}$ and $T_{2g}$ irreps couples to the ice states and at least $A_{1g}$ long-range quadrupolar order may apparently emerge from a disordered ice background.


\section{Materials}
\label{sec:materials}

The rare earth pyrochlores R$_2$M$_2$O$_7$ with rare earth magnetic ion, $R^{3+}$, and non-magnetic $M^{4+}$ exhibit a very wide range of magnetic phenomena. It is helpful to divide these materials into three classes depending on the nature of the single ion ground state: (i) the dipolar Kramers ions including those with crystal field ground state doublets whose effective spin one-half description has all three effective spin components transforming as dipoles, (ii) the non-Kramers ions with ground state doublets where one effective spin component is dipolar and the transverse components are quadupolar and (iii) the Kramers ions whose crystal field ground state doublets have mixed dipolar-octupolar character. As examples, class (i) includes the Yb$^{3+}$ and Er$^{3+}$ pyrochlores that are relatively well understood, class (ii) includes Pr$^{3+}$ and Tb$^{3+}$ while class (iii) includes Ce$^{3+}$ and Nd$^{3+}$. Broadly, the least understood of the pyrochlore materials are those with mixed character crystal field ground state doublets. We focus on the non-Kramers case, (ii), as these are quadrupolar active. 


Tb$_2$Ti$_2$O$_7$ is one of the most remarkable of the rare earth pyrochlore magnets. Upon cooling through the Curie scale, there is a Schottky anomaly. At lower temperatures, the thermodynamic properties depend sensitively on the degree to which titanium and terbium ions mix on the two pyrochlore sublattices. In the materials closest to stoichiometric, there is no clear sign of a phase transition from susceptibility and heat capacity. Above some small degree of stuffing Tb$_{2-x}$Ti$_{2+x}$O$_7$ for $x>0.0025$, there is a relatively sharp thermodynamic anomaly that has been attributed to the onset of quadrupolar long-range order~\cite{takatsuQuadrupoleOrderFrustrated2016,kadowakiCompositeSpinQuadrupole2015}. This picture is supported by experiments revealing the presence of a dispersive mode significantly below the energy of the lowest-lying crystal field excitations~\cite{guittenyAnisotropicPropagatingExcitations2013}. The latter feature is also shared by relatives Tb$_2$Sn$_2$O$_7$~\cite{petitSpinDynamicsOrdered2012} and Tb$_2$Hf$_2$O$_7$~\cite{sibilleCoulombSpinLiquid2017} of Tb$_2$Ti$_2$O$_7$. In this context, we mention also Tb$_2$Ir$_2$O$_7$ in which the iridium sublattice is magnetic and long-range ordered at low temperatures. Again there is an unusual inelastic signal in this material.

The magnetic Tb$^{3+}$ has $J=6$ and is therefore non-Kramers. The crystal field spectrum is 
separated from the first excited doublet by a gap of about $15$ K which is atypically small among this class of materials though still significantly larger than the exchange and dipolar scales $\sim 1 $K. We consider the action of dipolar and quadrupolar operators on the two lowest energy doublets. There are several works exploring the crystal field of Tb$_2$Ti$_2$O$_7$ both from first principles and by fitting the experimentally measured levels. To be concrete we consider the set of parameters from Zhang {\it et al.}~\cite{zhangNeutronSpectroscopicStudy2014} 
\begin{align}
& B_2^0 = -8.41835 \nonumber \\
& B_4^0 = 0.0476754 \nonumber \\
& B_4^3 = 0.681092 \nonumber \\
& B_6^0 = -0.000139008 \nonumber \\
& B_6^3 = -0.00581111 \nonumber \\
& B_6^6 = -0.00987507
\end{align} 
in Kelvin with crystal field Hamiltonian
\begin{equation}
H_{\rm CEF} = B_2^0 O_2^0 + B_4^0 O_4^0 + B_4^3 O_4^3 + B_6^0 O_6^0 + B_6^3 O_6^3 + B_6^6 O_6^6 
\end{equation}
where $O_{l}^{m}$ are the Stevens operators. The spectrum has two low-lying doublets separated by an energy of about $16$ K. The eigenstates of the four low-lying levels $n=1,\ldots,4$
\begin{equation}
\vert \psi_n \rangle = \sum_{m=-6}^{6} \alpha_n^m \vert  m \rangle 
\end{equation}
have nonzero amplitude in the states $\vert  m \rangle$ for $m=\pm 5, \pm 4, \pm 2, \pm 1$. The ground state doublet is principally $\vert G, \pm \rangle=\vert \pm 5 \rangle$ and the first excited doublet is  $\vert E, \pm \rangle = \vert \pm 4 \rangle$. As $\vert G, \pm \rangle$ and $\vert E, \pm \rangle$ are non-Kramers doublets they are Ising-like meaning that there are non-vanishing $J^z$ matrix elements $\langle p, \pm \vert J^z \vert p,\pm \rangle = \pm \alpha_p$ all other dipolar matrix elements equal to zero. There are dipolar matrix elements connecting the two doublets: in particular $\langle G, \pm \vert J^z \vert E,\mp \rangle$ and $\langle G, \pm \vert J^{x/y} \vert E,\pm \rangle$. So, if we wish to consider the lowest four states as belonging to an effective spin three-half $S^{\alpha}$ then there has to be term that both splits the spectra that does not commute with effective $S^z$. 

Matrix elements of the quadrupolar operators can introduce dynamics even into the ground states as all of $O_2^{2s}$, $O_2^{2c}$ and $O_2^{1s}$, $O_2^{1c}$ have off-diagonal matrix elements between $\vert p,\pm \rangle$ for fixed $p$ while $O_2^0$ is diagonal in this basis. 
In an effective spin one-half description, the effective transverse spin components encode quadrupolar components while the effective $z$ components are dipolar and hence Ising-like. When directly projected onto the non-Kramers ground state doublet the nine quadrupolar couplings are folded onto three exchange couplings. 

Magnetic interactions introduce a weak admixture of the two lowest-lying doublets. In addition to the aforementioned dipolar matrix elements connecting the doublets, all quadrupolar operators couple them \cite{hallasIntertwinedMagneticDipolar2020}. 

The pyrochlore magnets based on praseodymium include Pr$_2$M$_2$O$_7$ with M$=$Hf, Sn and Zr. Pr$^{3+}$ is non-Kramers with $J=4$. In these materials, the excited crystal field levels are well separated from the ground state doublet \cite{princepCrystalFieldStates2015,sibilleCandidateQuantumSpin2016,kimuraQuantumFluctuationsSpinicelike2013} so an effective spin one-half model is fully justified at low energies.
In this case also, matrix elements of the quadrupolar operators can introduce dynamics even into the ground doublet \cite{Martin2017}. 

Structural disorder is of particular significance in the non-Kramers cases as doublets are not protected by time reversal symmetry and are therefore sensitive to time reversal invariant perturbations including all varieties of structural disorder. The zirconates and hafnates are instances of such effects where static and dynamic disorder appear to play a crucial role in the low energy physics although interactions between the moments are significant \cite{sibilleCoulombSpinLiquid2017,Martin2017,Wen2017}. A simple picture within the doublet is of disorder modifying the quadrupolar terms at the single ion level which amounts to the appearance of transverse fields in an effective spin one-half model. The correct description of these systems therefore most likely involves the interplay of magnetic interactions with disordered transverse fields. The quadrupolar activity of these praseodymium-containing pyrochlore is borne out again by the appearance of low energy spectral weight in inelastic neutron scattering studies.

\section{Discussion and Conclusion}
\label{sec:discussion}

This work combines two central ingredients in modern frustrated magnetism---anisotropic couplings induced by spin-orbit coupling and lattices with geometrical frustration---but instead of magnetic moments we couple quadrupoles. In particular, we studied a model with pure quadrupolar degrees of freedom on a pyrochlore lattices: their symmetry-allowed couplings, the semi-classical ground states including order-by-disorder selection and classical quadrupolar liquids and their finite temperature stability. We highlight some findings.

Among our main results $-$ going beyond our investigation of the pyrochlore lattice problem $-$ is a characterization of the nature of quadrupoles for different spins. Although quadrupolar degrees of freedom are present for all $S\geq 1$, different spins have different constraints on the possible expectation values. A naive counting tells us that $S\geq 1$ has the full number of quadrupolar degrees of freedom ($4$ continuous parameters) albeit these may be tied to other multipolar components. However, the local spins may not be able to span to full space of quadrupolar states. $S=1$ is particularly constrained such that the purely biaxial configurations are forbidden. This profoundly affects the semi-classical phase diagram for this spin by mixing irreps. One might imagine that large values of the spin would have more freedom to explore the states. However, it is spin $3/2$ that is the most isotropic and therefore has the greatest freedom to explore the quadrupolar states. $S=2$ ground states are also free to explore the semi-classical states. Higher spins are biased towards uniaxial nematic states. 

In the pyrochlore model, although there are nine couplings - a forbiddingly large parameter space - there are great simplifications as (i) the dipolar easy plane phases are in some sense doubled and (ii)  this feature carries over to the model which can be thought of as two (generically coupled) dipolar copies. For this reason we can find semi-classical phase diagrams much like those of the dipolar model.

Aside from the semi-classical spin dependence of the ground states and the increased number of ground states, one major difference between the dipolar and quadrupolar models is the nature of order-by-disorder. First there are two $E$ irreps instead of one for the dipolar case. Secondly, the crystalline symmetries and the time reversal symmetry of the quadrupoles allows three-fold selection in the $E$ irreps. This is the generic behavior observed in both flavor wave calculations (quantum fluctuations) and Monte Carlo (thermal fluctuations). The differing spin lengths enter the story again here. For example, $S=1$ generally does not exhibit the semi-classical accidental degeneracy that is a pre-requisite for order-by-disorder. Instead there is direct energetic selection at the mean field level for this spin. The same is typically the case for $S\geq 5/2$. $S=3/2$ is, once again, special by having typical six-fold selection similar to the dipolar case.

The dipolar model has a varied set of classical spin liquids. These include the Ising ice regime, the Heisenberg and pseudo-Heisenberg points, and higher rank spin liquids. The quadrupolar model has  quadrupolar liquid analogs of all of these states and more and, in particular, we have demonstrated the presence of new physics in the form of a rank three quadrupolar liquid phase. 

We have addressed some of the possible couplings between the quadrupolar order phases and the better known dipolar phases on the pyrochlore lattice. We have brought the analysis to the point where a natural next step is to bring the connection to materials closer. The principal target is to understand Tb$_2$Ti$_2$O$_7$ in which quadrupolar order has been suggested when there is small terbium-titanium site mixing. If this is true the nature of the order must fit into the scheme developed here. Indeed, the whole family of terbium pyrochlores exhibits somewhat similar physics where the quadrupolar degrees of freedom are active and the crystal field gap is quite small. We expect this work to connect directly to these materials. 

%

In this connection it would be interesting to combine  dipolar and  quadrupolar couplings. We have partially considered this problem at the level of symmetry. The isotropic spin one billinear-biquadratic model on the pyrochlore lattice has been considered in Ref.~\cite{pohleAbundanceSpinLiquids2025} revealing the presence of many liquid phases. This work on the isotropic case highlights the possibility of dipolar long-range order even in the disordered and purely quadrupolar coupled regime as well as hinting at considerable additional richness in passing over to the spin-orbit coupled case. 

More broadly, our work is one of few examples where the effects of geometrical frustration and spin-orbit coupling have been explored with non-magnetic local degrees of freedom. As we have seen, the space of possibilities compared to traditional frustrated magnetism is considerably enlarged.

\begin{acknowledgments}
We thank Karlo Penc and Jeff Rau for useful discussions. P.M. acknowledges funding from the CNRS. This work was in part supported by the Deutsche Forschungsgemeinschaft  under the cluster of excellence ct.qmat (EXC-2147, project number 390858490).
\end{acknowledgments}

\bibliography{PyrochloreQuadrupoles,refs}

\appendix

\section{Conventions}

The primitive unit cell is an FCC lattice with 4-site basis, with primitive Bravais lattice vectors
\begin{equation}
    \bm{a}_1 = \frac{a_0}{2}[110], \quad
    \bm{a}_2 = \frac{a_0}{2}[101], \quad
    \bm{a}_3 = \frac{a_0}{2}[011],
\end{equation}
where $a_0$ is the edge length of the cubic conventional cell. 
Each tetrahedron has six corners whose easy-axis directions are labeled as
\begin{equation}
    \hat{\bm{z}}_1 = [\bar{1}\bar{1}\bar{1}],\quad 
    \hat{\bm{z}}_2 = [11\bar{1}],\quad 
    \hat{\bm{z}}_3 = [\bar{1}11],\quad 
    \hat{\bm{z}}_4 = [1\bar{1}1],\quad 
    \label{eq:sublattice_convention}
\end{equation}
The sublattice basis positions (tetrahedron corners) are given by the vectors
\begin{equation}
    \bm{c}_\mu = \frac{\hat{\bm{a}}_{\mu-1}}{2}, \quad (\mu=1\ldots 4).
\end{equation}
where $\bm{a}_0 \equiv [000]$. 

\section{Mapping Between Parameterizations}
\label{apx:parameter_maps}

We have provided three different parameterizations of the quadrupole-quadrupole interactions, \cref{eq:interaction_matrix_Jglo} in the global frame, and 
\cref{eq:interaction_matrix_Kloc,eq:interaction_matrix_Jloc} in the local frame. 
We define the following nine-component vectors of parameters
\begin{equation}
    \vec{J} = 
    \begin{pmatrix}
        J_1 \\[.3ex]
        J_2 \\[.3ex]
        J_3 \\[.3ex]
        J_4 \\[.3ex]
        J_5 \\[.3ex]
        J_6 \\[.3ex]
        J_7 \\[.3ex]
        J_8 \\[.3ex]
        J_9 
    \end{pmatrix},
    \quad
    \vec{J}_{\pm} = 
    \begin{pmatrix}
        J_{\pm}^{t_{2g}}\\[.3ex]
        J_{\pm\pm}^{t_{2g}}\\[.3ex]
        J_{zz} \\[.3ex]
        J_{z\pm}^{t_{2g}}\\[.3ex]
        J_{\pm}^{e_g}\\[.3ex]
        J_{\pm\pm}^{e_g}\\[.3ex]
        J_{z\pm}^{e_g}\\[.3ex]
        J_{\pm}'\\[.3ex]
        J_{\pm\pm}'
    \end{pmatrix},
    \quad
    \vec{K} = 
    \begin{pmatrix}
        K_1 \\[.3ex]
        K_2 \\[.3ex]
        K_3 \\[.3ex]
        K_4 \\[.3ex]
        K_5 \\[.3ex]
        K_6 \\[.3ex]
        K_7 \\[.3ex]
        K_8 \\[.3ex]
        K_9 
    \end{pmatrix},
    \quad
    \vec{K}_{\pm} = 
    \begin{pmatrix}
        K_{\pm}^{1}\\[.3ex]
        K_{\pm\pm}^{1}\\[.3ex]
        K_{zz} \\[.3ex]
        K_{z\pm}^{1}\\[.3ex]
        K_{\pm}^{2} \\[.3ex]
        K_{\pm\pm}^{2}\\[.3ex]
        K_{z\pm}^{2}\\[.3ex]
        K_{\pm}'\\[.3ex]
        K_{\pm\pm}'
    \end{pmatrix}.
\end{equation}
These are related by invertible linear transformations such as 
\begin{equation}
    J_n = \sum_{m=1}^9 [M_{J\leftarrow J_{\pm}}]_{nm} [J_{\pm}]_m\,,
\end{equation}
with
\begin{equation}
    M_{J\leftarrow J_{\pm}} = \begin{psmallmatrix}
         \frac{4}{3} & \frac{2}{3} & -\frac{1}{3} & \frac{2 \sqrt{2}}{3} & 0 & 0 & 0 & 0 & 0 \\
         -\frac{4}{3} & \frac{4}{3} & \frac{1}{3} & \frac{4 \sqrt{2}}{3} & 0 & 0 & 0 & 0 & 0 \\
         -\frac{2}{3} & -\frac{4}{3} & -\frac{1}{3} & \frac{2 \sqrt{2}}{3} & 0 & 0 & 0 & 0 & 0 \\
         -\frac{2}{3} & \frac{2}{3} & -\frac{1}{3} & -\frac{\sqrt{2}}{3} & 0 & 0 & 0 & 0 & 0 \\
         0 & 0 & 0 & 0 & 0 & 0 & -1 & -\sqrt{2} & -\sqrt{2} \\
         0 & 0 & 0 & 0 & 0 & 0 & -1 & 0 & \sqrt{2} \\
         0 & 0 & 0 & 0 & 0 & 0 & 0 & -\sqrt{2} & \sqrt{2} \\
         0 & 0 & 0 & 0 & -2 & 2 & 0 & 0 & 0 \\
         0 & 0 & 0 & 0 & -2 & -2 & 0 & 0 & 0 
    \end{psmallmatrix}
    \,,
\end{equation}
\begin{equation}
    M_{J\leftarrow K}
    =
    \begin{psmallmatrix}
         -\frac{2}{9} & \frac{4}{9 \sqrt{3}} & -\frac{4 \sqrt{2}}{9} & \frac{2 \sqrt{2}}{9} & -\frac{8 \sqrt{\frac{2}{3}}}{9} &
           -\frac{4}{9} & \frac{4}{9} & -\frac{8}{9 \sqrt{3}} & -\frac{1}{3} \\
         \frac{2}{9} & \frac{2}{9 \sqrt{3}} & \frac{4 \sqrt{2}}{9} & \frac{4 \sqrt{2}}{9} & -\frac{4 \sqrt{\frac{2}{3}}}{9} &
           \frac{4}{9} & \frac{8}{9} & -\frac{4}{9 \sqrt{3}} & \frac{1}{3} \\
         \frac{1}{9} & -\frac{5}{9 \sqrt{3}} & \frac{2 \sqrt{2}}{9} & \frac{2 \sqrt{2}}{9} & \frac{10 \sqrt{\frac{2}{3}}}{9} &
           \frac{2}{9} & \frac{4}{9} & \frac{10}{9 \sqrt{3}} & -\frac{1}{3} \\
         \frac{1}{9} & \frac{1}{9 \sqrt{3}} & \frac{2 \sqrt{2}}{9} & -\frac{\sqrt{2}}{9} & -\frac{2 \sqrt{\frac{2}{3}}}{9} &
           \frac{2}{9} & -\frac{2}{9} & -\frac{2}{9 \sqrt{3}} & -\frac{1}{3} \\
         \frac{1}{3} & \frac{1}{3 \sqrt{3}} & \frac{1}{3 \sqrt{2}} & \frac{\sqrt{2}}{3} & -\frac{1}{3 \sqrt{6}} & -\frac{1}{3} &
           -\frac{1}{3} & \frac{1}{3 \sqrt{3}} & 0 \\
         0 & -\frac{2}{3 \sqrt{3}} & 0 & \frac{\sqrt{2}}{3} & \frac{\sqrt{\frac{2}{3}}}{3} & 0 & -\frac{1}{3} & -\frac{2}{3 \sqrt{3}} &
           0 \\
         \frac{1}{3} & -\frac{1}{\sqrt{3}} & \frac{1}{3 \sqrt{2}} & 0 & \frac{1}{\sqrt{6}} & -\frac{1}{3} & 0 & -\frac{1}{\sqrt{3}} & 0
           \\
         \frac{2}{3} & \frac{2}{3 \sqrt{3}} & -\frac{2 \sqrt{2}}{3} & 0 & \frac{2 \sqrt{\frac{2}{3}}}{3} & \frac{1}{3} & 0 &
           -\frac{1}{3 \sqrt{3}} & 0 \\
         \frac{2}{3} & -\frac{2}{\sqrt{3}} & -\frac{2 \sqrt{2}}{3} & 0 & -2 \sqrt{\frac{2}{3}} & \frac{1}{3} & 0 & \frac{1}{\sqrt{3}} &
           0
    \end{psmallmatrix}
    \,,
\end{equation}
\begin{equation}
    M_{J\leftarrow K_\pm}
    =
    \begin{psmallmatrix}
        \frac{4}{9} & \frac{2}{9} & -\frac{1}{3} & -\frac{2 \sqrt{\frac{2}{3}}}{3} & \frac{8}{9} & \frac{4}{9} & \frac{4}{3 \sqrt{3}} &
           \frac{8 \sqrt{2}}{9} & -\frac{4 \sqrt{2}}{9} \\
         -\frac{4}{9} & \frac{4}{9} & \frac{1}{3} & -\frac{4 \sqrt{\frac{2}{3}}}{3} & -\frac{8}{9} & \frac{8}{9} & \frac{8}{3 \sqrt{3}} &
           -\frac{8 \sqrt{2}}{9} & -\frac{8 \sqrt{2}}{9} \\
         -\frac{2}{9} & -\frac{4}{9} & -\frac{1}{3} & -\frac{2 \sqrt{\frac{2}{3}}}{3} & -\frac{4}{9} & -\frac{8}{9} & \frac{4}{3 \sqrt{3}}
           & -\frac{4 \sqrt{2}}{9} & \frac{8 \sqrt{2}}{9} \\
         -\frac{2}{9} & \frac{2}{9} & -\frac{1}{3} & \frac{\sqrt{\frac{2}{3}}}{3} & -\frac{4}{9} & \frac{4}{9} & -\frac{2}{3 \sqrt{3}} &
           -\frac{4 \sqrt{2}}{9} & -\frac{4 \sqrt{2}}{9} \\
         -\frac{2}{3} & \frac{2}{3} & 0 & -\sqrt{\frac{2}{3}} & \frac{2}{3} & -\frac{2}{3} & -\frac{1}{\sqrt{3}} & -\frac{\sqrt{2}}{3} &
           -\frac{\sqrt{2}}{3} \\
         0 & -\frac{2}{3} & 0 & -\sqrt{\frac{2}{3}} & 0 & \frac{2}{3} & -\frac{1}{\sqrt{3}} & 0 & \frac{\sqrt{2}}{3} \\
         -\frac{2}{3} & -\frac{2}{3} & 0 & 0 & \frac{2}{3} & \frac{2}{3} & 0 & -\frac{\sqrt{2}}{3} & \frac{\sqrt{2}}{3} \\
         -\frac{4}{3} & \frac{4}{3} & 0 & 0 & -\frac{2}{3} & \frac{2}{3} & 0 & \frac{4 \sqrt{2}}{3} & \frac{4 \sqrt{2}}{3} \\
         -\frac{4}{3} & -\frac{4}{3} & 0 & 0 & -\frac{2}{3} & -\frac{2}{3} & 0 & \frac{4 \sqrt{2}}{3} & -\frac{4 \sqrt{2}}{3} \\
    \end{psmallmatrix}
    \,.
\end{equation}
The remaining transformations can all be obtained by, for example
\begin{equation}
    M_{K \leftarrow J_{\pm}} = M_{K\leftarrow J}M_{J\leftarrow J_{\pm}} 
    \quad\text{with}\quad
     M_{K\leftarrow J} = M_{J \leftarrow K}^{-1}.
\end{equation}

\section{Single-Ion Anisotropies}
\label{apx:single_ion_anisotropy}

For $S=1/2$ all quadrupole operators are zero.
Similarly, the symmetry-allowed single-ion dipole anisotropy term $(S_i^z)^2$ is a constant. 
For $S\geq 1$ quadrupoles can become active along with the dipolar single-ion anisotropy. 
In addition, there are single-ion quadrupole anisotropies allowed by the site symmetry group which may be non-zero.
These appear in \cref{eq:Hamiltonian-Q-5-generic} as diagonal elements $\mathcal{J}_{ii}^{ab}$.
For the pyrochlore lattice with site symmetry $D_{3d}$, these have the form in the local frame
\begin{align}
    H_{\text{SIA}} &= J_{zz}^{\text{SIA}} (Q^z)^2 
    \nonumber
    \\
    &+ J_{t_{2g},\pm}^{\text{SIA}} (Q_{t_{2g}}^+ Q_{t_{2g}}^- + Q_{t_{2g}}^- Q_{t_{2g}}^+)
    \nonumber
    \\
    &+ J_{e_{g},\pm}^{\text{SIA}} (Q_{e_g}^+ Q_{e_g}^- + Q_{e_g}^- Q_{e_g}^+)
    \nonumber
    \\
    &+J_{\text{off-diag},\pm}^{\text{SIA}} (Q_{t_{2g}}^+ Q_{e_g}^- + Q_{t_{2g}}^- Q_{e_g}^+)\,,
\end{align}
where all operators are on a single site. One of the three diagonal terms may be freely set to zero thanks to the operator identity $\Tr[Q^2] = \text{const.}$
The quantum quadrupole moment operator satisfies
\begin{align}
    &\quad 4\Tr[Q^2] = 
    \nonumber
    \\
    &\!\left([S^\alpha S^\beta + S^\beta S^\alpha] - \frac{2}{3}\vert \bm{S} \vert^2\delta^{\alpha\beta}\right)\!\left([S^\beta S^\alpha + S^\alpha S^\beta] - \frac{2}{3}\vert \bm{S} \vert^2\delta^{\beta\alpha}\right)
    \nonumber
    \\
    &\quad = \left(S^\alpha S^\beta + S^\beta S^\alpha\right)^2+ \frac{4}{9}\Tr[1]\vert \bm{S}\vert^4  - 2\cdot\frac{2}{3}\vert \bm{S}\vert^2 (\vert \bm{S}\vert^2 + \vert \bm{S}\vert^2)
    \nonumber
    \\
    &\quad = 2 \vert \bm{S} \vert^4 + \frac{4}{3}\vert \bm{S}\vert^4  - \frac{8}{3}\vert \bm{S} \vert^4+ 2(S^\alpha S^\beta S^\alpha S^\beta)
    \nonumber
    \\
    &\quad =  \frac{2}{3}\vert \bm{S}\vert^4  + 2(\vert \bm{S}\vert^4 -  \hbar \vert \bm{S}\vert^2)
    \nonumber
    \\
    &\quad =  \frac{8}{3} \vert \bm{S} \vert^4 -2 \hbar \vert \bm{S} \vert^2
    \nonumber
    \\
    &\quad =  \left(\frac{8}{3} [S(S+1)]^2 -2 [S(S+1)]\right)\hbar^2,
\end{align}
where 
\begin{align}
     S^\alpha S^\beta S^\alpha S^\beta &=S^\alpha S^\beta (S^\beta S^\alpha + i \hbar \epsilon_{\alpha\beta\gamma}S^\gamma)
    \nonumber
    \\
    &=  \vert \bm{S}\vert^4 + i \hbar \epsilon_{\alpha\beta\gamma}S^\alpha S^\beta S^\gamma
    \nonumber
    \\
    &=  \vert \bm{S}\vert^4 - \hbar^2 \vert \bm{S}\vert^2,
\end{align}
using the relation
\begin{align}
    \epsilon_{\alpha\beta\gamma} S^\alpha S^\beta S^\gamma &= S^x [S^y,S^z] + S^y [S^z,S^x] + S^z [S^x,S^y]
    \nonumber
    \\
    &=i\hbar \vert \bm{S} \vert^2.
\end{align}

\section{Local Frame Irrep Basis}
\label{apx:irrepslocal}

Here we provide a local frame basis for the irreducible representations which separates the $\vert m \vert = 1$ and $\vert m \vert = 2$ sectors, similar to how the basis provided in \cref{tab:irrep_table_t2g_eg_basis} separated instead the $t_{2g}$ and $e_g$ sectors.  
In terms of the local frame spin operators (where bars indicate local components) we use the notation
\begin{align}
    (\vert m \vert = 2) \qquad Q^1 & \equiv  \left( (S^{\bar{x}})^2 - (S^{\bar{y}})^2 \right) 
    \nonumber 
    \\
    (\vert m \vert = 0) \qquad Q^2 & \equiv \frac{1}{\sqrt{3}} \left( 2(S^{\bar{z}})^2 - (S^{\bar{x}})^2  - (S^{\bar{y}})^2 \right) 
    \nonumber 
    \\
    (\vert m \vert = 2) \qquad Q^3 & \equiv  \left(  S^{\bar{x}} S^{\bar{y}} + S^{\bar{y}} S^{\bar{x}} \right)   
    \nonumber
    \\
    (\vert m \vert = 1) \qquad Q^4 & \equiv  \left(  S^{\bar{x}} S^{\bar{z}} + S^{\bar{z}} S^{\bar{x}} \right)   
    \nonumber 
    \\
    (\vert m \vert = 1) \qquad Q^5 & \equiv  \left( S^{\bar{y}} S^{\bar{z}} + S^{\bar{z}} S^{\bar{y}}  \right)\,,
    \label{eq:quad_local_cubic_harmonic_basis}
\end{align}
which are the cubic harmonic equivalents (real and imaginary parts) of the spherical harmonic basis defined in \cref{eq:local_basis_raising_lowering}.
We begin with the $A_{1}$ irrep, for which the sole operator is given by
\begin{equation}
O_{A_{1}} = Q^2_1 + Q^2_2 + Q^2_3 + Q^2_4 \,,
\end{equation}
where the lower index is the sublattice index. 
The $E$ irrep multiplet can be nicely separated into two components which we label $A$ ($\vert m \vert = 2$) and $B$ ($\vert m \vert = 1$), 
\begin{align}
O_{E_A}^1 &=  Q^1_1 + Q^1_2 + Q^1_3 + Q^1_4 \nonumber\\
O_{E_A}^2 &=  Q^3_1 + Q^3_2 + Q^3_3 + Q^3_4 \nonumber\\
O_{E_B}^1 &=  Q^4_1 + Q^4_2 + Q^4_3 + Q^4_4 \nonumber\\
O_{E_B}^2 &=  Q^5_1 + Q^5_2 + Q^5_3 + Q^5_4 \,.
\end{align}
which is equivalent to \cref{eq:E_irreps_local}.
A similar local basis for the $T_{1}$ multiplets separating the $m = \pm$ and $m=\pm 2$ components is given by
\begin{align}
O_{T_1,A}^{1} &= Q^3_1 - Q^3_2 - Q^3_3 + Q^3_4 \nonumber\\
O_{T_1,A}^{2} &= 
 \frac{\sqrt{3}}{2} Q^1_1 
+ \frac{1}{2} Q^3_1 
+ \frac{\sqrt{3}}{2} Q^1_2 
+ \frac{1}{2} Q^3_2 
\nonumber\\
&- \frac{\sqrt{3}}{2} Q^1_3 
 - \frac{1}{2} Q^3_3 
 - \frac{\sqrt{3}}{2} Q^1_4 
 - \frac{1}{2} Q^3_4  
\nonumber\\
O_{T_1,A}^{3} &= 
  \frac{\sqrt{3}}{2} Q^1_1 
- \frac{1}{2} Q^3_1 
- \frac{\sqrt{3}}{2} Q^1_2 
+ \frac{1}{2} Q^3_2 
\nonumber\\
&+ \frac{\sqrt{3}}{2} Q^1_3 
 - \frac{1}{2} Q^3_3 
 - \frac{\sqrt{3}}{2} Q^1_4 
 + \frac{1}{2} Q^3_4 \,,
\nonumber\\[2ex]
O_{T_1,B}^{1} 
&= 
-Q^5_1 + Q^5_2 + Q^5_3 - Q^5_4 
\nonumber\\
O_{T_1,B}^{2} 
&= 
  \frac{\sqrt{3}}{2} Q^4_1 
- \frac{1}{2} Q^5_1 
+ \frac{\sqrt{3}}{2} Q^4_2 
- \frac{1}{2} Q^5_2 
\nonumber\\
&- \frac{\sqrt{3}}{2} Q^4_3 
 + \frac{1}{2} Q^5_3 
 - \frac{\sqrt{3}}{2} Q^4_4 
 + \frac{1}{2} Q^5_4  
\nonumber\\
O_{T_1,B}^{3} 
&= 
\frac{\sqrt{3}}{2} Q^4_1 + \frac{1}{2} Q^5_1 - \frac{\sqrt{3}}{2} Q^4_2 - \frac{1}{2} Q^5_2  \nonumber\\
&+ \frac{\sqrt{3}}{2} Q^4_3 + \frac{1}{2} Q^5_3 - \frac{\sqrt{3}}{2} Q^4_4 - \frac{1}{2} Q^5_4 \,.
\end{align}
Similarly, a local basis for the three $T_{2}$ multiplets is
\begin{align}
O_{T_2,A}^{1} &=  
\phantom{-}
\frac{1}{2} Q^1_1 - \frac{\sqrt{3}}{2} Q^3_1 + \frac{1}{2} Q^1_2 - \frac{\sqrt{3}}{2} Q^3_2 
\nonumber\\
&\phantom{-}- 
\frac{1}{2} Q^1_3 + \frac{\sqrt{3}}{2} Q^3_3 - \frac{1}{2} Q^1_4 + \frac{\sqrt{3}}{2} Q^3_4  \nonumber\\
O_{T_2,A}^{2} &=  
-\frac{1}{2} Q^1_1 - \frac{\sqrt{3}}{2} Q^3_1 + \frac{1}{2} Q^1_2 + \frac{\sqrt{3}}{2} Q^3_2 
\nonumber\\
&\phantom{-} - 
\frac{1}{2} Q^1_3 - \frac{\sqrt{3}}{2} Q^3_3 + \frac{1}{2} Q^1_4 + \frac{\sqrt{3}}{2} Q^3_4  \nonumber\\
O_{T_2,A}^{3} &=  
\phantom{-\frac{1}{2}}
Q^1_1 - Q^1_2 - Q^1_3 + Q^1_4 
\,,
\nonumber 
\\[2ex]
O_{T_2,B}^{1} &=  
-\frac{1}{2} Q^4_1 - \frac{\sqrt{3}}{2} Q^5_1 - \frac{1}{2} Q^4_2 - \frac{\sqrt{3}}{2} Q^5_2 
\nonumber\\
&\phantom{-} + 
\frac{1}{2} Q^4_3 + \frac{\sqrt{3}}{2} Q^5_3 + \frac{1}{2} Q^4_4 + \frac{\sqrt{3}}{2} Q^5_4 \nonumber \\
O_{T_2,B}^{2} &=  
\phantom{-}
\frac{1}{2} Q^4_1 - \frac{\sqrt{3}}{2} Q^5_1 - \frac{1}{2} Q^4_2 + \frac{\sqrt{3}}{2} Q^5_2 
\nonumber\\
&\phantom{-} + 
\frac{1}{2} Q^4_3 - \frac{\sqrt{3}}{2} Q^5_3 - \frac{1}{2} Q^4_4 + \frac{\sqrt{3}}{2} Q^5_4 
\nonumber \\
O_{T_2,B}^{3} &=  \phantom{-\frac{1}{2}} Q^4_1 + Q^4_2 + Q^4_3 - Q^4_4
\,,
\nonumber 
\\[2ex]
O_{T_{2},\text{ice}}^{1} &= - Q^2_1 -  Q^2_2 + Q^2_3 + Q^2_4 \nonumber\\
O_{T_{2},\text{ice}}^{2} &=  \phantom{-}Q^2_1 - Q^2_2 + Q^2_3 - Q^2_4 \nonumber\\
O_{T_{2},\text{ice}}^{3} &=  \phantom{-}Q^2_1 - Q^2_2 - Q^2_3 + Q^2_4 
\,.
\end{align}
These bases are used for labeling the mean field phase diagrams in \cref{fig:mean_field_phase_diagram_S1_vs_S32}.

\clearpage

\end{document}